\documentclass[superscriptaddress,longbibliography,prd,aps,nofootinbib,notitlepage, preprintnumbers, showkeys]{revtex4-2}
\usepackage[utf8]{inputenc}
\usepackage{graphicx,bm,bbm}
\usepackage{slashed,verbatim}
\usepackage{amssymb,graphicx,epstopdf}
\usepackage{slashed,subfigure}
\usepackage{caption}
\usepackage{amsmath,mathtools}
\usepackage{epstopdf,dcolumn}
\usepackage{enumerate}
\allowdisplaybreaks
\usepackage[normalem]{ulem}
\usepackage{cancel}
 \usepackage{graphicx}
\usepackage{subcaption}
\usepackage[colorlinks=true,linktocpage=true,linkcolor=blue,citecolor=blue]{hyperref}
\usepackage{orcidlink} 

\usepackage{titlesec}

\titleclass{\subsubsubsection}{straight}[\subsubsection]

\newcounter{subsubsubsection}[subsubsection]
\renewcommand\thesubsubsubsection{\thesubsubsection.\arabic{subsubsubsection}}

\titleformat{\subsubsubsection}
{\normalfont\normalsize\bfseries}
{\thesubsubsubsection}{1em}{}

\titlespacing*{\subsubsubsection}
{0pt}{3.25ex plus 1ex minus .2ex}{1.5ex}

\makeatletter
\newcommand\l@subsubsubsection{\@dottedtocline{4}{10em}{4em}}
\makeatother

\allowdisplaybreaks
\usepackage[displaymath, mathlines]{lineno}

\def\be {\begin{equation}}
\def\ee {\end{equation}}
\def\bea {\begin{eqnarray}}
\def\eea {\end{eqnarray}}
\def\bc {\begin{center}}
\def\ec {\end{center}}
\def\nn {\nonumber}

\def\gm {\gamma}

\def\mn {\mu\nu}

\def\({\left(}
\def\){\right)}
\def\[{\left[}
\def\]{\right]}

\newcommand \Tr{\operatorname{\text{Tr}}}

\definecolor{red}{rgb}{0.7,0,0}
\definecolor{green}{rgb}{0,0.5,0}
\newcommand{\revision}[1]{\textcolor{black}{#1}}

\def\slashed{\slash\!\!\!\!}

\def\sumint{\sum \!\!\!\!\!\!\!\!\int\,}

\DeclareGraphicsExtensions{.jpg,.pdf,.eps}

\begin{document}
	
\title{Beyond leading-logarithm photon production from two-loop diagrams in a hot QCD medium}

\author{Sumit\,\orcidlink{0000-0001-7137-6433}}
\email{sumit@ph.iitr.ac.in}
\affiliation{School of Physics, Beijing Institute of Technology, Beijing 102488, China}
 	
\author{Ritesh Ghosh\footnote{Alexander von Humboldt fellow}\,\orcidlink{0000-0002-6740-7038}}
\email{riteshghosh283@gmail.com}
\affiliation{Institute of Physics, Academia Sinica, Taipei 11529, Taiwan}
 	
\author{ Munshi G. Mustafa\footnote{Raja Ramanna Chair}\,\orcidlink{0000-0002-7874-6932}}
\email{munshigolam.mustafa@saha.ac.in}
\affiliation{Department of Physics, Murshidabad Maharaja Krishnanath University, Berhampore 742101, India}

\begin{abstract}
We investigate high-energy photon production from $2\leftrightarrow 2$ processes in a quark-gluon plasma by evaluating the imaginary part of the two-loop photon self-energies in thermal QCD. Working in the imaginary-time formalism, we derive analytical expressions for the leading-logarithmic and beyond-leading-logarithmic contributions to the photon production rate from $2\leftrightarrow 2$ processes. We retain the full Fermi--Dirac and Bose--Einstein statistical factors and compare the resulting Compton and annihilation rates with those obtained from kinetic theory and the closed-time-path formalism. The summed rates agree with the corresponding established results, providing a nontrivial consistency check of the calculation. Furthermore, when we consider the Maxwell-Boltzmann distribution for incoming particles, we find that quantum corrections are absent in the individual $2\leftrightarrow 2$ processes. However, the total rate agrees with that of full statistics. 
\end{abstract}
\keywords{Quark-gluon plasma, Thermal photons, Photon self-energy, Finite-temperature field theory, QCD}
\maketitle 
\vspace{-.2cm}
%
\section{Introduction}
\label{intro}
Electromagnetically interacting particles, including real photons~\cite{Kapusta:1991qp,Baier:1991em,Baier:1997xc,Aurenche:1998nw,Arnold:2001ba,Shuryak:1978ij,Kajantie:1981wg,Halzen:1981kz,Kajantie:1982nj,Sinha:1983jm,Hwa:1985xg,Staadt:1985uc,Neubert:1989hu,alam1996electromagnetic,Peitzmann:2001mz,Oliva:2017pri,Iatrakis:2016ugz,Hauksson:2017udm,Schafer:2019edr,vanHees:2011vb,Arnold:2001ms,Ali:2024xae,Srivastava:1996qd,Ghiglieri:2013gia} and virtual photons that subsequently decay into dileptons~\cite{Cleymans:1986na,Weldon:1990iw,Braaten:1990wp,Mustafa:1999dt,Mustafa:2002pb,Mustafa:1999cp,Peshier:1999dt,Strickland:1994rf,Aurenche:1999ec,Thoma:1997dk,Aurenche:2002pc,Aurenche:2002wq,Carrington:2007gt,Greiner:2010zg,Bandyopadhyay:2015wua,Kasmaei:2018oag,Wang:2023fst,Churchill:2023vpt}, serve as powerful probes of the thermodynamic properties of strongly interacting matter formed in ultra-relativistic nucleus-nucleus collisions~\cite{Braun-Munzinger:2015hba,Busza:2018rrf,Nagle:2018nvi}. Their significance arises from the nature of electromagnetic interactions, which are sufficiently strong to generate measurable signals yet weak enough for emitted photons and leptons to escape the hot, dense medium with minimal subsequent interactions. The emission characteristics of both real and virtual photons depend critically on the size of the thermalized system. In a sufficiently large medium, photons undergo repeated scattering and eventually attain thermal equilibrium, resulting in a momentum distribution described by Planck's law. Under such conditions, the emission follows a black-body radiation spectrum, determined solely by the temperature and surface area of the emitting source, independent of its microscopic structure. In contrast, the systems created in relativistic heavy-ion collisions are much smaller than the photon mean free path. Consequently, most photons escape the medium without significant rescattering or thermalization. The medium's microscopic dynamics therefore govern the photon production rate and are directly related to the imaginary part of the photon self-energy. As a result, the observed spectra of photons and dileptons carry valuable information about the properties and dynamics of the constituents of the hot, strongly interacting matter from which they originate.

For most practical applications, photon and dilepton production rates can be evaluated within the framework of classical kinetic theory. In this approach, the photon emission rate can be expressed in terms of the equilibration rate of photons in a thermal medium~\cite{Ruuskanen:1992hh,Gutbrod1993ParticlePI,Ruuskanen:1994sa}. The emission rate is determined by the kinetic processes of photon production and absorption, which are governed by the squared matrix elements of the relevant microscopic interactions occurring in the medium.
A more general and fundamentally quantum-mechanical description was introduced by Feinberg in Ref.~\cite{Finberg:1976}, where electromagnetic emission rates were shown to be directly related to the thermal correlation function of the electromagnetic current. This formulation provides a field-theoretical description of electromagnetic emission that does not rely on a specific microscopic scattering picture. In particular, the relation between the emission rate and the thermal current correlator is nonperturbative in principle and remains valid irrespective of the strength of the interactions among the constituents of the medium. More generally, the production rate of any weakly interacting probe propagating through a thermal medium whose constituents may interact strongly among themselves can be expressed in terms of the discontinuities, or equivalently the imaginary parts, of the probe's self-energy~\cite{Gutbrod1993ParticlePI,Weldon:1990iw,McLerran,Gale:1990pn,Kobes:1985kc,Kobes:1986za}. The explicit strength of the interactions among the medium constituents does not alter this general relationship.
\begin{figure}[htb]
\begin{center}
\includegraphics[scale=0.41]{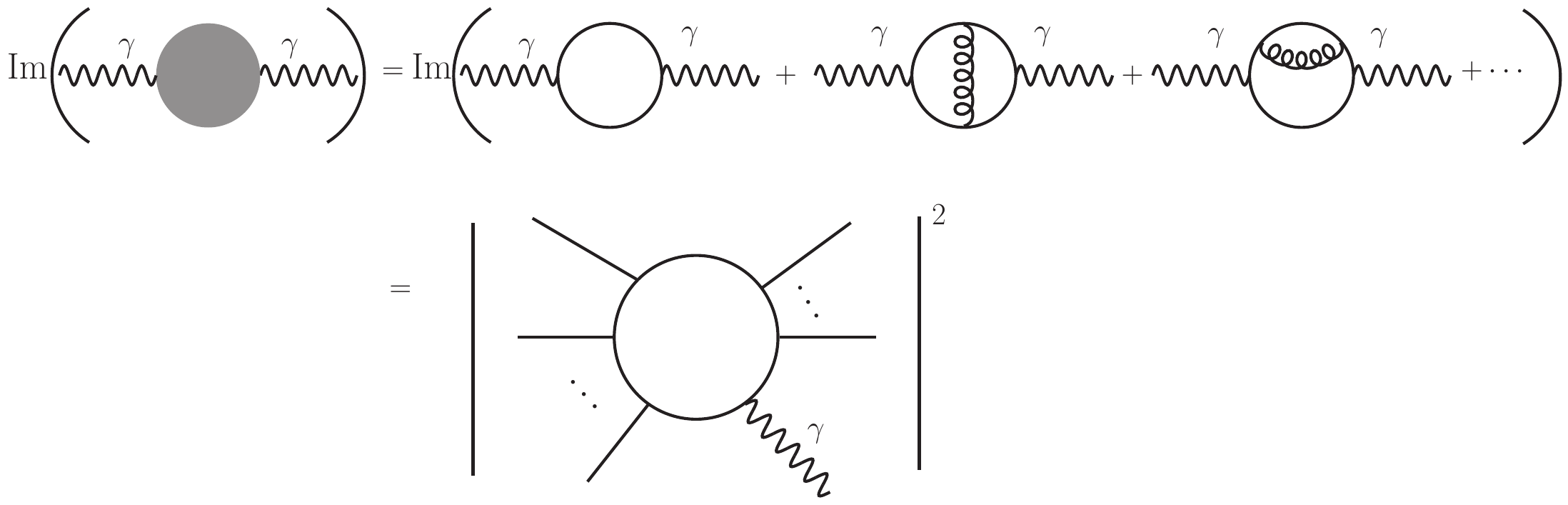}
\end{center}
\caption{Schematic representation of the optical theorem in thermal field theory.}
\label{cutkowski}
\end{figure}
The differential real photon emission rate in thermal QCD following the imaginary part of the photon self-energy is given as
\be
E\frac{dR}{d^4x d^3 \mathbf{   p}} = \frac{1 }{(2\pi)^3} \, n_B(E) \, {\Im}\Big[\Pi^{\mu}_{\mu}(E,\mathbf{ p})\Big]   , \label{ph3}
\ee
where $\Im$ denotes the imaginary part, $E$ is the photon energy, $p=|\mathbf p|$ is the magnitude of the three momentum $\mathbf{  p}$ and $n_B(E)$ is Bose-Einstein distribution.
The real photon emission rate is accurate up to ${\cal O}(e^2)$ in the electromagnetic coupling, as it neglects the rescattering of photons after their production. In contrast, it is, in principle, valid for all orders in the strong interaction, although practical calculations are restricted to a finite loop order. Therefore, determining the photon emission rate requires evaluating the imaginary part of the photon self-energy. The Cutkosky thermal cutting rules~\cite{Das:1997gg,Weldon:1983jn,Kobes:1985kc,Kobes:1986za,Gelis:1997zv} provide a systematic way to obtain this imaginary part by relating higher-order loop diagrams to lower-order physical processes. This formalism is equivalent to relativistic kinetic theory, where the emission rate is expressed in terms of microscopic production and absorption processes, incorporating thermal distribution functions, quantum statistical factors, phase-space integrals, degeneracy factors, and the corresponding squared matrix elements~\cite{Ruuskanen:1992hh,Gutbrod1993ParticlePI,Ruuskanen:1994sa,Gale:1987ki}. The relation between the imaginary part of the photon self-energy and the squared amplitudes for photon-producing processes is illustrated schematically in Fig.~\ref{cutkowski}.
The differential production rate for a particle $X$ with four momentum $P\equiv \left ( E, \bm{  p}\right )$ can be written in kinetic theory as
\bea
\frac{dR}{d^4x d^3\mathbf{  p}} &=& \frac{1}{2E(2\pi)^3} \int \frac{d^3\bm{  p}_1}{2E_1(2\pi)^3} n_1(E_1) \cdots  \frac{d^3\mathbf{  p}_m}{2E_m(2\pi)^3}  n_m(E_m)
 \int \frac{d^3{\mathbf{  p}}^{\,\, \prime}_1}{2E^\prime_1(2\pi)^3}  \left (1\pm n^\prime_1(E^\prime_1)\right ) \cdots  \frac{d^3\mathbf{  p}^{\,\,\prime}_n}{2E^\prime_n(2\pi)^3} \nonumber\\
 && \times  \left (1\pm n^\prime_n(E^\prime_n)\right )  (2\pi)^4 \delta\left(\sum_{i=1}^{m}P_i-\sum_{f=1}^{n}P^\prime_f-P\right)  \left|{\cal M}\right |^2 \, , \label{ph4}
\eea
where ${\cal M}$ is the matrix element of the basic process of $m$ particles $\rightarrow n$ particles $+X$.  $n_i(E_i)$ and $n^\prime_f(E^\prime_f )$  are  a Bose-Einstein or Fermi-Dirac distribution for initial state particle $i$ and final state particle $f$ but excluding the produced particle $X$, respectively.  $P_i$ is the four-momentum of the initial particle $i$, whereas
$P^\prime_f$ is that for final state particle $f$ except the produced particle $X$. The factor $ \left (1\pm n^\prime_f(E_f^\prime)\right ) $ is either a Bose-enhancement with $(+)$ sign or a Pauli-suppression with $(-)$ sign for each particle in the final state,  excluding the produced particle $X$.

The equilibrium real photon rate from $ 2\leftrightarrow 2$ processes has been obtained in Refs.~\cite{Kapusta:1991qp,Baier:1991em} following Eq.~\eqref{ph4} in kinetic theory. In Ref.~\cite{Baier:1997xc}, it is obtained for a chemically equilibrating QGP using Eq.~\eqref{ph3} within the closed-time-path formalism, and it reduces to the equilibrium rate when the quark and gluon fugacities saturate to unity. In this paper, our main objective is to calculate the equilibrium photon rate from $ 2\leftrightarrow 2$ processes following the imaginary-time formalism using the imaginary part of the photon self-energy as given in Eq.~\eqref{ph3} from two-loop diagrams in thermal QCD and compare the results with those obtained in kinetic theory~\cite{Kapusta:1991qp,Baier:1991em} and the closed-time-path formalism~\cite{Baier:1997xc}. We note that the two-loop diagrams in the imaginary-time formalism have previously been used in different contexts, viz., to obtain the imaginary part and to regulate the infrared divergences of two-loop vector boson self-energies semi-analytically in thermal QCD. Here, our purpose is to analytically compute the hard-photon production rate from the two-loop diagrams within the imaginary time formalism at leading logarithm (LL) and beyond leading logarithm (BLL).

The paper is organized as follows. In Sec.~\ref{qcd_2l}, we compute the imaginary part of the topologies $I$ and $II$ of two-loop photon self-energies. In Sec.~\ref{phot_rate}, we obtain photon rates for Compton and annihilation processes, and compare the rate with the kinetic theory~\cite{Kapusta:1991qp,Baier:1991em} and the closed-time-path~\cite{Baier:1997xc} results. We conclude in Sec.~\ref{conc}. A few short appendices follow. In the interest of quantitative accuracy and reproducibility, we present the detailed calculations underlying our analysis. Given the technical nature of the subject, a rigorous and transparent treatment is essential.

\section{TWO-LOOP PHOTON SELF-ENERGY IN THERMAL QCD}
\label{qcd_2l}
\subsection{Topology $I$}
\label{topo1}

\begin{figure}[t]
\begin{center}
\includegraphics[scale=0.43]{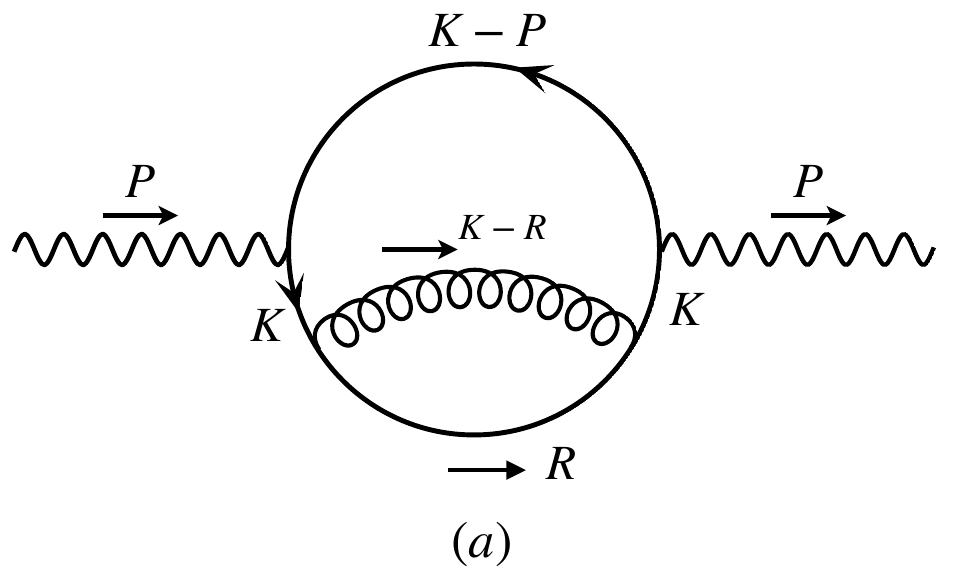}
\includegraphics[scale=0.43]{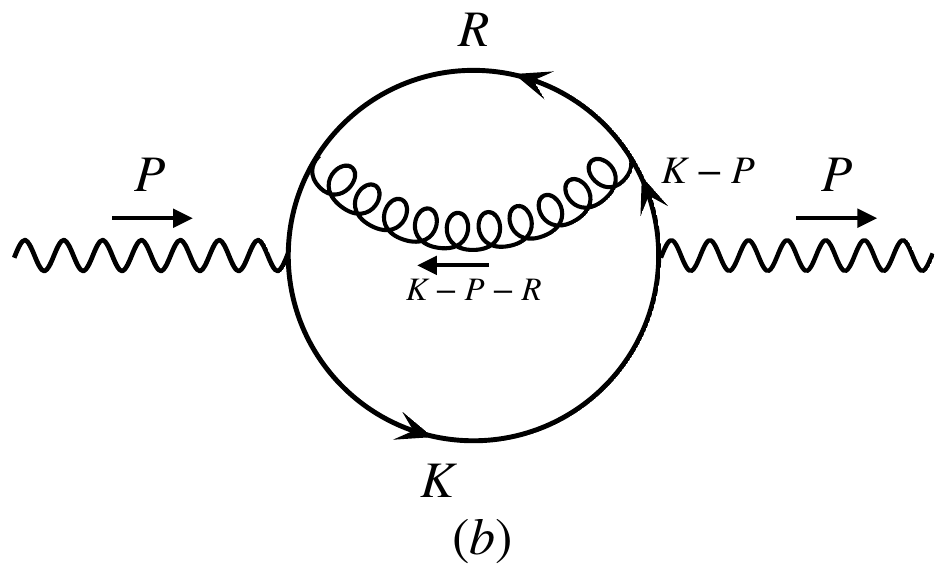}
\caption{ Two-loop Feynman diagrams contributing to the photon self-energy. The right panel (b) is obtained from the left panel (a) by interchanging the gluon insertions on the quark line.}
\label{photon_self_1}
\end{center}
\end{figure}

 We first evaluate the two-loop contribution to the photon self-energy corresponding to the left diagram shown in Fig.~\ref{photon_self_1}, in which a virtual gluon is exchanged between two points on the quark loop. 
 The photon couples to the quark line through the electromagnetic vertex $-ie\gm_\mu$, while the gluon exchange is described by the quark-gluon vertices $-ig \gm_\alpha t_a$ and the gluon propagator, where $t_a$ are the generators of the $SU(N_c)$ color group in the fundamental representation. We can write the photon self-energy as
\bea
i\Pi_{\mn}^{I}(P)& =&- \sum_f q_f^2  \int_{K}\int_{R} \Tr\bigg[(-ie\gm_\mu)S(K)\bigg\{(-ig \gm_\alpha t_a) S(R)(-i g \gm_\beta t_b)D_{ab}^{\alpha\beta}(R-K)\bigg\}S(K)(-i e \gm_\nu)S(K-P)\bigg] \nonumber \\
&=&- \frac{5}{9}  \int_{K}\int_{R} \Tr\bigg[(-ie\gm_\mu) \frac{i\slashed{K}}{K^2} (-ig \gm_\alpha t_a) \frac{i\slashed{R}}{R^2}  (-i g \gm_{\beta} t_b) \left( \frac{-i\delta_{ab}g^{\alpha\beta}}{(R-K)^2} \right)  \frac{i\slashed{K}}{K^2} (-i e \gm_{\nu}) \frac{i(\slashed{K}-\slashed{P})}{(P-K)^2}\bigg], \label{eq1}
\eea
where $D_{ab}^{\alpha\beta}(R-K)=\frac{-i\delta_{ab}g^{\alpha\beta}}{(R-K)^2}$ is free gluon propagator in the Feynman gauge, $S(X)=\frac{i}{\slashed{X}}$ is the free fermion propagator, and $\sum_f q^2_f$ denotes the sum over the squared fractional quark charges. The overall minus sign arises from the closed fermion loop. We consider two massless quark flavours, $u$ and $d$, for which $\sum_{f=u,d}q_f^2=5/9$. The corresponding fermionic sum-integrals are defined as
\bea
\hspace{-.5cm}\int_K &\equiv& T\sum_{k_0=(2n+1)\pi iT} \int_{\bm k}, \quad \int_R \equiv T\sum_{r_0=(2m+1)\pi iT} \int_{\bm r}, \quad \rm{where}, \quad \int_{\bm k} \equiv \int\frac{d^3\bm k}{(2\pi)^3}, \quad \int_{\bm r} \equiv \int\frac{d^3\bm r}{(2\pi)^3}, \quad n,m\in\mathbb{Z}.
\label{eq:sum_integral_definition}
\eea
\text
The right diagram in Fig.~\ref{photon_self_1} is the symmetric counterpart, obtained by interchanging the gluon insertions on the quark line, and gives an identical contribution~\cite{Majumder:2001iy}. We therefore evaluate the left diagram and multiply the result by
a factor of two. Contracting the Lorentz indices and collecting the coupling and color factors, the photon self-energy reduces to
\bea
\Pi^{\mu\, I}_{\mu}(P)&=& \frac{10}{9} e^2g^2  \int_{K}\int_{R} \Tr(t_a t_b) \delta_{ab} 
\frac{\Tr\big [\gm_\mu \slashed{K}  \gm_\alpha  \slashed{R}   \gm^\alpha \slashed{K} \gm^\mu(\slashed{K}-\slashed{P}) \big] }{K^4Q^2R^2(K-R)^2} \, ,\label{eq2}
\eea
where $Q=(K-P)$.
%
The color trace and the Dirac trace can be evaluated as
\begin{subequations}
\begin{align}
\Tr(t_a t_b)\delta_{ab} &= \frac{1}{2}\delta_{ab}\delta_{ab}= \frac{N_c^2-1}{2} = 4 \,\,\,\,\, {\rm {for\, \, QCD}} ,\label{eq3a} \\
\Tr\big [\gm_\mu \slashed{K}  \gm_\alpha  \slashed{R}   \gm^\alpha \slashed{K} \gm^\mu \slashed{Q} \big] &= 
16 \big [ 2 (K\cdot R) (K \cdot Q) -K^2 (Q \cdot R) \big ] \, . \label{eq3b}
\end{align}
\end{subequations}
Substituting Eq.~\eqref{eq3a} and Eq.~\eqref{eq3b} into Eq.~\eqref{eq2}, we obtain
\bea
\hspace{-.7cm}\Pi^{\mu\, I}_{\mu}(P)&=&\frac{640}{9} e^2g^2 T^{2} \sumint_{k_0} \sumint_{r_0} \frac{\big [ 2 (K\cdot R) (K \cdot Q) -K^2 (Q \cdot R) \big ] } {K^4Q^2R^2(K-R)^2} \nonumber \\
&=& \frac{640}{9} e^2g^2 T^{2} \sumint_{k_0}\sumint_{r_0} \bigg [\frac{2 \{k_0^2 q_0 r_0 -({\bf{k}}\cdot {\bf{r}})k_0q_0 - ( {\bf{k}}\cdot {\bf{q}}) k_0 r_0 +({\bf{k}}\cdot  {\bf{r}})( {\bf{k}}\cdot  {\bf{q}})\} }  {K^4Q^2R^2(K-R)^2}- \frac{r_0q_0- {\bf{r}}\cdot  {\bf{q}}}{K^2Q^2R^2(K-R)^2}\bigg]. 
\label{eq4}
\eea
In order to evaluate the Matsubara sums, we first perform the fermionic frequency sum over $r_0$ using the Saclay representation~\cite{Pisarski:1987wc}. The required mixed fermion-boson~\cite{Haque:2024gva} sum-integrals are 
\begin{subequations}
\begin{align}
T\sum_{r_0=(2n+1)\pi i T}  \Delta_F(R) \Delta_B(K-R)&= -\sum_{s_1,s_2=\pm 1}\frac{s_1s_2}{4E_rE_{k-r}}
\frac{1-n_F(s_1E_r)+n_B(s_2E_{k-r})}{k_0-s_1E_r-s_2E_{k-r}}, \label{eq5a}\\
T\sum_{r_0=(2n+1)\pi i T} r_0 \Delta_F(R) \Delta_B(K-R)&= -\sum_{s_1,s_2=\pm 1}\frac{s_2}{4E_{k-r}}
\frac{1-n_F(s_1E_r)+n_B(s_2E_{k-r})}{k_0-s_1E_r-s_2E_{k-r}} \, .\label{eq5b}
\end{align}
\end{subequations}
Here the subscripts $B$ and $F$ denote bosonic and fermionic propagators, respectively. Using Eqs.~\eqref{eq5a} and Eq.~\eqref{eq5b}, the self-energy can be written as
\bea
\Pi^{\mu\, I}_{\mu}&=&\frac{640}{9}e^2g^2 \, T \sumint_{k_0} \int \frac{d^3r}{(2\pi)^3} \bigg [ \frac{- 2k_0^2q_0}{K^4Q^2}  \frac{1}{4E_{k-r}} 
\bigg \{ \frac{1-n_F(E_r)+n_B(E_{k-r})}{k_0-E_r-E_{k-r}} +\frac{n_F(E_r)+n_B(E_{k-r})}{k_0+E_r-E_{k-r}} \nonumber \\
&+& \frac{n_F(E_r)+n_B(E_{k-r})}{k_0-E_r+E_{k-r}}+\frac{1-n_F(E_r)+n_B(E_{k-r})}{k_0+E_r+E_{k-r}}\bigg\}
+ \frac{2(  \mathbf k\cdot   \mathbf r)k_0q_0}{K^4Q^2}  \frac{1}{4E_rE_{k-r}} 
\bigg \{ \frac{1-n_F(E_r)+n_B(E_{k-r})}{k_0-E_r-E_{k-r}}  \nonumber \\
&+& \frac{n_F(E_r)+n_B(E_{k-r})}{k_0-E_r+E_{k-r}} 
- \frac{n_F(E_r)+n_B(E_{k-r})}{k_0+E_r-E_{k-r}}-\frac{1-n_F(E_r)+n_B(E_{k-r})}{k_0+E_r+E_{k-r}}\bigg\} +
 \frac{2(  \mathbf k\cdot   \mathbf q)k_0}{K^4Q^2}  \frac{1}{4E_{k-r}}  \nonumber \\ 
&\times& \bigg \{ \frac{1-n_F(E_r)+n_B(E_{k-r})}{k_0-E_r-E_{k-r}}
+\frac{n_F(E_r)+n_B(E_{k-r})}{k_0+E_r-E_{k-r}} 
+ \frac{n_F(E_r)+n_B(E_{k-r})}{k_0-E_r+E_{k-r}}+\frac{1-n_F(E_r)+n_B(E_{k-r})}{k_0+E_r+E_{k-r}}\bigg\}\nonumber \\
&-&\frac{2(  \mathbf k\cdot   \mathbf r)(  \mathbf k\cdot   \mathbf q)}{K^4Q^2}  \frac{1}{4E_rE_{k-r}}
 \bigg \{ \frac{1-n_F(E_r)+n_B(E_{k-r})}{k_0-E_r-E_{k-r}}  + \frac{n_F(E_r)+n_B(E_{k-r})}{k_0-E_r+E_{k-r}} 
- \frac{n_F(E_r)+n_B(E_{k-r})}{k_0+E_r-E_{k-r}}\nonumber \\
&-& \frac{1-n_F(E_r) +n_B(E_{k-r})}{k_0+E_r+E_{k-r}}\bigg\}
+ \frac{q_0}{K^2Q^2}  \frac{1}{4E_{k-r}} 
\bigg \{ \frac{1-n_F(E_r)+n_B(E_{k-r})}{k_0-E_r-E_{k-r}} +\frac{n_F(E_r)+n_B(E_{k-r})}{k_0+E_r-E_{k-r}} \nonumber \\
&+& \frac{n_F(E_r)+n_B(E_{k-r})}{k_0-E_r+E_{k-r}}+\frac{1-n_F(E_r)+n_B(E_{k-r})}{k_0+E_r+E_{k-r}}\bigg\} 
-\frac{(  \mathbf r\cdot   \mathbf q)}{K^2Q^2}  \frac{1}{4E_rE_{k-r}} \bigg \{ \frac{1-n_F(E_r)+n_B(E_{k-r})}{k_0-E_r-E_{k-r}}  \nonumber \\
& +& \frac{n_F(E_r)+n_B(E_{k-r})}{k_0-E_r+E_{k-r}} 
- \frac{n_F(E_r)+n_B(E_{k-r})}{k_0+E_r-E_{k-r}}
- \frac{1-n_F(E_r) +n_B(E_{k-r})}{k_0+E_r+E_{k-r}}\bigg\}
\bigg]. \label{eq6}
\eea

We evaluate the remaining fermionic Matsubara sum in Eq.~\eqref{eq6} using the contour method described in Appendix~\ref{matsubara_I}. We discard the terms containing delta functions or their derivatives whose arguments do not correspond to the kinematic conditions for the Compton-scattering or annihilation processes considered here. Substituting Eqs.~\eqref{eq12a}--\eqref{eq12r} from Appendix~\ref{imaginary_I}, we obtain the imaginary part of the photon self-energy as
\bea
{{\Im}}\Pi^{\mu\, I}_{\mu}  &=&  -  \frac{640\pi}{9}e^2g^2 \int\frac{d^3k}{(2\pi)^3}\int\frac{d^3r}{(2\pi)^3}  \bigg[ \big(1-n_F(E_r)+n_B(E_{k-r})\big)  \delta(E_r+E_{k-r}-E_q-p_0)  \mathcal W_-(E_r+E_{k-r})  \nonumber\\
&\times&  \bigg[ -\frac{(E_r+E_{k-r})^2}  {4E_{k-r}\big((E_r+E_{k-r})^2-E_k^2\big)^2}  + \frac{(  \mathbf k\cdot  \mathbf r)(E_r+E_{k-r})}  {4E_rE_{k-r}\big((E_r+E_{k-r})^2-E_k^2\big)^2}  + \frac{(  \mathbf k\cdot  \mathbf q)(E_r+E_{k-r})}  {4E_qE_{k-r}\big((E_r+E_{k-r})^2-E_k^2\big)^2} \nonumber \\
&-& \frac{(  \mathbf k\cdot  \mathbf r)(  \mathbf k\cdot  \mathbf q)}  {4E_rE_{k-r}E_q\big((E_r+E_{k-r})^2-E_k^2\big)^2} + \frac{1}  {8E_{k-r}\big((E_r+E_{k-r})^2-E_k^2\big)} - \frac{  \mathbf r\cdot  \mathbf q}  {8E_rE_{k-r}E_q\big((E_r+E_{k-r})^2-E_k^2\big)}
\bigg] \nonumber\\
&+& \big(n_F(E_r)+n_B(E_{k-r})\big)  \delta(E_{k-r}+E_q-E_r-p_0) \mathcal W_+(E_{k-r}-E_r) \nonumber\\
&\times& \bigg[ -\frac{(E_{k-r}-E_r)^2}  {4E_{k-r}\big((E_{k-r}-E_r)^2-E_k^2\big)^2}   - \frac{(  \mathbf k\cdot  \mathbf r)(E_{k-r}-E_r)} {4E_rE_{k-r}\big((E_{k-r}-E_r)^2-E_k^2\big)^2} -\frac{(  \mathbf k\cdot  \mathbf q)(E_{k-r}-E_r)} {4E_qE_{k-r}\big((E_{k-r}-E_r)^2-E_k^2\big)^2} \nonumber \\ 
&-& \frac{(  \mathbf k\cdot  \mathbf r)(  \mathbf k\cdot  \mathbf q)}  {4E_rE_{k-r}E_q\big((E_{k-r}-E_r)^2-E_k^2\big)^2} +  \frac{1}  {8E_{k-r}\big((E_{k-r}-E_r)^2-E_k^2\big)}  - \frac{  \mathbf r\cdot  \mathbf q}  {8E_rE_{k-r}E_q\big((E_{k-r}-E_r)^2-E_k^2\big)} \bigg] \nonumber\\
&+&  \big(n_F(E_r)+n_B(E_{k-r})\big)  \delta(E_r+E_q-E_{k-r}-p_0)  \mathcal W_+(E_r-E_{k-r}) \times \bigg[ -\frac{(E_r-E_{k-r})^2}  {4E_{k-r}\big((E_r-E_{k-r})^2-E_k^2\big)^2} \nonumber\\ 
&+& \frac{(  \mathbf k\cdot  \mathbf r)(E_r-E_{k-r})} {4E_rE_{k-r}\big((E_r-E_{k-r})^2-E_k^2\big)^2} - \frac{(  \mathbf k\cdot  \mathbf q)(E_r-E_{k-r})} {4E_qE_{k-r}\big((E_r-E_{k-r})^2-E_k^2\big)^2} + \frac{(  \mathbf k\cdot  \mathbf r)(  \mathbf k\cdot  \mathbf q)}  {4E_rE_{k-r}E_q\big((E_r-E_{k-r})^2-E_k^2\big)^2} \nonumber \\
&+&  \frac{1}  {8E_{k-r}\big((E_r-E_{k-r})^2-E_k^2\big)}  + \frac{  \mathbf r\cdot  \mathbf q}  {8E_rE_{k-r}E_q\big((E_r-E_{k-r})^2-E_k^2\big)} \bigg]  \bigg],
\label{eq14_simplified}
\eea
where, $\mathcal W_{\pm}(X) \equiv \left(\frac{1}{2}-n_F(X)\right) \pm \left(\frac{1}{2}-n_F(E_q)\right).$ In deriving this expression, we have also used the identity  $\delta(x)=\delta(-x)$.

\begin{figure}[htb]
	\centering
	\includegraphics[scale=0.52]{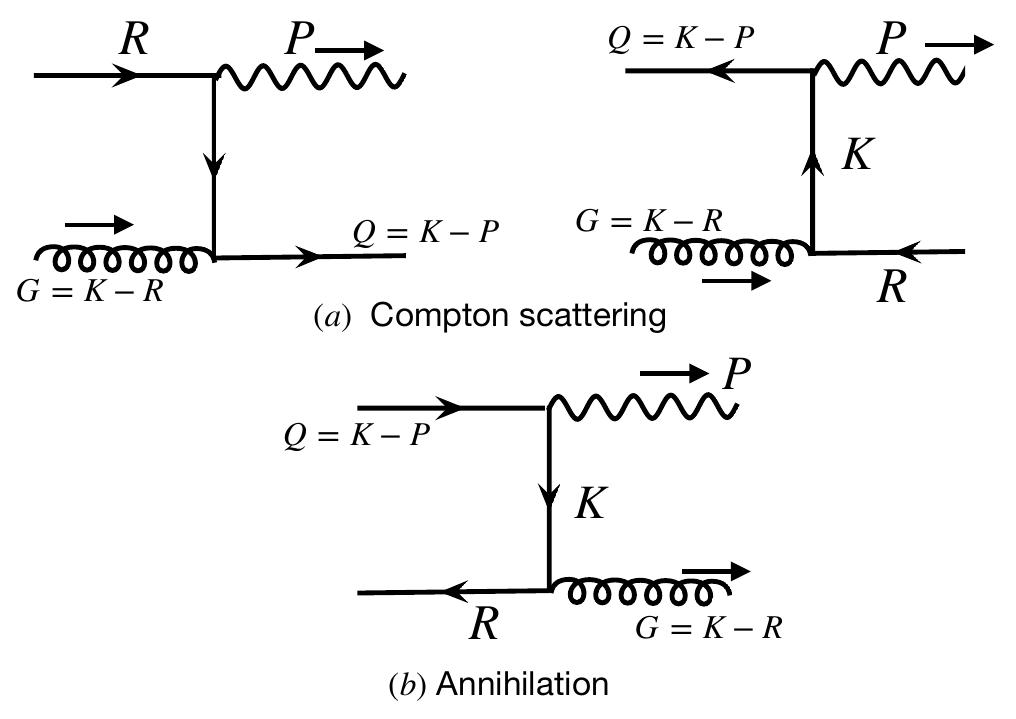}
	\caption{The $2\leftrightarrow2$ photon-production processes: (a) Compton scattering, $q(\bar q)g \rightarrow q(\bar q)\gamma$, and (b) quark-antiquark annihilation, $q\bar q \rightarrow g\gamma$.}
	\label{fig:scatterings}
\end{figure}
The energy-conserving delta functions in Eq.~\eqref{eq14_simplified} arising from physical cuts correspond to various leading-order scattering processes as displayed in Fig.~\ref{fig:scatterings}. The delta-function constraints allow the products of thermal distribution functions to be simplified as
\begin{subequations}
\bea
\big(1-n_F(E_r)+n_B(E_g)\big)\mathcal W_-(E_r+E_g) &=& \left(e^{\beta p_0}-1\right) n_F(E_r)n_B(E_g)\big(1-n_F(E_q)\big), \\
\big(n_F(E_r)+n_B(E_g)\big)\mathcal W_+(E_g-E_r) &=& \left(e^{\beta p_0}-1\right) \big(1-n_F(E_r)\big)n_B(E_g)n_F(E_q), \\
\big(n_F(E_r)+n_B(E_g)\big)\mathcal W_+(E_r-E_g) &=& \left(e^{\beta p_0}-1\right) n_F(E_r)\big(1+n_B(E_g)\big)n_F(E_q). 
\eea
\end{subequations}
Here, $E_g\equiv E_{k-r}$. Thus Eq.~\eqref{eq14_simplified} becomes
\bea
\hspace{-2.5cm}{{\Im}}\Pi^{\mu\, I}_{\mu} &=& -\frac{640\pi}{9}e^2g^2 \left(e^{\beta p_0}-1\right) \int\frac{d^3k}{(2\pi)^3} \int\frac{d^3r}{(2\pi)^3} \bigg[n_F(E_r)n_B(E_g)\big(1-n_F(E_q)\big) \delta(E_r+E_g-E_q-p_0) \mathcal{B}_1 + n_B(E_g) \nonumber\\
&\times &\big(1-n_F(E_r)\big)n_F(E_q) \delta(E_g+E_q-E_r-p_0) \mathcal{B}_2 +  n_F(E_r)\big(1+n_B(E_g)\big)n_F(E_q) \delta(E_r+E_q-E_g-p_0) \mathcal{B}_3 \bigg].
\label{eq14_thermal_simplified}
\eea
where the functions $\mathcal{B}_1, \mathcal{B}_2$ and $\mathcal{B}_3$ in Eq.~\eqref{eq14_thermal_simplified} are given by
\begin{subequations}
\bea
\mathcal{B}_1 &=& -\frac{(E_r+E_g)^2} {4E_g\big((E_r+E_g)^2-E_k^2\big)^2} + \frac{(  \mathbf k\cdot  \mathbf r)(E_r+E_g)} {4E_rE_g\big((E_r+E_g)^2-E_k^2\big)^2} + \frac{(  \mathbf k\cdot  \mathbf q)(E_r+E_g)} {4E_qE_g\big((E_r+E_g)^2-E_k^2\big)^2} \nonumber\\ 
&- &\frac{(  \mathbf k\cdot  \mathbf r)(  \mathbf k\cdot  \mathbf q)} {4E_rE_gE_q\big((E_r+E_g)^2-E_k^2\big)^2} + \frac{1} {8E_g\big((E_r+E_g)^2-E_k^2\big)} - \frac{  \mathbf r\cdot  \mathbf q} {8E_rE_gE_q\big((E_r+E_g)^2-E_k^2\big)},  \\
\mathcal{B}_2 &=& -\frac{(E_g-E_r)^2} {4E_g\big((E_g-E_r)^2-E_k^2\big)^2} - \frac{(  \mathbf k\cdot  \mathbf r)(E_g-E_r)} {4E_rE_g\big((E_g-E_r)^2-E_k^2\big)^2} - \frac{(  \mathbf k\cdot  \mathbf q)(E_g-E_r)} {4E_qE_g\big((E_g-E_r)^2-E_k^2\big)^2} \nonumber\\
&- &\frac{(  \mathbf k\cdot  \mathbf r)(  \mathbf k\cdot  \mathbf q)} {4E_rE_gE_q\big((E_g-E_r)^2-E_k^2\big)^2} + \frac{1} {8E_g\big((E_g-E_r)^2-E_k^2\big)} - \frac{  \mathbf r\cdot  \mathbf q} {8E_rE_gE_q\big((E_g-E_r)^2-E_k^2\big)}, \\
\mathcal{B}_3 &=& -\frac{(E_r-E_g)^2} {4E_g\big((E_r-E_g)^2-E_k^2\big)^2} +\frac{(  \mathbf k\cdot  \mathbf r)(E_r-E_g)} {4E_rE_g\big((E_r-E_g)^2-E_k^2\big)^2} - \frac{(  \mathbf k\cdot  \mathbf q)(E_r-E_g)} {4E_qE_g\big((E_r-E_g)^2-E_k^2\big)^2} \nonumber\\
&+& \frac{(  \mathbf k\cdot  \mathbf r)(  \mathbf k\cdot  \mathbf q)} {4E_rE_gE_q\big((E_r-E_g)^2-E_k^2\big)^2} + \frac{1} {8E_g\big((E_r-E_g)^2-E_k^2\big)} + \frac{  \mathbf r\cdot  \mathbf q} {8E_rE_gE_q\big((E_r-E_g)^2-E_k^2\big)}.
\eea
\end{subequations}

We now reduce the momentum integrals using the energy-conserving delta functions. For massless internal particles, $E_k =k, \, E_r=r, \, E_q=|\mathbf k-\mathbf p|, \, E_g=|\mathbf k-\mathbf r|.$  We choose $\mathbf p$ along the $z$-axis and define $p=|\mathbf p|$.
The full angular measure is $d^3k\,d^3r = k^2dk\,d\Omega_k\, r^2dr\,d\Omega_r$. The polar angle of ${\bf k}$ is measured with respect to ${\bf p}$, and we then define $u \equiv \cos\theta_{kp} = \frac{\mathbf k\cdot\mathbf p}{kp},$ as $E_q^2=k^2+p^2-2kp\,u, \, \, du = -\frac{E_q}{kp}\,dE_q$. The condition $u\in[-1,1]$ gives the corresponding range for $E_{q}$ as $|k-p|\leq E_q\leq k+p.$ \\
Now, after fixing ${\bf p}\parallel\hat{\bf z}$, the system is invariant under a common rotation around the $z$-axis, and thus the overall azimuthal angle of ${\bf k}$ gives a factor $2\pi$. For the ${\bf r}$-integration, it is convenient to choose the polar axis along ${\bf k}$, because the gluon energy is $E_g=|{\bf k}-{\bf r}|$. Thus we define $\, v \equiv \cos\theta_{kr} = \frac{\mathbf k\cdot\mathbf r}{kr}.$ then, we get $E_g^2=k^2+r^2-2kr\,v, \, \,  dv = -\frac{E_g}{kr}\,dE_g.$ and the condition $v\in[-1,1]$ gives the corresponding range for $E_{g}$ as $|k-r|\leq E_g\leq k+r.$ The remaining azimuthal angle, denoted by $\phi$, is the azimuth of ${\bf r}$ around the ${\bf k}$-axis. Therefore, $ d\Omega_k\,d\Omega_r = (2\pi\,du)(d\phi\,dv).$ Consequently,
\bea
\int_k\int_r &\equiv& \int\hspace{-.1cm}\frac{d^3k}{(2\pi)^3} \int\hspace{-.1cm}\frac{d^3r}{(2\pi)^3}=\frac{1}{(2\pi)^6} \hspace{-.1cm}\int\hspace{-.1cm} k^2dk\,(2\pi\,du) \int\hspace{-.1cm} r^2dr\,(d\phi\,dv) 
\eea
Since $E_{g}$ depends only on $v$, the remaining $\phi$ integration can be done trivially and gives a factor of $2\pi$. Thus, we get,  
\bea
\int_k\int_r &=& \frac{1}{16\pi^4p} \int_0^\infty \hspace{-.1cm}dk \int_0^\infty \hspace{-.1cm} dr \int_{|k-p|}^{k+p}dE_q \int_{|k-r|}^{k+r}dE_g\, rE_qE_g .
\eea
The scalar products can be written as $ K_{q}= \mathbf k\cdot\mathbf q = \mathbf k\cdot(\mathbf k-\mathbf p) = \frac{k^2-p^2+E_q^2}{2}, \, \mathbf k\cdot\mathbf r = \frac{k^2+r^2-E_g^2}{2}.$ Since $\mathbf r\cdot\mathbf q$ appears only linearly in Eq.~\eqref{eq14_thermal_simplified}, the azimuthal average gives $\mathbf r\cdot\mathbf q \longrightarrow \frac{(\mathbf k\cdot\mathbf r)(\mathbf k\cdot\mathbf q)}{k^2}.$ Thus, after the azimuthal integration, $<\mathbf r\cdot\mathbf q>_{\phi} = \frac{1}{k^2} \left( \frac{k^2+r^2-E_g^2}{2} \right) \left( \frac{k^2-p^2+E_q^2}{2} \right). $ The three physical delta functions, using $E_r=r$, become
\bea
D_1&=\delta(r+E_g-E_q-p_0),\qquad D_2=\delta(E_g+E_q-r-p_0),\qquad D_3=\delta(r+E_q-E_g-p_0). 
\eea
Each delta function is now a function of the integration variable $E_g$. Hence, we use delta functions to perform the $dE_g$ integration. The general identity used is $\delta(f(x)) = \sum_i \frac{\delta(x-x_i)} {\left|f'(x_i)\right|}, $ where $x_i$ are the roots of $f(x_i)=0.$ 
Therefore, the delta functions fix $E_g$ as $E_g^{(1)} = E_q+p_0-r, \,  E_g^{(2)} = r+p_0-E_q, \, E_g^{(3)} = r+E_q-p_0.$ Since each delta function is linear in $E_g$, the Jacobian is unity. Thus, for any function $F_i(k,r,E_q,E_g)$, we have
\bea
\int_{|k-r|}^{k+r}dE_g\, E_g\,D_i\,F_i(k,r,E_q,E_g) & = & E_g^{(i)} F_i(k,r,E_q,E_g^{(i)}) \Theta\!\left(E_g^{(i)}-|k-r|\right) \Theta\!\left(k+r-E_g^{(i)}\right) \Theta\!\left(E_g^{(i)}\right). 
\eea
Therefore,
\begin{equation}
\int_k\int_r D_i\,F_i = \frac{1}{16\pi^4p} \int_0^\infty dk \int_0^\infty dr \int_{|k-p|}^{k+p}dE_q\, rE_qE_g^{(i)} \Theta_i F_i(k,r,E_q,E_g^{(i)}),
\label{phase_space_measure}
\end{equation}
where the $\Theta_i $ are defined as  $\Theta_i \equiv \Theta\!\left(E_g^{(i)}-|k-r|\right) \Theta\!\left(k+r-E_g^{(i)}\right) \Theta\!\left(E_g^{(i)}\right). $ In Eq.~\eqref{phase_space_measure} the $F_i(k,r,E_q,E_g^{(i)})$ are defined as
\begin{subequations}
\bea
F_1(k,r,E_q,E_g) &=&  \left(e^{\beta p_0}-1\right) n_F(r)n_B(E_g)\big(1-n_F(E_q)\big) \mathcal B_1(k,r,E_q,E_g), \\
F_2(k,r,E_q,E_g) &=&  \left(e^{\beta p_0}-1\right) \big(1-n_F(r)\big)n_B(E_g)n_F(E_q) \mathcal B_2(k,r,E_q,E_g),
\\
F_3(k,r,E_q,E_g) &=&  \left(e^{\beta p_0}-1\right) n_F(r)\big(1+n_B(E_g)\big)n_F(E_q) \mathcal B_3(k,r,E_q,E_g).
\label{definition_Fi}
\eea
\end{subequations}
where $\mathcal B_i$ are given by
\begin{subequations}
\bea
\mathcal B_1 &=& -\frac{(r+E_g)^2}{4E_g\big((r+E_g)^2-E_k^2\big)^2} + \frac{(\mathbf k\cdot\mathbf r)(r+E_g)} {4rE_g\big((r+E_g)^2-E_k^2\big)^2} + \frac{(\mathbf k\cdot\mathbf q)(r+E_g)} {4E_qE_g\big((r+E_g)^2-E_k^2\big)^2} \nonumber\\
&-& \frac{(\mathbf k\cdot\mathbf r)(\mathbf k\cdot\mathbf q)} {4rE_gE_q\big((r+E_g)^2-E_k^2\big)^2} + \frac{1}{8E_g\big((r+E_g)^2-E_k^2\big)} - \frac{\mathbf r\cdot\mathbf q} {8rE_gE_q\big((r+E_g)^2-E_k^2\big)}, \\
\mathcal B_2 &=& -\frac{(E_g-r)^2}{4E_g\big((E_g-r)^2-E_k^2\big)^2} - \frac{(\mathbf k\cdot\mathbf r)(E_g-r)} {4rE_g\big((E_g-r)^2-E_k^2\big)^2} - \frac{(\mathbf k\cdot\mathbf q)(E_g-r)} {4E_qE_g\big((E_g-r)^2-E_k^2\big)^2} \nonumber\\
&-& \frac{(\mathbf k\cdot\mathbf r)(\mathbf k\cdot\mathbf q)} {4rE_gE_q\big((E_g-r)^2-E_k^2\big)^2} + \frac{1}{8E_g\big((E_g-r)^2-E_k^2\big)} - \frac{\mathbf r\cdot\mathbf q} {8rE_gE_q\big((E_g-r)^2-E_k^2\big)}, \\
\mathcal B_3 &=& -\frac{(r-E_g)^2}{4E_g\big((r-E_g)^2-E_k^2\big)^2} + \frac{(\mathbf k\cdot\mathbf r)(r-E_g)} {4rE_g\big((r-E_g)^2-E_k^2\big)^2} - \frac{(\mathbf k\cdot\mathbf q)(r-E_g)} {4E_qE_g\big((r-E_g)^2-E_k^2\big)^2} \nonumber\\
&+& \frac{(\mathbf k\cdot\mathbf r)(\mathbf k\cdot\mathbf q)} {4rE_gE_q\big((r-E_g)^2-E_k^2\big)^2} + \frac{1}{8E_g\big((r-E_g)^2-E_k^2\big)} + \frac{\mathbf r\cdot\mathbf q} {8rE_gE_q\big((r-E_g)^2-E_k^2\big)}.
\eea
\end{subequations}
After using the delta functions, the final reduced form is
\bea
\hspace{-0cm}{\Im}\Pi^{\mu I}_{\mu} &=& -\frac{640\pi}{9}e^2g^2 \frac{\left(e^{\beta p_0}-1\right)}{16\pi^4p} \int_0^\infty dk
\int_0^\infty dr \int_{|k-p|}^{k+p}dE_q\, rE_q \bigg[ E_g^{(1)}\Theta_1\, n_F(r)n_B(E_g^{(1)})\big(1-n_F(E_q)\big)   \nonumber\\
&\times& \mathcal B_1(k,r,E_q,E_g^{(1)}) + E_g^{(2)}\Theta_2\, \big(1-n_F(r)\big)n_B(E_g^{(2)})n_F(E_q) \mathcal B_2(k,r,E_q,E_g^{(2)}) + E_g^{(3)}\Theta_3\, n_F(r) \nonumber\\
&\times& \big(1+n_B(E_g^{(3)})\big)n_F(E_q) \mathcal B_3(k,r,E_q,E_g^{(3)}) \bigg].
\label{eq22m}
\eea
For a real photon, $p_0=p$, thus, we have $E_g^{(1)} = E_q+p-r, \, E_g^{(2)} = r+p-E_q, \, E_g^{(3)} = r+E_q-p. $
Also 
\begin{subequations}
\bea
\Theta_1 &=& \Theta\!\left(E_q+p-r-|k-r|\right) \Theta\!\left(k+2r-E_q-p\right) \Theta\!\left(E_q+p-r\right), \\
\Theta_2 &=& \Theta\!\left(r+p-E_q-|k-r|\right) \Theta\!\left(k+E_q-p\right) \Theta\!\left(r+p-E_q\right), \\
\Theta_3 &=& \Theta\!\left(r+E_q-p-|k-r|\right) \Theta\!\left(k-E_q+p\right) \Theta\!\left(r+E_q-p\right). 
\eea
\end{subequations}
The theta-function restriction is equivalent to imposing $|k-r|\leq E_g^{(i)}\leq k+r, \, E_g^{(i)}>0.$ Therefore, the three delta functions give the following conditions :
\begin{subequations}
\bea
\Theta_1 &=& \Theta(E_q+p-k) \Theta\!\left(r-\frac{E_q+p-k}{2}\right) \Theta\!\left(\frac{E_q+p+k}{2}-r\right), \,\,\, \Rightarrow  \quad \frac{E_q+p-k}{2} \leq r\leq \frac{E_q+p+k}{2} . \\
\Theta_2 &=& \Theta(k+E_q-p) \Theta(k+p-E_q) \Theta\!\left(r-\frac{k+E_q-p}{2}\right), \qquad \Rightarrow \qquad r\geq \frac{k+E_q-p}{2}.   \\
\Theta_3 &=& \Theta(k+p-E_q) \Theta(k+E_q-p) \Theta\!\left(r-\frac{k+p-E_q}{2}\right), \qquad \Rightarrow \qquad r\geq \frac{k+p-E_q}{2}.
\eea
\end{subequations}
Using the theta-function restrictions, Eq.~\eqref{eq22m} becomes
\bea
{\Im}\Pi^{\mu I}_{\mu} &=& -\frac{640\pi}{9}e^2g^2  \frac{\left(e^{\beta p}-1\right)}{16\pi^4p} \int_0^\infty dk \int_{|k-p|}^{k+p}dE_q\, E_q \bigg[ \int_{\frac{E_q+p-k}{2}}^{\frac{E_q+p+k}{2}} dr\, rE_g^{(1)} n_F(r)n_B(E_g^{(1)})\big(1-n_F(E_q)\big)  \nonumber\\
&\times& \mathcal B_1(k,r,E_q,E_g^{(1)}) +\int_{\frac{k+E_q-p}{2}}^{\infty} dr\, rE_g^{(2)} \big(1-n_F(r)\big)n_B(E_g^{(2)})n_F(E_q) \mathcal B_2(k,r,E_q,E_g^{(2)}) \nonumber\\
&+& \int_{\frac{k+p-E_q}{2}}^{\infty} dr\, rE_g^{(3)} n_F(r)\big(1+n_B(E_g^{(3)})\big)n_F(E_q) \mathcal B_3(k,r,E_q,E_g^{(3)}) \bigg].
\eea
Now we define $S=E_q+p, \, A=p-E_q$. Then, the three roots obtained from the delta functions are $E_g^{(1)}=S-r, \, E_g^{(2)}=r+A, \, E_g^{(3)}=r-A.$ The corresponding $r$-limits for the three channels are
\bea
\frac{S-k}{2}\leq r\leq \frac{S+k}{2},  \qquad r\geq L_2\equiv \frac{k-A}{2} = \frac{k+E_q-p}{2}, \qquad r\geq L_3\equiv \frac{k+A}{2} = \frac{k+p-E_q}{2}, 
\eea
Now, for the first channel, using $\Delta_S=S^2-k^2$ and $E_g^{(1)}=S-r$, we have $\mathbf k\cdot\mathbf r = \frac{k^2+r^2-(E_g^{(1)})^2}{2} = Sr-\frac{\Delta_S}{2}.$ Then,
\bea
rE_g^{(1)}\mathcal B_1(k,r,E_q,E_g^{(1)}) = \alpha_1 r+\gamma_1, \qquad  \alpha_1 = \frac{1}{8\Delta_S} \left( 1-\frac{S K_q}{E_q k^2} \right), \qquad \gamma_1 = -\frac{S}{8\Delta_S} + \frac{K_q}{8E_q\Delta_S} + \frac{K_q}{16E_qk^2}.
\eea
The hard-photon condition $p\gg T$ does not imply that every incoming thermal particle has an energy much larger than $T$. We therefore retain the exact Fermi--Dirac and Bose--Einstein distributions in all three physical cuts. With $S=p+E_q$ and $A=p-E_q$, their statistical weights are $w_1 = n_F(r)n_B(S-r)\big[1-n_F(E_q)\big], \, w_2 = \big[1-n_F(r)\big]n_B(r+A)n_F(E_q), \, w_3 = n_F(r)\big[1+n_B(r-A)\big]n_F(E_q).$ Using $1-n_F(z)=e^{\beta z}n_F(z)$ and $1+n_B(z)=e^{\beta z}n_B(z)$, together with energy conservation, these weights can be rewritten exactly as $ w_1 = e^{-\beta p}n_F(E_q)\big[1-n_F(r)\big]\big[1+n_B(S-r)\big], \, w_2 = e^{-\beta p}\big[1-n_F(E_q)\big]n_F(r)\big[1+n_B(r+A)\big], \, w_3 = e^{-\beta p}\big[1-n_F(E_q)\big]n_B(r-A)\big[1-n_F(r)\big].$ For the first physical channel, the exact thermal integral is
\bea
I_1 &=& e^{-\beta p}n_F(E_q)\int_{(S-k)/2}^{(S+k)/2}dr\,(\alpha_1r+\gamma_1)\big[1-n_F(r)\big]\big[1+n_B(S-r)\big].
\label{eq:topI_I1_full_statistics}
\eea 
For the second channel, let $\Delta_A=A^2-k^2$. Since $E_g^{(2)}=r+A$, one has $\mathbf k\cdot\mathbf r=-Ar-\Delta_A/2$, and the kinematic coefficients are
\bea
rE_g^{(2)}\mathcal B_2 &=& \alpha_2r+\gamma_2, \qquad \alpha_2=\frac{1}{8\Delta_A}\left(1+\frac{AK_q}{E_qk^2}\right), \qquad \gamma_2=\frac{A}{8\Delta_A}+\frac{K_q}{8E_q\Delta_A}+\frac{K_q}{16E_qk^2}.
\eea
For the third channel, $E_g^{(3)}=r-A$ and $\mathbf k\cdot\mathbf r=Ar-\Delta_A/2$, giving $rE_g^{(3)}\mathcal B_3 = \alpha_3r+\gamma_3, \, \alpha_3=\alpha_2, \, \gamma_3=-\gamma_2.$ The common lower thermal-energy boundary is given by the kinematically generated variable $\ell \equiv \frac{k+E_q-p}{2}=L_2=L_3-A, \, \hat{s}\equiv4p\ell.$ Thus, $\hat{s}$ is an auxiliary energy-squared variable parametrizing the lower thermal-energy boundary. Using the exact thermal weights above, and changing variables from $r$ to $E_g=r-A$ in the third channel, gives
\begin{subequations}
\bea
I_2 &=& e^{-\beta p}\big[1-n_F(E_q)\big]\int_\ell^\infty dr\,(\alpha_2r+\gamma_2)n_F(r)\big[1+n_B(r+A)\big], \\
I_3 &=& e^{-\beta p}\big[1-n_F(E_q)\big]\int_\ell^\infty dE_g\,\big[\alpha_3(E_g+A)+\gamma_3\big]n_B(E_g)\big[1-n_F(E_g+A)\big].
\eea
\end{subequations}
We introduce the full-statistics thermal moments, for $m=0,1$,
\bea
\Psi_C^{(m)}(\ell,A) &\equiv& \int_\ell^\infty dr\,r^m n_F(r)\big[1+n_B(r+A)\big], \quad \Psi_{\rm Ann}^{(m)}(\ell,A) \equiv \int_\ell^\infty dE_g\,E_g^m n_B(E_g)\big[1-n_F(E_g+A)\big].
\label{eq:topI_full_thermal_moments}
\eea
Here, the subscripts $`C$' and `${\rm Ann}$' denote the Compton and annihilation sectors, respectively, whereas $A=p-E_q$ is the kinematic energy-shift variable. Defining $c(A)\equiv1-n_F(A)=1/(1+e^{-A/T})$, the exact identities $n_F(r)\big[1+n_B(r+A)\big] = c(A)\big[n_F(r)+n_B(r+A)\big], \, n_B(E_g)\big[1-n_F(E_g+A)\big] = c(A)\big[n_B(E_g)+n_F(E_g+A)\big]$ allow all four moments to be evaluated in closed form. Using Eq.~\eqref{eq:topI_full_thermal_moments}, and the exact statistical identities mentioned above, together with the shifts $y=r+A$ and $y=E_g+A$ where required, all four full-statistics moments can be evaluated analytically. For the Compton channel, the exact thermal moments can be written as
\begin{subequations}
\label{Fermi_integrals}
\bea
\hspace{-1cm}\Psi_C^{(0)}(\ell,A) &=& Tc(A)\ln\left[\frac{1+e^{-\ell/T}} {1-e^{-(\ell+A)/T}}\right], \\
\Psi_C^{(1)}(\ell,A)  &=& \ell\Psi_C^{(0)}(\ell,A) +T^2c(A)\left[ {\rm Li}_2\!\left(e^{-(\ell+A)/T}\right) -{\rm Li}_2\!\left(-e^{-\ell/T}\right) \right].
\eea
\end{subequations}
For the annihilation channel, one similarly obtains
\begin{subequations}
\label{Bose_integrals}
\bea
\hspace{-2cm} \Psi_{\rm Ann}^{(0)}(\ell,A) &=& Tc(A)\ln\left[\frac{1+e^{-(\ell+A)/T}} {1-e^{-\ell/T}}\right], \\
\Psi_{\rm Ann}^{(1)}(\ell,A) &=& \ell\Psi_{\rm Ann}^{(0)}(\ell,A) +T^2c(A)\left[ {\rm Li}_2\!\left(e^{-\ell/T}\right) -{\rm Li}_2\!\left(-e^{-(\ell+A)/T}\right) \right].
\eea
\end{subequations}
These closed-form expressions are valid in the interior of the physical domain, $\ell>0$ and $\ell+A>0$, which ensures that all thermal-energy arguments are positive. Boundary configurations are defined by the corresponding one-sided limits, $\ell\to0^{+}$ or $\ell+A\to0^{+}$. Because the exact statistical weights contain the shifted energies $r+A$ and $E_g+A$, the moments depend separately on both $\ell$ and $A$, rather than only on $\hat{s}=4p\ell$. The $r$ and $E_g$ integrations in $I_2$ and $I_3$ are thus contained in the moments above. 
Combining the three channels, the imaginary part can be written as
\bea
\Im \Pi^{\mu{\rm I}}_{\ \ \mu}(P) = -\frac{640\pi}{9}e^2g^2 \left(e^{\beta p}-1\right) \frac{1}{16\pi^4p} \left[ \mathcal I_1+\mathcal I_2+\mathcal I_3 \right], \qquad \qquad \mathcal I_i \equiv \int_0^\infty dk \int_{|k-p|}^{k+p}dE_q\,E_q I_i. 
\label{eq:topI_ImPi_I_relation}
\eea
At fixed $k$, we have $\ell=\frac{k+E_q-p}{2}.$  Then $E_q=p-k+2\ell, \, dE_q=2d\ell.$  The original range $|k-p|\leq E_q\leq k+p $ becomes $\ell_-(k)\leq \ell\leq k, $ where $\ell_-(k)= \begin{cases} 0, & 0\leq k\leq p,\\ k-p, & k\geq p. \end{cases} $. \, Thus, we can write
\bea
\int_0^\infty dk \int_{|k-p|}^{k+p}dE_q = \int_0^p dk\int_0^k2d\ell + \int_p^\infty dk\int_{k-p}^k2d\ell.
\eea
Using the $r$ and $E_{g}$-integrated thermal functions defined in Eqs.~\eqref{Fermi_integrals} and \eqref{Bose_integrals}, the remaining $k, E_q$ phase space can be written in terms of $\ell$. With $\hat{s}=4p\ell$, one has $\ell=\frac{\hat{s}}{4p}, \, dE_q=\frac{d\hat{s}}{2p}.$ For $(0\leq k\leq p)$, the $\hat{s}$-range is  $0\leq \hat{s}\leq 4pk,$ whereas for $(k\geq p)$,  $4p(k-p)\leq \hat{s}\leq 4pk.$ For compactness, we define the phase-space operator
\bea
\mathcal P[f] \equiv \int_0^p dk \int_0^{4pk}\frac{d\hat{s}}{2p}\,f(k,\hat{s}) + \int_p^\infty dk \int_{4p(k-p)}^{4pk}\frac{d\hat{s}}{2p}\,f(k,\hat{s}).
\eea
Equivalently, in terms of $\ell$,
\bea
\mathcal P[f] &=& \int_0^p dk\int_0^k2d\ell\,f(k,4p\ell) + \int_p^\infty dk\int_{k-p}^k2d\ell\,f(k,4p\ell).
\label{Projection_l_operator}
\eea
 Equivalently, by interchanging the order of integration and combining the two kinematic sectors at fixed $\ell$, the phase-space operator can then be written as
\bea
\mathcal P[f] = 2\int_0^\infty d\ell \int_\ell^{p+\ell} dk\, f(k,4p\ell).
\label{Projection_fixed_l_operator}
\eea
This form makes the small-$\ell$ behavior explicit and will be used later. The variable $\hat{s}=4p\ell$ provides a convenient parametrization of the remaining phase space, but it should not by itself be identified with the virtuality of the repeated uncut quark propagator. For the second and third physical channels, this virtuality is $-\Delta_A = k^2-A^2 = 4\ell(k-\ell).$ The hard region is therefore defined by $4\ell(k-\ell) \geq k_c^2.$ Introducing $a \equiv \frac{k_c^2}{4}, \, \ell_c\equiv\frac{a}{p} =\frac{k_c^2}{4p},$ the hard condition gives $k \geq \ell+\frac{a}{\ell}.$ Combining this restriction with the unrestricted kinematic upper boundary $k\leq p+\ell$, gives the exact hard domain interval as $\ell\geq\ell_c, \, \ell+\frac{a}{\ell} \leq k\leq p+\ell$. The scale $\ell_c=k_c^2/(4p)$ is absent from the unrestricted phase space. It arises after imposing the hard-virtuality condition and requiring the resulting $k$-integration interval to be nonempty. For clarity, the operator ${\cal P}$ denotes the complete unrestricted phase space only defined in Eq.~\eqref{Projection_fixed_l_operator}. For the second and third physical channels, its hard-domain counterpart is
\bea
{\cal P}_{\rm h}[f] &\equiv& 2\int_{\ell_c}^{\infty}d\ell \int_{\ell+a/\ell}^{p+\ell}dk\, f(k,4p\ell).
\label{eq:Projection_hard_fixed_l_operator}
\eea
The remaining kinematic quantities are defined as $ E_q  =  p-k+\frac{\hat{s}}{2p}, \, A=p-E_q = k-\frac{\hat{s}}{2p}, \, S=p+E_q = 2p-k+\frac{\hat{s}}{2p},\, K_q = \frac{k^2-p^2+E_q^2}{2} = \frac{k^2-AS}{2}.$ In the second channel, the lower limit is $ L_2 = \ell$. In the third channel, $L_3=\frac{k+p-E_q}{2}, \, A=p-E_q, $ so that $L_3-A = \ell = \frac{\hat{s}}{4p}.$ From this point onward, the $r$- and $E_g$ integrations have already been carried out and are contained in the thermal functions $\Psi_C^{(i)}(\ell, A)$ and $\Psi_{\rm Ann}^{(i)}(\ell, A)$, with $i=0,1$, obtained in Eq.~\eqref{Fermi_integrals} and Eq.~\eqref{Bose_integrals}. Therefore, the remaining integrations are only over the phase-space variables $k$ and $E_q$, or equivalently $k$ and $\ell$. Before writing the remaining phase-space contributions, it is useful to simplify the kinematic coefficients. For the first channel, the reduced coefficients satisfy $E_q\alpha_1=A/(16k^2)$ and $E_q\gamma_1=-(k^2+AS)/(32k^2)$. Therefore, its exact contribution is
\bea
E_qI_1 &=& \frac{e^{-\beta p}n_F(E_q)}{32k^2}\int_{(S-k)/2}^{(S+k)/2}dr\,\big[2Ar-k^2-AS\big]\big[1-n_F(r)\big]\big[1+n_B(S-r)\big].
\label{eq:topI_EqI1_reduced}
\eea
Both statistical factors remain inside the $r$-integral. Its leading hard-photon contribution will be extracted together with the finite terms in the BLL calculation. For the second and third channels, the kinematic coefficients are given by:
\bea
E_q\alpha_2 &=& -\frac{S}{16k^2}, \qquad E_q\gamma_2=-\frac{k^2+AS}{32k^2}, \qquad E_q\alpha_3 = -\frac{S}{16k^2}, \qquad E_q(\alpha_3A+\gamma_3)=\frac{k^2-AS}{32k^2}.
\label{eq:topI_alpha3_gamma3_reduced}
\eea
Substituting the thermal moments into the second and third channel integrals, using the reduced coefficients in Eq.~\eqref{eq:topI_alpha3_gamma3_reduced}, and applying $\mathcal P$, the remaining phase-space contributions are
\begin{subequations}
\label{H_definitions}
\bea
\mathcal I_1 &=& \mathcal P[E_qI_1], \\
\mathcal I_2 &=& -e^{-\beta p}\mathcal P\left[\big[1-n_F(E_q)\big]\left\{\frac{S}{16k^2}\Psi_C^{(1)}(\ell,A)+\frac{k^2+AS}{32k^2}\Psi_C^{(0)}(\ell,A)\right\}\right], \\
\mathcal I_3 &=& -e^{-\beta p}\mathcal P\left[\big[1-n_F(E_q)\big]\left\{\frac{S}{16k^2}\Psi_{\rm Ann}^{(1)}(\ell,A)-\frac{k^2-AS}{32k^2}\Psi_{\rm Ann}^{(0)}(\ell,A)\right\}\right].
\eea
\end{subequations}
Here, the quantities ${\cal I}_i$ are defined over the complete phase space using ${\cal P}$. For the second and third physical cuts,
the hard-domain operator ${\cal P}_{\rm h}$ is imposed when extracting their LL and BLL hard contributions. The first cut involves the
different exchanged-quark virtuality $\Delta_S=S^2-k^2$; consequently, its hard restriction, whenever required, must be imposed separately.
No approximations to the thermal distributions have been made in the expressions above.
\subsubsection{Leading-logarithmic contribution}
We now extract the logarithmic sensitivity to the exchanged-quark cutoff in the hierarchy $p\gg T\gg k_c$. We define $L \equiv \ln\left({4pT}/{k_c^2}\right).$ Here, LL denotes the coefficient of the cutoff logarithm at fixed $p/T$. Terms independent of $k_c$, including terms proportional to $\ln(p/T)$, are retained in the subsequent BLL calculation. Using the exact hard domain as described above, we now extract the leading logarithmic contributions. The first physical channel does not generate a leading-power logarithmic dependence on the exchanged-quark cutoff. Although its exact thermal integral can contain endpoint logarithms, these are integrable and do not produce a term proportional to $\ln k_c$. Consequently,  
\bea
{\cal I}_1^{\rm LL}&=&0.
\label{eq:topI_I1_LL_zero}
\eea
For the second and third channels, the cutoff logarithm is fixed by the overlap region $k,\ell,|A|\ll T\ll p$. In this region, $E_q=p+{\cal O}(T)$, $S=2p+{\cal O}(T)$, and $1-n_F(E_q)=1+{\cal O}(e^{-p/T})$ at the retained leading order in the hard-photon expansion. The thermal moments must nevertheless be evaluated with the exact quantum distributions. The identity
\bea
n_F(r)\big[1+n_B(r)\big] &=& n_B(r)\big[1-n_F(r)\big]=\frac{1}{2\sinh(r/T)}
\label{eq:topI_common_quantum_weight}
\eea
gives the common first moment
\bea
M_Q &\equiv& \Psi_C^{(1)}(0,0)=\Psi_{\rm Ann}^{(1)}(0,0)=\int_0^\infty dr\,\frac{r}{2\sinh(r/T)}=\frac{\pi^2T^2}{8}.
\label{eq:topI_common_quantum_moment}
\eea
At the retained leading-power accuracy, the terms involving $\Psi_C^{(0)}$ and $\Psi_{\rm Ann}^{(0)}$ do not generate a term proportional to $\ln k_c$; after the $k$ integration, they contribute finite BLL terms and, at most, power-suppressed cutoff dependence. Therefore, the logarithmic coefficient follows from the $S\Psi_C^{(1)}/(16k^2)$ term in ${\cal I}_2$ and the $S\Psi_{\rm Ann}^{(1)}/(16k^2)$ term in ${\cal I}_3$. For $\ell\gg T$, the thermal moments are exponentially suppressed. Hence, the effective upper scale of the logarithmic $\ell$-integration may be chosen as $T$; changing it by a numerical factor affects only the finite BLL contribution. Thus, for either singular channel, where ${\cal I}_{2,3}$ denotes either ${\cal I}_2$ or ${\cal I}_3$ separately, retaining the $S\Psi_C^{(1)}/(16k^2)$ or $S\Psi_{\rm Ann}^{(1)}/(16k^2)$ part gives
\bea
\left.{\cal I}_{2,3}\right|_{\ln k_c} &=& -2e^{-\beta p} \int_{\ell_c}^{T}d\ell \int_{\ell+a/\ell}^{p+\ell}dk\, \frac{2p}{16k^2}M_Q  = -\frac{p}{4}e^{-\beta p}M_Q \left. \int_{\ell_c}^{T}d\ell \left[ \frac{\ell}{\ell^2+a} -\frac{1}{p+\ell} \right] \right|_{\ln k_c} \nn\\
&=& -\frac{p}{4}e^{-\beta p}M_Q \left. \int_{\ell_c}^{T}d\ell\, \frac{\ell}{\ell^2+a} \right|_{\ln k_c}.
\label{eq:topI_H2a_LL_exact}
\eea
The factor $p/4$ follows from the factor $2$ in ${\cal P}_{\rm h}$ and $S/(16k^2)\simeq2p/(16k^2)$. The omitted contribution satisfies
\bea
\int_{\ell_c}^{T}\frac{d\ell}{p+\ell} &=& \ln\left(\frac{p+T}{p+\ell_c}\right) = {\cal O}\left(\frac{T}{p}\right),
\eea
and therefore contains no cutoff logarithm. It is suppressed by $T/p$, with only power-suppressed dependence on $k_c$ through $\ell_c$. Although the exact hard domain starts at $\ell_c=a/p$, the logarithm is accumulated in the interval $\sqrt{a}\lesssim\ell\lesssim T$, where $\ell/(\ell^2+a)\simeq1/\ell$. Explicitly,
\bea
\int_{\ell_c}^{T}d\ell\, \frac{\ell}{\ell^2+a} &=& \frac{1}{2} \ln\left(\frac{T^2+a}{\ell_c^2+a}\right) = \ln\left(\frac{2T}{k_c}\right)
+ {\cal O}\left(\frac{a}{T^2},\frac{a}{p^2}\right).
\label{eq:topI_H2a_LL_log_integral}
\eea
Using $2\ln(2T/k_c)=L-\ln(p/T)$ and assigning the cutoff-independent part to the BLL result, the Compton contribution is
\bea
\mathcal I_{\rm Comp}^{\rm LL} &=& \mathcal I_1^{\rm LL}+\mathcal I_2^{\rm LL}=-\frac{pM_Q}{8}e^{-\beta p}L=-\frac{p\pi^2T^2}{64}e^{-\beta p}L.
\label{eq:topI_HComp_LL_result}
\eea
The annihilation contribution has the same logarithmic coefficient,
\bea
\mathcal I_{\rm Ann}^{\rm LL} &=& \mathcal I_3^{\rm LL}=-\frac{p\pi^2T^2}{64}e^{-\beta p}L=\mathcal I_{\rm Comp}^{\rm LL}.
\label{eq:topI_HAnn_LL_result}
\eea
Substituting these results into Eq.~\eqref{eq:topI_ImPi_I_relation}, and using $ (e^{\beta p}-1)e^{-p/T}=1-e^{-p/T} $, yields
\begin{subequations}
\label{eq:topI_full_LL_individual_self_energies}
\bea \left.\Im\Pi^{\mu I}_{\ \ \mu}(P)\right|_{\rm Comp}^{\rm LL} &=& \frac{5e^2g^2}{72\pi}T^2\left(1-e^{-p/T}\right)L,
\label{eq:topI_ImPi_Comp_LL} \\
\left.\Im\Pi^{\mu I}_{\ \ \mu}(P)\right|_{\rm Ann}^{\rm LL} &=& \frac{5e^2g^2}{72\pi}T^2\left(1-e^{-p/T}\right)L.
\label{eq:topI_ImPi_Ann_LL}
\eea
\end{subequations}
Thus, the total leading-logarithmic hard contribution to the imaginary part of the photon self-energy is
\bea
\left. \Im \Pi^{\mu{\rm I}}_{\ \ \mu}(P) \right|_{\rm hard}^{\rm LL} = \frac{5e^2g^2}{36\pi}\, T^2 \left(1-e^{-p/T}\right) \ln\left(\frac{4pT}{k_c^2}\right). 
\label{eq:topI_ImPi_total_LL}
\eea
The equality of the individual Compton and annihilation coefficients follows from the common quantum-statistical moment $M_Q$. 

For comparison, applying the Maxwell--Boltzmann (MB) approximation to the incoming thermal distributions, while retaining the final-state statistical factors, gives the zero-boundary first moments
\bea
M_C^{\rm MB} &\equiv& \int_0^\infty dr\,r\,n_F(r)=\frac{\pi^2T^2}{12}, \qquad M_{\rm Ann}^{\rm MB}\equiv\int_0^\infty dr\,r\,n_B(r)=\frac{\pi^2T^2}{6}.
\label{eq:topI_MB_LL_moments}
\eea
The corresponding LL phase-space contributions are
\begin{subequations}
\label{eq:topI_MB_LL_phase_space}
\bea
{\cal I}_{\rm Comp,MB}^{\rm LL} &=& -\frac{pM_C^{\rm MB}}{8}e^{-\beta p}L =-\frac{p\pi^2T^2}{96}e^{-\beta p}L,
\label{eq:topI_HComp_MB_LL_result}\\
{\cal I}_{\rm Ann,MB}^{\rm LL} &=& -\frac{pM_{\rm Ann}^{\rm MB}}{8}e^{-\beta p}L =-\frac{p\pi^2T^2}{48}e^{-\beta p}L =2{\cal I}_{\rm Comp,MB}^{\rm LL}.
\label{eq:topI_HAnn_MB_LL_result}
\eea
\end{subequations}
Substitution into Eq.~\eqref{eq:topI_ImPi_I_relation} gives
\begin{subequations}
\label{eq:topI_MB_LL_individual_self_energies}
\bea
\left.\Im\Pi^{\mu{\rm I}}_{\ \ \mu}(P)\right|_{\rm Comp,MB}^{\rm LL} &=& \frac{5e^2g^2}{108\pi}T^2\big(1-e^{-p/T}\big)L,
\label{eq:topI_ImPi_Comp_MB_LL}\\
\left.\Im\Pi^{\mu{\rm I}}_{\ \ \mu}(P)\right|_{\rm Ann,MB}^{\rm LL} &=& \frac{5e^2g^2}{54\pi}T^2\big(1-e^{-p/T}\big)L.
\label{eq:topI_ImPi_Ann_MB_LL}
\eea
\end{subequations}
Therefore, the full quantum distributions give equal Compton and annihilation LL coefficients, whereas the incoming-distribution MB approximation gives ${\cal I}_{\rm Ann,MB}^{\rm LL} =2{\cal I}_{\rm Comp,MB}^{\rm LL}$. Nevertheless, adding the two MB channel contributions reproduces the full-statistics hard LL self-energy in Eq.~\eqref{eq:topI_ImPi_total_LL}, as follows:
\bea
M_C^{\rm MB}+M_{\rm Ann}^{\rm MB} &=& \frac{\pi^2T^2}{4}=2M_Q.
\label{eq:topI_LL_moment_sum_identity}
\eea
\subsubsection{Beyond leading-logarithmic contribution}
The leading-logarithmic calculation fixes only the coefficient of $\ln(4pT/k_c^2)$. To obtain the hard $2\leftrightarrow2$ contribution through BLL accuracy, we now retain the cutoff-independent terms accompanying the leading logarithm at leading order in the hard-photon expansion. Throughout this subsection, we use $p\gg T\gg k_c, \, a\equiv {k_c^2}/{4}, \, \ell_c\equiv {a}/{p}={k_c^2}/{4p}, \, L\equiv\ln\left({4pT}/{k_c^2}\right)=\ln\left({pT}/{a}\right).$ Here $k_c$ is the hard-soft separation scale, whereas $\ell_c$ is the lower $\ell$ boundary induced by the hard-virtuality condition. By BLL accuracy, we mean the leading logarithm together with the finite terms of order $pT^2e^{-p/T}$; terms suppressed by $T/p$ or $k_c/T$ are omitted. Starting from the exact thermal moments and the hard-domain operator derived above, we introduce $x\equiv k-\ell, \, k=x+\ell, \, E_q=p-x+\ell, \, A=x-\ell, \, S=2p-x+\ell=2p-A.$ Since $dk=dx$, the exchanged-quark virtuality and exact hard domain become $-\Delta_A=k^2-A^2=4\ell x, \, \ell\geq {a}/{p}, \,  {a}/{\ell}\leq x\leq p.$ Accordingly, Eq.~\eqref{eq:Projection_hard_fixed_l_operator} takes the form
\bea
{\cal P}_{\rm h}[F] &=& 2\int_{a/p}^{\infty}d\ell\int_{a/\ell}^{p}dx\,F(x+\ell,4p\ell).
\label{eq:topI_full_BLL_x_operator}
\eea
At the retained leading-power accuracy, one may set $1-n_F(E_q)=1$ in the second and third cuts. Outside the upper endpoint layer, $p-x\gg T$, one has $E_q=p-x+\ell\gg T$ and hence $n_F(E_q)$ is exponentially suppressed. In particular, for $x,\ell={\cal O}(T)$, the correction combines with the overall factor $e^{-p/T}$ to give a contribution of order $e^{-2p/T}$. The replacement is not pointwise valid in the boundary layer $x=p-y$, with $y={\cal O}(T)$, where $E_q=y+\ell={\cal O}(T)$. Nevertheless, this layer has width $\Delta x={\cal O}(T)$, and its kernels carry no leading $p$ enhancement because $k=x+\ell={\cal O}(p)$. Its contribution is therefore at most ${\cal O}(T^3e^{-p/T})$, which is suppressed by $T/p$ relative to the retained ${\cal O}(pT^2e^{-p/T})$ result.
\subsubsubsection{Compton process} 
The Compton contribution is ${\cal I}_{\rm Comp}={\cal I}_1+{\cal I}_2$. The first cut does not generate a leading-power logarithmic dependence on the exchanged-quark cutoff. Defining $r_\pm=(S\pm k)/2$, the statistical factors may be replaced by unity in the leading-power part of the $r$ integral. Since $r_+-r_-=k$ and $r_++r_-=S$, one obtains
\bea
\int_{r_-}^{r_+}dr\,\big[2Ar-k^2-AS\big] &=& A(r_+^2-r_-^2)-(k^2+AS)(r_+-r_-)=AkS-k(k^2+AS)=-k^3.
\eea
Consequently, $E_qI_1=-e^{-p/T}k\,n_F(E_q)/32$ at leading power. Using the exact fixed-$E_q$ interval $|p-E_q|\leq k\leq p+E_q$, for which $\int_{|p-E_q|}^{p+E_q}dk\,k = 2pE_q,$ gives
\bea
{\cal I}_1 &=& -\frac{p}{32}e^{-p/T}\int_0^\infty dE_q\,2E_qn_F(E_q)+{\cal O}\!\left(e^{-p/T}T^3\right)=-\frac{p\pi^2T^2}{192}e^{-p/T}+{\cal O}\!\left(e^{-p/T}T^3\right).
\label{eq:topI_full_BLL_I1}
\eea 
The regions in which either statistical factor differs appreciably from unity are confined to endpoint layers of width ${\cal O}(T)$. Their contribution is ${\cal O}\!\left(T^3e^{-p/T}\right)$ and is therefore suppressed by $T/p$ relative to the retained ${\cal O}\!\left(pT^2e^{-p/T}\right)$ term. The region $E_q>p$ is additionally exponentially suppressed by $e^{-p/T}n_F(E_q)\lesssim e^{-2p/T}$. The use of the unrestricted $k$ interval in Eq.~\eqref{eq:topI_full_BLL_I1} does not affect the result at the retained accuracy. For the first cut, the hard condition $\Delta_S=S^2-k^2\geq k_c^2$ gives the exact domain $E_q \geq \ell_c=\frac{k_c^2}{4p}, \, |p-E_q|\leq k\leq\sqrt{(p+E_q)^2-k_c^2}.$ Indeed, the exact hard-domain $k$ integral is
\bea
\int_{|p-E_q|}^{\sqrt{(p+E_q)^2-k_c^2}}dk\,k &=& 2pE_q-\frac{k_c^2}{2},
\label{eq:topI_full_BLL_I1_hard_k_integral}
\eea
whereas the unrestricted result is $2pE_q$. Consequently, the hard-boundary correction satisfies
\bea
\delta{\cal I}_1 &=& {\cal O}\left[ e^{-p/T}\left(k_c^2T+p\ell_c^2\right) \right] ={\cal O}\left(k_c^2T e^{-p/T}\right),
\eea
and is suppressed by $k_c^2/(pT)$ relative to the retained ${\cal O}(pT^2e^{-p/T})$ contribution. Any integrable endpoint logarithms in the exact $r$ integral can only multiply this power-suppressed correction and do not generate a leading-power $\ln k_c$ term. For the second cut, the exact identities $k^2+AS = 2pA+4\ell x, \,\, S=2p-A$ give
\bea
\frac{S}{16k^2}\Psi_C^{(1)}+\frac{k^2+AS}{32k^2}\Psi_C^{(0)} &=& \frac{p}{16(x+\ell)^2}\left(2\Psi_C^{(1)}+A\Psi_C^{(0)}\right)+\frac{-A\Psi_C^{(1)}+2\ell x\Psi_C^{(0)}}{16(x+\ell)^2},
\label{eq:topI_full_BLL_C_kernel_split}
\eea
where each moment on the right-hand side has arguments $(\ell,A)=(\ell,x-\ell)$. The second term is subleading in the hard-photon expansion, as we now show explicitly. We denote its contribution by
\bea
\delta{\cal I}_{2} &=& -\frac{e^{-p/T}}{8} \int_{a/p}^{\infty}d\ell\int_{a/\ell}^{p}dx\, \frac{-(x-\ell)\Psi_C^{(1)}(\ell,x-\ell) +2\ell x\Psi_C^{(0)}(\ell,x-\ell)} {(x+\ell)^2}.
\label{eq:topI_full_BLL_I2_subleading}
\eea
Since $r\geq\ell$, the Fermi factor $n_F(r)$ restricts the thermally relevant $\ell$ region to $\ell={\cal O}(T)$. Consequently, $\Psi_C^{(0)}={\cal O}(T)$ and $\Psi_C^{(1)}={\cal O}(T^2)$. The region $x={\cal O}(T)$ therefore contributes ${\cal O}(T^3e^{-p/T})$. 
For $x\gg T$, the Bose-enhancement factor $1+n_B(r+x-\ell)$ approaches unity exponentially, so that
\bea
\Psi_C^{(m)}(\ell,x-\ell) &=& \Phi_F^{(m)}(4p\ell) +{\cal O}\!\left(T^{m+1}e^{-x/T}\right), \qquad \Phi_F^{(m)}(4p\ell)\equiv \int_{\ell}^{\infty}dr\,r^m n_F(r).
\eea
Hence, for $\ell={\cal O}(T)$,
\bea
\frac{-(x-\ell)\Psi_C^{(1)}+2\ell x\Psi_C^{(0)}} {(x+\ell)^2} &=& \frac{2\ell\Phi_F^{(0)}(4p\ell) -\Phi_F^{(1)}(4p\ell)}{x} +{\cal O}\!\left(\frac{T^3}{x^2}, \frac{T^2}{x}e^{-x/T}\right).
\label{eq:topI_full_BLL_I2_subleading_large_x}
\eea
Writing $b_F(\ell) \equiv 2\ell\Phi_F^{(0)}(4p\ell)-\Phi_F^{(1)}(4p\ell),$ the coefficient of the apparent large-$x$ logarithm vanishes after the $\ell$ integration. Indeed, interchanging the order of the $\ell$ and $r$ integrations gives
\bea
\int_0^\infty d\ell\,b_F(\ell) &=& \int_0^\infty dr\,r^2n_F(r) -\int_0^\infty dr\,r^2n_F(r)=0.
\label{eq:topI_full_BLL_I2_subleading_cancellation}
\eea
Equivalently, at fixed $x$, the exact hard boundary is $\ell\geq a/x$. Since $b_F(0)=-\pi^2T^2/12$ is finite, the exact boundary changes the preceding cancellation only by
\bea
\int_{a/x}^{\infty}d\ell\,b_F(\ell) &=& -\int_0^{a/x}d\ell\,b_F(\ell) ={\cal O}\left(\frac{aT^2}{x}\right).
\label{eq:topI_full_BLL_I2_boundary_correction}
\eea
After including the remaining factor $1/x$, this boundary correction behaves as $aT^2/x^2$. Its integration over $T\lesssim x\leq p$ therefore gives
\bea
\int_T^p dx\,\frac{aT^2}{x^2} &=& aT^2\left(\frac{1}{T}-\frac{1}{p}\right) ={\cal O}(aT),
\label{eq:topI_full_BLL_I2_boundary_correction_final}
\eea
and hence contributes only ${\cal O}(aT e^{-p/T})$ to $\delta{\cal I}_2$. The remaining terms in Eq.~\eqref{eq:topI_full_BLL_I2_subleading_large_x} fall at least as $x^{-2}$ and contribute ${\cal O}(T^3e^{-p/T})$. Therefore,
$\delta{\cal I}_2$ is suppressed by $T/p$, together with additional cutoff-power suppression, relative to the retained ${\cal O}(pT^2e^{-p/T})$ contribution. Keeping only the leading power part, we obtain
\bea
{\cal I}_{2,p} &=& -\frac{p}{8}e^{-p/T} \int_{a/p}^{\infty}d\ell\int_{a/\ell}^{p}dx\, \frac{2\Psi_C^{(1)}(\ell,x-\ell) +(x-\ell)\Psi_C^{(0)}(\ell,x-\ell)} {(x+\ell)^2}.
\label{eq:topI_full_BLL_I2p}
\eea
Now, it is useful to combine the two moments into a single thermal integral:
\bea
2\Psi_C^{(1)}(\ell,x-\ell) +(x-\ell)\Psi_C^{(0)}(\ell,x-\ell) &=& \int_\ell^\infty dr\, (2r+x-\ell)W_C(r;\ell,x).
\label{eq:topI_full_BLL_C_combination}
\eea
where $W_C(r;\ell,x) \equiv n_F(r)\left[1+n_B(r+x-\ell)\right]$. To evaluate this integral, we use the convergent thermal-distribution expansions as
\bea
n_F(r) &=& \frac{1}{e^{r/T}+1} =\sum_{n=1}^{\infty}(-1)^{n+1}e^{-nr/T},\quad 1+n_B(r+x-\ell) = \frac{1}{1-e^{-(r+x-\ell)/T}} =\sum_{m=0}^{\infty}e^{-m(r+x-\ell)/T}.
\label{eq:topI_full_BLL_thermal_series}
\eea
Their product is therefore
\bea
W_C(r;\ell,x) &=& \sum_{n=1}^{\infty}\sum_{m=0}^{\infty} (-1)^{n+1} \exp\left[ -\frac{(n+m)r}{T} -\frac{m(x-\ell)}{T} \right].
\label{eq:topI_full_BLL_WC_series}
\eea
Defining $N\equiv n+m$, the $r$ integral associated with each coefficient $(-1)^{n+1}$ becomes
\bea
e^{-m(x-\ell)/T}\int_\ell^\infty dr\,(2r+x-\ell)e^{-Nr/T} &=& e^{-n\ell/T-mx/T}\left[\frac{T(x+\ell)}{N}+\frac{2T^2}{N^2}\right].
\label{eq:topI_full_BLL_r_integral}
\eea
Defining $F_{nm}(\ell,x) \equiv \frac{e^{-n\ell/T-mx/T}}{\ell+x},$ one finds the total-derivative identity
\bea
e^{-n\ell/T-mx/T}\left[\frac{T}{N(\ell+x)}+\frac{2T^2}{N^2(\ell+x)^2}\right] &=& -\frac{T^2}{N^2}\left(\partial_\ell+\partial_x\right)F_{nm}(\ell,x).
\label{eq:topI_full_BLL_total_derivative}
\eea
Consequently, the two-dimensional hard-domain integral reduces to its boundary. We define
\bea
B_{nm} &\equiv& -\int_{a/p}^{\infty}d\ell\int_{a/\ell}^{p}dx\,\left(\partial_\ell+\partial_x\right)F_{nm}(\ell,x) = \int_{a/p}^{\infty}\frac{d\ell}{\ell}\exp\left[-\frac{n\ell}{T}-\frac{ma}{\ell T}\right]-e^{-mp/T}\int_{a/p}^{\infty}d\ell\,\frac{e^{-n\ell/T}}{\ell+p}.
\label{eq:topI_full_BLL_Bnm_exact}
\eea
Hereafter, ${\cal O}_{\rm pow}$ denotes terms suppressed by at least one positive power of $T/p$ or $k_c/T$, possibly accompanied by logarithms. Exponentially suppressed corrections are displayed separately. The first and second terms on the right-hand side arise from the cutoff-induced hyperbolic boundary $x=a/\ell$ and the upper boundary $x=p$, respectively. For $m=0$, the first integral can be expressed in terms of the exponential integral. For $m\geq1$, its lower limit can be extended to zero, while the upper-boundary contribution and the correction generated by this extension are exponentially suppressed in $p/T$. Consequently,
\begin{subequations}
\label{eq:topI_full_BLL_Bnm_asymptotic}
\bea
B_{n0} &=& L-\ln n-\gamma_E+{\cal O}_{\rm pow}, \\
B_{nm} &=& 2K_0\left(\frac{2\sqrt{mna}}{T}\right) +{\cal O}\!\left(e^{-mp/T}\right) = \ln\left(\frac{T^2}{a}\right)-\ln(mn)-2\gamma_E +{\cal O}_{\rm pow} +{\cal O}\!\left(e^{-mp/T}\right), \qquad m,n\geq1.\nonumber\\ \label{eq.75b}
\eea
\end{subequations}
Here $K_0$ denotes the modified Bessel function of the second kind. The small-$a$ expansion in Eq.~\eqref{eq.75b} is taken at fixed $m$ and
$n$. When these expressions are inserted into the double sums below, the terms are grouped at fixed $N=m+n$; this specifies the summation
prescription for the conditionally convergent expressions. The $m=0$ part represents the unity in $1+n_B$, whereas $m\geq1$ is the genuine Bose-enhancement contribution. We now introduce the quantum-statistical constant
\bea
{\cal S}_Q &\equiv& \sum_{m,n=1}^{\infty}\frac{(-1)^{m-1}\ln(mn)}{(m+n)^2}.
\label{eq:topI_full_BLL_SQ_double}
\eea
Since $\ln(mn)/(m+n)^2$ is symmetric under $m\leftrightarrow n$, interchanging the summation indices shows that the same constant is obtained with the alternating factor $(-1)^{n+1}$. Grouping the terms at fixed $m+n$ gives the equivalent one-dimensional representation
\bea
{\cal S}_Q &=& \frac12\sum_{q=1}^{\infty}\frac{1}{q^2}\ln\left[\frac{2q}{4^q}\binom{2q}{q}\right]=0.260797352203946\ldots\, .
\label{eq:topI_full_BLL_SQ_single}
\eea
The required elementary sums follow from the Dirichlet eta function,
\bea
\eta(s) &\equiv& \sum_{n=1}^{\infty}\frac{(-1)^{n+1}}{n^s} =\left(1-2^{1-s}\right)\zeta(s), \qquad \eta(2) \equiv \sum_{n=1}^{\infty}\frac{(-1)^{n+1}}{n^2}=\frac{\zeta(2)}{2}, \quad \eta'(2)=\frac{\ln2}{2}\zeta(2)+\frac12\zeta'(2), \nonumber \\
\sum_{m,n=1}^{\infty}\frac{(-1)^{n+1}}{(m+n)^2} &=& \frac{\zeta(2)}{4}, \quad \sum_{m,n=1}^{\infty}\frac{(-1)^m}{(m+n)^2}= -\frac{\zeta(2)}{4}.
\label{eq:topI_full_BLL_elementary_sums}
\eea
Equations~\eqref{eq:topI_full_BLL_I2p}--\eqref{eq:topI_full_BLL_Bnm_asymptotic} give
\bea
{\cal I}_{2,p} &=& -\frac{pT^2}{8}e^{-p/T}J_C, \qquad \text{where} \qquad J_C\equiv\sum_{n=1}^{\infty}\sum_{m=0}^{\infty}\frac{(-1)^{n+1}}{(n+m)^2}B_{nm}.
\label{eq:topI_full_BLL_JC_definition}
\eea
Separating $m=0$ and $m\geq1$ yields
\bea
J_C^{(m=0)} &=& \frac{\zeta(2)}{2}\left[L-\gamma_E+\ln2\right]+\frac12\zeta'(2), \qquad J_C^{(m\geq1)} = \frac{\zeta(2)}{4}\left[\ln\left(\frac{T^2}{a}\right)-2\gamma_E\right]-{\cal S}_Q.
\eea
Using $\ln(T^2/a)=L-\ln(p/T)$ and $\zeta(2)=\pi^2/6$, their sum is
\bea
J_C &=& \frac{\pi^2}{8}\left[L-\frac13\ln\left(\frac{p}{T}\right)+\frac23\ln2-\frac43\gamma_E+\frac23\frac{\zeta'(2)}{\zeta(2)}-\frac{8}{\pi^2}{\cal S}_Q\right].
\label{eq:topI_full_BLL_JC_result}
\eea
Combining ${\cal I}_{2,p}$ with the first-cut contribution in Eq.~\eqref{eq:topI_full_BLL_I1}, while omitting $\delta{\cal I}_2$, whose power suppression was established in Eqs.~\eqref{eq:topI_full_BLL_I2_subleading}--\eqref{eq:topI_full_BLL_I2_boundary_correction_final}, gives
\bea
{\cal I}_{\rm Comp}^{\rm BLL} &=& -\frac{p\pi^2T^2}{64}e^{-p/T}\left[L-\frac13\ln\left(\frac{p}{T}\right)+C_C\right], \nonumber\\
 C_C &\equiv &\frac13+\frac23\ln2-\frac43\gamma_E+\frac23\frac{\zeta'(2)}{\zeta(2)}-\frac{8}{\pi^2}{\cal S}_Q=-0.565557799553834\ldots\, .
\label{eq:topI_full_BLL_CC}
\eea
\subsubsubsection{Annihilation process}
The annihilation contribution is ${\cal I}_{\rm Ann}={\cal I}_3$. Using $k=x+\ell$, $A=x-\ell$, $S=2p-A$, and $k^2-AS=2(x^2+\ell^2)-2pA$, the third-cut kernel becomes
\bea
\frac{S}{16k^2}\Psi_{\rm Ann}^{(1)} -\frac{k^2-AS}{32k^2}\Psi_{\rm Ann}^{(0)} &=& \frac{p}{16(x+\ell)^2} \left[2\Psi_{\rm Ann}^{(1)}+A\Psi_{\rm Ann}^{(0)}\right]  -\frac{A\Psi_{\rm Ann}^{(1)}+(x^2+\ell^2)\Psi_{\rm Ann}^{(0)}} {16(x+\ell)^2}.
\label{eq:topI_full_BLL_A_kernel_split}
\eea
All moments on the right-hand side have arguments $(\ell,A)=(\ell,x-\ell)$. After making the leading-power replacement $1-n_F(E_q)=1$ discussed above, we write
\bea
{\cal I}_3 &=& {\cal I}_{3,p}+\delta{\cal I}_{3,1}+\Delta{\cal I}_3 +{\cal O}\!\left(T^3e^{-p/T}\right),
\label{eq:topI_full_BLL_I3_decomposition}
\eea
where
\begin{subequations}
\label{eq:topI_full_BLL_I3_parts}
\bea
{\cal I}_{3,p} &\equiv& -\frac{p}{8}e^{-p/T} \int_{\ell_c}^{\infty}d\ell\int_{a/\ell}^{p}dx\, \frac{2\Psi_{\rm Ann}^{(1)}+(x-\ell)\Psi_{\rm Ann}^{(0)}}{(x+\ell)^2},
\label{eq:topI_full_BLL_I3p} \\
\delta{\cal I}_{3,1} &\equiv& \frac{e^{-p/T}}{8} \int_{\ell_c}^{\infty}d\ell\int_{a/\ell}^{p}dx\, \frac{(x-\ell)\Psi_{\rm Ann}^{(1)}}{(x+\ell)^2}, \qquad \Delta{\cal I}_3 \equiv \frac{e^{-p/T}}{8} \int_{\ell_c}^{\infty}d\ell\int_{a/\ell}^{p}dx\, \frac{x^2+\ell^2}{(x+\ell)^2}\Psi_{\rm Ann}^{(0)}.
\eea
\end{subequations}
The two terms without an explicit factor of $p$ require separate estimation: $\delta{\cal I}_{3,1}$ is subleading, whereas $\Delta{\cal I}_3$ contributes at the retained order. Since $0\leq1-n_F\leq1$, the exact annihilation moments satisfy
\bea
0\leq\Psi_{\rm Ann}^{(j)}(\ell,x-\ell) &\leq& \Phi_B^{(j)}(4p\ell) \equiv\int_\ell^\infty dE_g\,E_g^j n_B(E_g), \qquad j=0,1.
\label{eq:topI_full_BLL_A_moment_bound}
\eea
Here $\Phi_B^{(1)}(0)=\pi^2T^2/6$, while
\bea
\Phi_B^{(0)}(4p\ell) &=& -T\ln\left(1-e^{-\ell/T}\right) =T\ln\left(\frac{T}{\ell}\right)+{\cal O}(\ell), \qquad \ell\ll T.
\label{eq:topI_full_BLL_A_Bose_endpoint}
\eea
Thus, the zeroth moment has an integrable logarithmic endpoint singularity, and both moments are exponentially suppressed for $\ell\gg T$. For the term proportional to $\Psi_{\rm Ann}^{(1)}$, using $|x-\ell|/(x+\ell)^2\leq1/(x+\ell)$ and enlarging the integration domain gives the cutoff-independent bound
\bea
\left|\delta{\cal I}_{3,1}\right| &\leq& \frac{e^{-p/T}}{8} \int_0^\infty d\ell\,\Phi_B^{(1)}(4p\ell) \int_0^p\frac{dx}{x+\ell} = \frac{e^{-p/T}}{8} \int_0^\infty d\ell\,\Phi_B^{(1)}(4p\ell) \ln\left(1+\frac{p}{\ell}\right) \nonumber\\
&=& {\cal O}\!\left( T^3e^{-p/T}\left[1+\ln\left(\frac{p}{T}\right)\right] \right).
\label{eq:topI_full_BLL_A_Psi1_subleading}
\eea
The $\ell$ integral is finite at the origin. Therefore, $\delta{\cal I}_{3,1}$ is suppressed by $T/p$, up to a logarithm, relative to the retained contribution, without any enhancement as $k_c\to0$. To evaluate $\Delta{\cal I}_3$, first note that $E_g\geq\ell$ implies $E_g+x-\ell\geq x$. Consequently,
\bea
0\leq \Phi_B^{(0)}(4p\ell)-\Psi_{\rm Ann}^{(0)}(\ell,x-\ell) &=& \int_\ell^\infty dE_g\, n_B(E_g)n_F(E_g+x-\ell) \leq e^{-x/T}\Phi_B^{(0)}(4p\ell).
\label{eq:topI_full_BLL_A_Psi0_bound}
\eea
Replacing $\Psi_{\rm Ann}^{(0)}$ by $\Phi_B^{(0)}$ in $\Delta{\cal I}_3$ therefore changes the result by at most ${\cal O}(T^3e^{-p/T})$. Moreover,
${x^2+\ell^2}/{(x+\ell)^2} = 1-{2\ell x}/{(x+\ell)^2},$ and the correction to replacing this ratio by unity obeys
\bea
\int_0^\infty d\ell\,\Phi_B^{(0)}(4p\ell) \int_0^p dx\,\frac{2\ell x}{(x+\ell)^2} &\leq& 2\int_0^\infty d\ell\,\ell\Phi_B^{(0)}(4p\ell) \ln\left(1+\frac{p}{\ell}\right) = {\cal O}\!\left( T^3\left[1+\ln\left(\frac{p}{T}\right)\right] \right).
\label{eq:topI_full_BLL_A_geometry_bound}
\eea
Keeping the exact hard boundary in the leading term, we obtain
\bea
\Delta{\cal I}_3 &=& \frac{e^{-p/T}}{8} \int_{\ell_c}^\infty d\ell\, \left(p-\frac{a}{\ell}\right)\Phi_B^{(0)}(4p\ell)  +{\cal O}\!\left( T^3e^{-p/T}\left[1+\ln\left(\frac{p}{T}\right)\right] \right).
\label{eq:topI_full_BLL_DeltaI3_hard}
\eea
The cutoff corrections to replacing this integral by $p\int_0^\infty d\ell\,\Phi_B^{(0)}(4p\ell)$ are power suppressed. Indeed, Eq.~\eqref{eq:topI_full_BLL_A_Bose_endpoint}, together with $\ell_c=a/p$ and $\ln(T/\ell_c)=L$, gives
\bea
p\int_0^{\ell_c}d\ell\,\Phi_B^{(0)}(4p\ell) &=& {\cal O}\!\left(aT[1+L]\right), \qquad a\int_{\ell_c}^{\infty}\frac{d\ell}{\ell}\, \Phi_B^{(0)}(4p\ell) = {\cal O}\!\left(aT[1+L^2]\right).
\label{eq:topI_full_BLL_DeltaI3_boundary}
\eea
Interchanging the order of the thermal integrations yields
\bea
\int_0^\infty d\ell\,\Phi_B^{(0)}(4p\ell) &=& \int_0^\infty dE_g\,n_B(E_g)\int_0^{E_g}d\ell =\int_0^\infty dE_g\,E_g n_B(E_g) =\frac{\pi^2T^2}{6}.
\eea
Thus,
\bea
\Delta{\cal I}_3 &=& \frac{p\pi^2T^2}{48}e^{-p/T}  +{\cal O}\!\left( e^{-p/T} \left\{ T^3\left[1+\ln\left(\frac{p}{T}\right)\right] + aT[1+L^2] \right\} \right).
\label{eq:topI_full_BLL_DeltaI3}
\eea
This term contributes at BLL order because its $x$ integration extends over an interval of order $p$, yielding a contribution
of order $pT^2e^{-p/T}$. We now evaluate ${\cal I}_{3,p}$. Combining the two moments gives
\bea
2\Psi_{\rm Ann}^{(1)}(\ell,x-\ell) +(x-\ell)\Psi_{\rm Ann}^{(0)}(\ell,x-\ell) &=& \int_\ell^\infty dE_g\, (2E_g+x-\ell)W_{\rm Ann}(E_g;\ell,x),
\label{eq:topI_full_BLL_A_combination}
\eea
where $W_{\rm Ann}(E_g;\ell,x)\equiv n_B(E_g)[1-n_F(E_g+x-\ell)]$. For $E_g>0$ and $E_g+x-\ell>0$, the convergent thermal expansions are
\bea 
\hspace{-.7cm} n_B(E_g) &=& \frac{1}{e^{E_g/T}-1} =\sum_{n=1}^{\infty}e^{-nE_g/T}, \,\, 1-n_F(E_g+x-\ell) = \frac{1}{1+e^{-(E_g+x-\ell)/T}} =\sum_{m=0}^{\infty}(-1)^m e^{-m(E_g+x-\ell)/T}.
\label{eq:topI_full_BLL_A_thermal_series}
\eea
These conditions hold in the interior of the hard domain. The product is therefore
\bea
W_{\rm Ann}(E_g;\ell,x) &=& \sum_{n=1}^{\infty}\sum_{m=0}^{\infty} (-1)^m \exp\left[ -\frac{(n+m)E_g}{T} -\frac{m(x-\ell)}{T} \right].
\label{eq:topI_full_BLL_WA_series}
\eea
The $m=0$ term represents the unity in the final-state Pauli-blocking factor, while $m\geq1$ gives its correction. With $N=n+m$, the thermal-energy integral associated with each coefficient $(-1)^m$ is
\bea
e^{-m(x-\ell)/T} \int_\ell^\infty dE_g\, (2E_g+x-\ell)e^{-NE_g/T} &=& e^{-n\ell/T-mx/T} \left[ \frac{T(x+\ell)}{N}+\frac{2T^2}{N^2} \right].
\label{eq:topI_full_BLL_A_energy_integral}
\eea
Thus, the total-derivative identity and hard-domain boundary function derived for the Compton process, Eqs.~\eqref{eq:topI_full_BLL_total_derivative} and \eqref{eq:topI_full_BLL_Bnm_exact}, apply here as well. Only the statistical coefficient changes. Hence,
\bea
{\cal I}_{3,p} &=& -\frac{pT^2}{8}e^{-p/T}J_{\rm Ann}, \qquad J_{\rm Ann}\equiv \sum_{n=1}^{\infty}\sum_{m=0}^{\infty} \frac{(-1)^m}{(n+m)^2}B_{nm}.
\label{eq:topI_full_BLL_JA_definition}
\eea
We use the same boundary functions $B_{nm}$ and the same grouping at fixed $N=n+m$ as in the Compton calculation. In particular, the finite-cutoff expression is understood before extracting its BLL terms using Eq.~\eqref{eq:topI_full_BLL_Bnm_asymptotic}. The required sums are
\bea
\sum_{n=1}^{\infty}\frac{\ln n}{n^2} &=& -\zeta'(2), \qquad  \sum_{m,n=1}^{\infty}\frac{(-1)^m}{(m+n)^2} = -\frac{\zeta(2)}{4}, \qquad \sum_{m,n=1}^{\infty} \frac{(-1)^m\ln(mn)}{(m+n)^2} = -{\cal S}_Q,
\label{eq:topI_full_BLL_A_required_sums}
\eea
where ${\cal S}_Q$ is the constant defined in Eq.~\eqref{eq:topI_full_BLL_SQ_double}. The last sign follows directly from $(-1)^m=-(-1)^{m-1}$. Separating $m=0$ and $m\geq1$, and retaining terms through BLL accuracy, gives
\bea
J_{\rm Ann}^{(m=0)} &=& \sum_{n=1}^{\infty}\frac{L-\ln n-\gamma_E}{n^2} =\zeta(2)(L-\gamma_E)+\zeta'(2), \qquad J_{\rm Ann}^{(m\geq1)} = -\frac{\zeta(2)}{4} \left[ \ln\left(\frac{T^2}{a}\right)-2\gamma_E \right]+{\cal S}_Q.
\label{eq:topI_full_BLL_JA_sectors}
\eea
Using $\ln(T^2/a)=L-\ln(p/T)$ and $\zeta(2)=\pi^2/6$, their sum is
\bea
J_{\rm Ann} &=& \frac{\pi^2}{8} \left[ L+\frac13\ln\left(\frac{p}{T}\right) -\frac23\gamma_E +\frac43\frac{\zeta'(2)}{\zeta(2)} +\frac{8}{\pi^2}{\cal S}_Q \right]
\label{eq:topI_full_BLL_JA_result}
\eea
at the retained accuracy. Combining ${\cal I}_{3,p}$ with $\Delta{\cal I}_3$ from Eq.~\eqref{eq:topI_full_BLL_DeltaI3}, and omitting the power-suppressed $\delta{\cal I}_{3,1}$, we obtain
\bea
{\cal I}_{\rm Ann}^{\rm BLL} &=& -\frac{p\pi^2T^2}{64}e^{-p/T} \left[ L+\frac13\ln\left(\frac{p}{T}\right)+C_{\rm Ann} \right],
\label{eq:topI_BLL_annihilation_single}
\eea
where
\bea
C_{\rm Ann} &\equiv& -\frac43-\frac23\gamma_E +\frac43\frac{\zeta'(2)}{\zeta(2)} +\frac{8}{\pi^2}{\cal S}_Q =-2.26669739606500\ldots\, .
\label{eq:topI_full_BLL_CA}
\eea
The term $-4/3$ in $C_{\rm Ann}$ comes from the positive contribution $\Delta{\cal I}_3=p\pi^2T^2e^{-p/T}/48$. The coefficient of $L$ is $ -p\pi^2T^2e^{-p/T}/64$, in agreement with the annihilation LL result. Substituting Eqs.~\eqref{eq:topI_full_BLL_CC} and \eqref{eq:topI_BLL_annihilation_single} into Eq.~\eqref{eq:topI_ImPi_I_relation} gives the individual full-statistics BLL self-energies:
\begin{subequations}
\label{eq:topI_full_BLL_individual_self_energies}
\bea
\left.\Im\Pi^{\mu I}_{\ \ \mu}(P)\right|_{\rm Comp}^{\rm BLL} &=& \frac{5e^2g^2}{72\pi}T^2\left(1-e^{-p/T}\right) \left[ L-\frac13\ln\left(\frac{p}{T}\right)+C_C \right],
\label{eq:topI_ImPi_Comp_BLL}\\
\left.\Im\Pi^{\mu I}_{\ \ \mu}(P)\right|_{\rm Ann}^{\rm BLL} &=& \frac{5e^2g^2}{72\pi}T^2\left(1-e^{-p/T}\right) \left[ L+\frac13\ln\left(\frac{p}{T}\right)+C_{\rm Ann} \right].
\label{eq:topI_ImPi_Ann_BLL}
\eea
\end{subequations}
Adding the two phase-space contributions gives
\bea 
{\cal I}_{\rm hard}^{\rm BLL} &=& -\frac{p\pi^2T^2}{32}e^{-p/T} \left[ L+\frac13\ln2-\frac12-\gamma_E +\frac{\zeta'(2)}{\zeta(2)}
\right].
\label{eq:topI_full_BLL_total_I}
\eea
Consequently,
\bea
\left.\Im\Pi^{\mu I}_{\ \ \mu}(P)\right|_{\rm hard}^{\rm BLL} &=& \frac{5e^2g^2}{36\pi}T^2\left(1-e^{-p/T}\right) \left[ L+\frac13\ln2-\frac12-\gamma_E +\frac{\zeta'(2)}{\zeta(2)} \right].
\label{eq:topI_ImPi_total_BLL}
\eea
The channel-dependent terms $\mp\frac13\ln(p/T)$ and $\mp8{\cal S}_Q/\pi^2$ cancel in the sum, whereas the terms proportional to $L$ add. Thus, the hard Compton and annihilation contributions retain the same infrared cutoff dependence.

For comparison, applying the Maxwell--Boltzmann approximation only to the incoming distributions gives the final individual BLL self-energies
\begin{subequations}
\label{eq:topI_MB_BLL_individual_self_energies}
\bea
\left.\Im\Pi^{\mu I}_{\ \ \mu}(P)\right|_{\rm Comp,MB}^{\rm BLL} &=& \frac{5e^2g^2}{108\pi}T^2\left(1-e^{-p/T}\right) \left[ L+\ln2+\frac12-\gamma_E+\frac{\zeta'(2)}{\zeta(2)} \right],
\label{eq:topI_ImPi_Comp_MB_BLL} \\
\left.\Im\Pi^{\mu I}_{\ \ \mu}(P)\right|_{\rm Ann,MB}^{\rm BLL} &=& \frac{5e^2g^2}{54\pi}T^2\left(1-e^{-p/T}\right) \left[ L-1-\gamma_E+\frac{\zeta'(2)}{\zeta(2)} \right].
\label{eq:topI_ImPi_Ann_MB_BLL}
\eea
\end{subequations}
Adding the two MB channel contributions reproduces the full-statistics hard BLL self-energy obtained above. Thus, the MB approximation changes the relative contributions of the two physical channels but leaves their summed BLL self-energy unchanged.
\subsection{Topology $II$}
\label{topo2}
\begin{figure}[htb]
	\centering
	\includegraphics[scale=0.45]{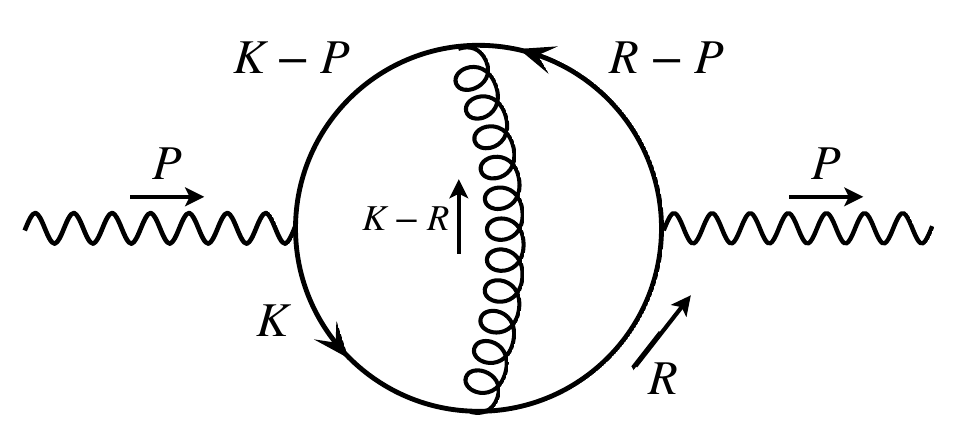}
	\caption{Two-loop photon self-energy for Topology $II$ }
	\label{photon_self_2}
\end{figure}
We next consider the contribution from topology~$II$, shown in Fig.~\ref{photon_self_2}. The photon self-energy is
\bea
\hspace{-.5cm} i\Pi^{II}_{\mu\nu}(P) &=& -\sum_f q_f^2 \int_K\int_R \Tr\Big[ (-ie\gamma_\mu)S(K) (-ig\gamma_\alpha t_a)S(R) (-ie\gamma_\nu)S(R-P) (-ig\gamma_\beta t_b)S(K-P) D^{\alpha\beta}_{ab}(K-R) \Big] \nonumber\\
&=& -\frac{5}{9} \int_K\int_R \Tr\Bigg[ (-ie\gamma_\mu) \frac{i\slashed K}{K^2} (-ig\gamma_\alpha t_a) \frac{i\slashed R}{R^2} (-ie\gamma_\nu) \frac{i(\slashed R-\slashed P)}{(R-P)^2} (-ig\gamma_\beta t_b) \frac{i(\slashed K-\slashed P)}{(K-P)^2} \left( \frac{-i\delta_{ab}g^{\alpha\beta}}{(K-R)^2} \right) \Bigg].
\label{eq:TopologyII}
\eea
Using the standard form of free gluon and quark propagators, the self-energy can be written as
\bea
\Pi^{\mu\,II}_{\mu}(P) &=& \frac{5}{9} e^2g^2 \int_K\int_R \Tr[t_at_b]\delta_{ab} \frac{ \Tr\!\left[ \gamma^\mu \slashed K
\gamma^\alpha \slashed R \gamma_\mu(\slashed R-\slashed P) \gamma_\alpha(\slashed K-\slashed P) \right] }{ K^2R^2(R-P)^2(K-P)^2(K-R)^2 }.
\label{eq:TopologyII_trace}
\eea
Let us choose the momentum exchanges as, $L=R-P,\, G=K-R,\, Q=K-P$. Therefore,
\bea
{\rm Tr}\left[ \gamma_\mu\slashed K \gamma_\alpha\slashed R \gamma^\mu(\slashed R-\slashed P) \gamma^\alpha\slashed Q \right]
&=& - 32\,(K\cdot L)(R\cdot Q) \, , \label{eq_Trace_2}
\eea
Thus, using Eqs.~\eqref{eq3a} and \eqref{eq_Trace_2}, one gets
\bea
\hspace{-1cm}\Pi^{\mu\,II}_{\mu}(P) &=& -\frac{640}{9}e^2g^2 T\sum_{k_0}T\sum_{r_0} \int_{\bf k}\int_{\bf r} \frac{ (K\cdot L)(R\cdot Q) }{ K^2Q^2R^2L^2G^2 } = -\frac{640}{9}e^2g^2 T\sum_{k_0}T\sum_{r_0} \int_{\bf k}\int_{\bf r} \frac{ (A_0+A_1r_0+A_2r_0^2)}{ K^2Q^2R^2L^2G^2. }
\label{eq:topII_after_trace}
\eea
with $A_{0}, A_{1}$ and $A_{2}$ given by
\bea
A_0 = k_0p_0(\mathbf r\cdot\mathbf q)+(\mathbf k\cdot\mathbf l)(\mathbf r\cdot\mathbf q), \qquad A_1 =-\left[ k_0p_0q_0+k_0(\mathbf r\cdot\mathbf q) +q_0(\mathbf k\cdot\mathbf l) \right], \qquad A_2 = k_0q_0 .
\eea
where $\mathbf l=\mathbf r-\mathbf p,\, \mathbf q=\mathbf k-\mathbf p,\, q_0=k_0-p_0.$ We evaluate the $r_0$ Matsubara sum using the Saclay method~\cite{Pisarski:1987wc}; the required sums~\cite{Haque:2024gva} are
{\footnotesize
\begin{subequations}
\bea
T\sum_{r_0} \Delta_F(R)\Delta_F(P-R)\Delta_B(K-R) &=& \sum_{s,s_1,s_2=\pm} \frac{ss_1s_2}{8E_rE_lE_g}  \left[  \frac{1-n_F(sE_r)+n_B(s_2E_g)} {k_0-sE_r-s_2E_g} - \frac{1-n_F(sE_r)-n_F(s_1E_l)} {p_0-sE_r - s_1E_l} \right]\nonumber \\
&\times&  \frac{1}{k_0-p_0-s_2E_g + s_1E_l} \, ,  
\label{eq:R0_sum} \\
T\sum_{r_0} r_{0} \, \Delta_{F}(R) \Delta_{F}(P-R) \Delta_{B}(K-R) &=& \sum_{s, s_1,s_2=\pm} \frac{s_1s_2}{8 E_lE_g}  \left[  \frac{1-n_F(sE_r)+n_B(s_2E_g)} {k_0-sE_r-s_2E_g} - \frac{1-n_F(sE_r)-n_F(s_1E_l)} {p_0-sE_r - s_1E_l} \right]\nonumber \\
&\times&  \frac{1}{k_0-p_0-s_2E_g + s_1E_l} \, ,  
\label{eq:R1_sum} \\
T\sum_{r_0} r_{0}^{2} \,\Delta_{F}(R) \Delta_{F}(P-R) \Delta_{B}(K-R) &=& \sum_{s,s_1,s_2=\pm} \frac{ss_1s_2E_r}{8E_lE_g} \left[\frac{ 1-n_F(sE_r)+n_B(s_2E_g) }{ k_0-sE_r-s_2E_g } - \frac{1-n_F(sE_r)-n_F(s_1E_l) }{ p_0-sE_r-s_1E_l } \right] \nonumber \\
&\times &\frac{1}{k_0-p_0-s_2E_g + s_1E_l} + \sum_{s_1,s_2=\pm} \frac{s_1s_2}{4E_lE_g} \frac{n_F(s_1E_l)+n_B(s_2E_g) }{ k_0-p_0-s_2E_g + s_1E_l }.
\label{eq:R2_sum} 
\eea 
\end{subequations} }
Here we have, $E_r=|\mathbf r|,\, E_l=|\mathbf r-\mathbf p|,\, E_g=|\mathbf k-\mathbf r|$. Using the Eqs.~\eqref{eq:R0_sum}, \eqref{eq:R1_sum}, \eqref{eq:R2_sum} we can do the $r_{0}$ frequency sum and one obtains 
\bea
\Pi^{\mu\,II}_{\mu}(P) &=& -\frac{640}{9}e^2g^2 T\sum_{k_0} \int_{\bf k}\int_{\bf r} \frac{1}{K^2Q^2} \Bigg\{ \sum_{s,s_1,s_2=\pm} \frac{s_1s_2}{8E_lE_g} \frac{\mathcal M_s}{k_0-p_0-s_2E_g + s_1E_l} \bigg[\frac{1-n_F(sE_r)+n_B(s_2E_g)} {k_0-sE_r-s_2E_g}  \nonumber \\
&-& \frac{1-n_F(sE_r)-n_F(s_1E_l)} {p_0-sE_r-s_1E_l}\bigg] + \sum_{s_1,s_2=\pm} \frac{s_1s_2A_2}{4E_lE_g} \frac{n_F(s_1E_l)+n_B(s_2E_g)} {k_0-p_0-s_2E_g+s_1E_l} \Bigg\}.
\label{eq:topII_after_r0_sum_compact}
\eea
where
\bea
\mathcal M_s = A_1+sE_rA_2+\frac{sA_0}{E_r}, \qquad \mathcal M_+=\mathcal M_{s=+1}, \qquad \mathcal M_-=\mathcal M_{s=-1}.
\label{M_def}
\eea
Using the following relations $n_F(-E_r)=1-n_F(E_r), \, n_F(-E_l)=1-n_F(E_l), \, n_B(-E_g)=-\left[1+n_B(E_g)\right]$ and doing the above sum over $s, s_1$ and $s_2$, the photon self-energy in Eq.~\eqref{eq:topII_after_trace} can be written as
{\footnotesize
\bea
\Pi^{\mu\,II}_{\mu}(P) &=&- \frac{640}{9}e^2g^2 T\sum_{k_0} \int_{\bf k}\int_{\bf r} \frac{1}{K^2Q^2} \Bigg\{ \frac{1}{8E_lE_g}
\Bigg\{ \frac{1}{k_0-p_0-E_g+E_l} \Bigg\{ \mathcal M_+ \bigg[ \frac{1-n_F(E_r)+n_B(E_g)} {k_0-E_r-E_g}  - \frac{1-n_F(E_r)-n_F(E_l)} {p_0-E_r-E_l} \bigg] \nonumber \\
&+& \mathcal M_- \bigg[ \frac{n_F(E_r)+n_B(E_g)} {k_0+E_r-E_g} - \frac{n_F(E_r)-n_F(E_l)} {p_0+E_r-E_l} \bigg] \Bigg\} + \frac{1}{k_0-p_0+E_g+E_l} \Bigg\{ \mathcal M_+ \bigg[ \frac{n_F(E_r)+n_B(E_g)} {k_0-E_r+E_g} \nonumber \\
&+& \frac{1-n_F(E_r)-n_F(E_l)} {p_0-E_r-E_l} \bigg]  + \mathcal M_- \bigg[ \frac{1+n_B(E_g)-n_F(E_r)} {k_0+E_r+E_g} + \frac{n_F(E_r)-n_F(E_l)} {p_0+E_r-E_l} \bigg] \Bigg\} - \frac{1}{k_0-p_0-E_g-E_l} \nonumber \\
&\times&  \Bigg\{ \mathcal M_+ \bigg[ \frac{1-n_F(E_r)+n_B(E_g)} {k_0-E_r-E_g} + \frac{n_F(E_r)-n_F(E_l)} {p_0-E_r+E_l} \bigg] +\mathcal M_- \bigg[ \frac{n_F(E_r)+n_B(E_g)} {k_0+E_r-E_g} + \frac{1-n_F(E_r)-n_F(E_l)} {p_0+E_r+E_l} \bigg] \Bigg\} \nonumber\\
&-& \frac{1}{k_0-p_0+E_g-E_l} \Bigg\{ \mathcal M_+ \bigg[ \frac{n_F(E_r)+n_B(E_g)} {k_0-E_r+E_g} - \frac{n_F(E_r)-n_F(E_l)} {p_0-E_r+E_l} \bigg] + \mathcal M_- \bigg[ \frac{1+n_B(E_g)-n_F(E_r)} {k_0+E_r+E_g} \nonumber \\
&-& \frac{1-n_F(E_r)-n_F(E_l)} {p_0+E_r+E_l} \bigg] \Bigg\} \Bigg\} + \frac{A_2}{4E_lE_g} \Bigg\{ \big[n_F(E_l)+n_B(E_g)\big] \bigg[ \frac{1}{k_0-p_0-E_g+E_l} - \frac{1}{k_0-p_0+E_g-E_l} \bigg] \nonumber\\
&+& \big[1+n_B(E_g)-n_F(E_l)\big] \bigg[ \frac{1}{k_0-p_0+E_g+E_l} - \frac{1}{k_0-p_0-E_g-E_l} \bigg] \Bigg\} \Bigg\}.
\label{eq:topII_after_r0_sum}
\eea }
For later convenience, we define the shorthand notation for the statistical factors as,
\begin{align}
W(x)&\equiv \frac{1}{2}-n_F(x), \qquad W_q\equiv \frac{1}{2}-n_F(E_q), \nonumber \\
\qquad F_1 & \equiv 1-n_F(E_r)+n_B(E_g), \qquad F_2\equiv n_F(E_r)+n_B(E_g), \qquad G_1\equiv 1-n_F(E_r)-n_F(E_l), \nonumber\\
G_2&\equiv  n_F(E_r)-n_F(E_l), \qquad H_1 \equiv n_F(E_l)+n_B(E_g),\qquad H_2\equiv 1+n_B(E_g)-n_F(E_l).
\end{align}

The remaining fermionic Matsubara sum over $k_0$ is evaluated by contour integration; details are given in Appendices~\ref{matsubara_II} and~\ref{imaginary_II}. We retain only the residue terms that generate physical $2\leftrightarrow2$ cuts. Using Eqs.~\eqref{eq87a}--\eqref{eq87pp} and the corresponding imaginary parts in Eqs.~\eqref{91a}--\eqref{91pp}, the photon self-energy in Eq.~\eqref{eq:topII_after_r0_sum} becomes:
\bea
\hspace{-0.7cm}\Im \Pi^{\mu\,II}_{\mu}(P) &=& -\frac{320\pi}{9}e^2g^2 \int\frac{d^3k}{(2\pi)^3} \int\frac{d^3r}{(2\pi)^3} \frac{1}{8E_lE_g} \Bigg\{ \delta(E_r+E_g-E_q-p_0)\,{\cal C}_1+\delta(E_g+E_q-E_r-p_0)\,{\cal C}_2 +{\cal C}_3\nonumber\\
&\times& \delta(E_r+E_q-E_g-p_0)   +\delta(k+E_l-E_g-p_0)\,{\cal C}_4+\delta(E_g+E_l-k-p_0)\,{\cal C}_5 +\delta(k+E_g-E_l-p_0)\,{\cal C}_6 \Bigg\}, \nonumber\\
\label{eq:top:2:after_im_parts}
\eea
where we have,  
\bea
{\cal C}_1 &=& F_1\left[W(E_r+E_g)-W_q\right] \frac{2E_l\,{\cal N}_1} {\left[(E_r+E_g)^2-k^2\right]\left[(E_q-E_g)^2-E_l^2\right]},
\nonumber\\
{\cal C}_2 &=& -F_2\left[W(E_g-E_r)+W_q\right] \frac{2E_l\,{\cal N}_2} {\left[(E_g-E_r)^2-k^2\right]\left[(E_g+E_q)^2-E_l^2\right]}, \nonumber\\
{\cal C}_3 &=& -F_2\left[W(E_r-E_g)+W_q\right] \frac{2E_l\,{\cal N}_3} {\left[(E_r-E_g)^2-k^2\right]\left[(E_q-E_g)^2-E_l^2\right]},
\nonumber\\
{\cal C}_4 &=& \frac{H_1\left[W(k)-W(E_g-E_l)\right]}{E_q^2-(E_l-E_g)^2} \Bigg[ \frac{{\cal N}_{4a}}{(k-E_g-E_r)} +\frac{{\cal N}_{4b}}{(k+E_r-E_g)} -2{(E_l-E_g)} \Bigg], \nonumber\\
{\cal C}_5 &=& \frac{H_2\left[W(-E_g-E_l)-W(-k)\right]}{E_q^2-(E_l+E_g)^2} \Bigg[ -\frac{{\cal N}_{5a}}{(k+E_r-E_g)} -\frac{{\cal N}_{5b}}{(k-E_r-E_g)} +2{(E_g+E_l)} \Bigg], \nonumber\\[2mm]
{\cal C}_6 &=& \frac{H_1\left[W(E_l-E_g)-W(k)\right]}{E_q^2-(E_l-E_g)^2} \Bigg[ \frac{{\cal N}_{6a}}{(k+E_g-E_r)} +\frac{{\cal N}_{6b}}{(k+E_g+E_r)} +2{(E_l-E_g)} \Bigg],
\label{eq:topII_Ci_simplified}
\eea
with 
\bea
{\cal N}_1&\equiv& (E_r+E_g)(E_q-E_g)-(\mathbf{k}\cdot\mathbf{l}) -\frac{(E_r+E_g)(\mathbf{r}\cdot\mathbf{q})}{E_rE_q}(E_q-E_g)
+\frac{(\mathbf{k}\cdot\mathbf{l})(\mathbf{r}\cdot\mathbf{q})}{E_rE_q}, \nonumber\\
{\cal N}_2&\equiv& (E_g-E_r)(E_g+E_q)+(\mathbf{k}\cdot\mathbf{l}) -\frac{(E_g-E_r)(\mathbf{r}\cdot\mathbf{q})}{E_rE_q}(E_g+E_q)
-\frac{(\mathbf{k}\cdot\mathbf{l})(\mathbf{r}\cdot\mathbf{q})}{E_rE_q}, \nonumber\\
{\cal N}_3&\equiv& (E_r-E_g)(E_q-E_g)+(\mathbf{k}\cdot\mathbf{l}) +\frac{(E_r-E_g)(\mathbf{r}\cdot\mathbf{q})}{E_rE_q}(E_q-E_g) +\frac{(\mathbf{k}\cdot\mathbf{l})(\mathbf{r}\cdot\mathbf{q})}{E_rE_q}, \nonumber\\
{\cal N}_{4a}&\equiv& (E_l-E_g)(k+E_l-E_g-E_r) +\frac{\mathbf{k}\cdot\mathbf{l}}{k}(E_l-E_g) +\frac{\mathbf{r}\cdot\mathbf{q}}{E_r}(k+E_l-E_g-E_r) +\frac{(\mathbf{k}\cdot\mathbf{l})(\mathbf{r}\cdot\mathbf{q})}{kE_r}, \nonumber\\
{\cal N}_{4b}&\equiv& (E_l-E_g)(E_r+k+E_l-E_g) +\frac{\mathbf{k}\cdot\mathbf{l}}{k}(E_l-E_g) -\frac{\mathbf{r}\cdot\mathbf{q}}{E_r}(E_r+k+E_l-E_g) -\frac{(\mathbf{k}\cdot\mathbf{l})(\mathbf{r}\cdot\mathbf{q})}{kE_r}, \nonumber\\
{\cal N}_{5a}&\equiv& (E_l+E_g)(E_r-E_l-E_g+k) +\frac{\mathbf{k}\cdot\mathbf{l}}{k}(E_l+E_g) +\frac{\mathbf{r}\cdot\mathbf{q}}{E_r}(E_r-E_l-E_g+k) +\frac{(\mathbf{k}\cdot\mathbf{l})(\mathbf{r}\cdot\mathbf{q})}{kE_r}, \nonumber\\
{\cal N}_{5b}&\equiv& -(E_l+E_g)(E_r+E_l+E_g-k) +\frac{\mathbf{k}\cdot\mathbf{l}}{k}(E_l+E_g) +\frac{\mathbf{r}\cdot\mathbf{q}}{E_r}(E_r+E_l+E_g-k) -\frac{(\mathbf{k}\cdot\mathbf{l})(\mathbf{r}\cdot\mathbf{q})}{kE_r}, \nonumber\\
{\cal N}_{6a}&\equiv& (E_l-E_g)(E_r-k-E_g+E_l) -\frac{\mathbf{k}\cdot\mathbf{l}}{k}(E_l-E_g) -\frac{\mathbf{r}\cdot\mathbf{q}}{E_r}(E_r-k-E_g+E_l) +\frac{(\mathbf{k}\cdot\mathbf{l})(\mathbf{r}\cdot\mathbf{q})}{kE_r}, \nonumber\\
{\cal N}_{6b}&\equiv& -(E_l-E_g)(E_r+k+E_g-E_l) -\frac{\mathbf{k}\cdot\mathbf{l}}{k}(E_l-E_g) -\frac{\mathbf{r}\cdot\mathbf{q}}{E_r}(E_r+k+E_g-E_l) -\frac{(\mathbf{k}\cdot\mathbf{l})(\mathbf{r}\cdot\mathbf{q})}{kE_r}. 
\label{eq:topII_N_def}
\eea
We can do the reduction of the photon self-energy in Eq.~\eqref{eq:top:2:after_im_parts} further from six physical cuts to three physical cuts since the two-loop momenta $(\mathbf k)$ and $(\mathbf r)$ are dummy variables integrated over the same phase space. Therefore the double integral is invariant under the exchange, ${\cal R}:\, \mathbf{k}\leftrightarrow\mathbf{r},\, k\leftrightarrow E_r,\, E_q\leftrightarrow E_l,\, E_g\to E_g.$ Here $E_g=|\mathbf k-\mathbf r|$ is invariant, whereas $E_q=|\mathbf k-\mathbf p|$ and $E_l=|\mathbf r-\mathbf p|$ are interchanged. Consequently, the physical cuts transform as
\bea
{\cal R}[D_4] &=& {\cal R}\!\left[\delta(k+E_l-E_g-p_0)\right] = \delta(E_r+E_q-E_g-p_0) , \nonumber\\
{\cal R}[D_5] &=& {\cal R}\!\left[\delta(E_g+E_l-k-p_0)\right] = \delta(E_g+E_q-E_r-p_0), \nonumber\\
{\cal R}[D_6] &=& {\cal R}\!\left[\delta(k+E_g-E_l-p_0)\right] = \delta(E_r+E_g-E_q-p_0).
\label{eq:topII_cut_relabel_map}
\eea
Since the six-cut expression in Eq.~\eqref{eq:top:2:after_im_parts} has the common factor $1/(8E_lE_g)$ outside, the relabelled-cut contributions carry an extra factor $E_l/E_q$. Therefore, we define
\bea
{\cal F}_1^{II} &=& {\cal C}_1+\frac{E_l}{E_q}\,{\cal R}\!\left[{\cal C}_6\right], \qquad {\cal F}_2^{II} = {\cal C}_2+\frac{E_l}{E_q}\,{\cal R}\!\left[{\cal C}_5\right], \qquad {\cal F}_3^{II} = {\cal C}_3+\frac{E_l}{E_q}\,{\cal R}\!\left[{\cal C}_4\right].
\label{eq:topII_Fi_P_def}
\eea
Thus, the exact three-cut form is
\bea
\Im \Pi_{\mu}^{\mu\,II}(P) &=& -\frac{320\pi}{9}e^2g^2 \int\frac{d^3k}{(2\pi)^3} \int\frac{d^3r}{(2\pi)^3} \frac{1}{8E_lE_g}
\left[ \delta(E_r+E_g-E_q-p_0){\cal F}_1^{II} +\delta(E_g+E_q-E_r-p_0){\cal F}_2^{II} \right. \nonumber\\
&+& \left. \delta(E_r+E_q-E_g-p_0){\cal F}_3^{II} \right].
\label{eq:topII_threecut}
\eea
In order to write the three-cut coefficients without the abstract map ${\cal R}$, let us define the relabelled thermal factors
\bea
\widetilde{H}_1&\equiv&n_F(E_q)+n_B(E_g),\qquad \widetilde{H}_2\equiv1+n_B(E_g)-n_F(E_q).
\label{eq:topII_tilde_thermal}
\eea
The relabelled contributions are given as 
\begin{subequations}
\bea
\widetilde{\cal C}_1^{(k)} &=& -\frac{\widetilde{H}_1\left[W(E_r)-W(E_q-E_g)\right]}{E_l^2-(E_q-E_g)^2} \Bigg[ \frac{\widetilde{\cal N}_{1a}}{(E_r+E_g-k)} +\frac{\widetilde{\cal N}_{1b}}{(E_r+E_g+k)} +2{(E_q-E_g)} \Bigg], \\
\widetilde{\cal C}_2^{(k)} &=& \frac{\widetilde{H}_2\left[W(-E_r)-W(-E_g-E_q)\right]}{E_l^2-(E_q+E_g)^2} \Bigg[ \frac{\widetilde{\cal N}_{2a}}{(E_r+k-E_g)} +\frac{\widetilde{\cal N}_{2b}}{(E_r-k-E_g)} -2({E_g+E_q}) \Bigg], \\
\widetilde{\cal C}_3^{(k)} &=& \frac{\widetilde{H}_1\left[W(E_r)-W(E_g-E_q)\right]}{E_l^2-(E_q-E_g)^2} \Bigg[ \frac{\widetilde{\cal N}_{3a}}{(E_r-E_g-k)} +\frac{\widetilde{\cal N}_{3b}}{(E_r+k-E_g)} -2({E_q-E_g})\Bigg],
\label{eq:Ci_tilde}
\eea
\end{subequations}
where we have, 
\bea
\widetilde{\cal N}_{1a} &\equiv& (E_q-E_g)(k-E_r-E_g+E_q) -\frac{\mathbf{r}\cdot\mathbf{q}}{E_r}(E_q-E_g) -\frac{\mathbf{k}\cdot\mathbf{l}}{k}(k-E_r-E_g+E_q) +\frac{(\mathbf{k}\cdot\mathbf{l})(\mathbf{r}\cdot\mathbf{q})}{kE_r}, \nonumber\\
\widetilde{\cal N}_{1b} &\equiv& -(E_q-E_g)(k+E_r+E_g-E_q) -\frac{\mathbf{r}\cdot\mathbf{q}}{E_r}(E_q-E_g) -\frac{\mathbf{k}\cdot\mathbf{l}}{k}(k+E_r+E_g-E_q) -\frac{(\mathbf{k}\cdot\mathbf{l})(\mathbf{r}\cdot\mathbf{q})}{kE_r}, \nonumber \\
\widetilde{\cal N}_{2a} &\equiv& (E_q+E_g)(k-E_q-E_g+E_r) +\frac{\mathbf{r}\cdot\mathbf{q}}{E_r}(E_q+E_g) +\frac{\mathbf{k}\cdot\mathbf{l}}{k}(k-E_q-E_g+E_r) +\frac{(\mathbf{k}\cdot\mathbf{l})(\mathbf{r}\cdot\mathbf{q})}{kE_r}, \nonumber\\ 
\widetilde{\cal N}_{2b} &\equiv& -(E_q+E_g)(k+E_q+E_g-E_r) +\frac{\mathbf{r}\cdot\mathbf{q}}{E_r}(E_q+E_g) +\frac{\mathbf{k}\cdot\mathbf{l}}{k}(k+E_q+E_g-E_r) -\frac{(\mathbf{k}\cdot\mathbf{l})(\mathbf{r}\cdot\mathbf{q})}{kE_r}, \nonumber \\
\widetilde{\cal N}_{3a} &\equiv& (E_q-E_g)(E_r+E_q-E_g-k) +\frac{\mathbf{r}\cdot\mathbf{q}}{E_r}(E_q-E_g) +\frac{\mathbf{k}\cdot\mathbf{l}}{k}(E_r+E_q-E_g-k) +\frac{(\mathbf{k}\cdot\mathbf{l})(\mathbf{r}\cdot\mathbf{q})}{kE_r}, \nonumber\\
\widetilde{\cal N}_{3b} &\equiv& (E_q-E_g)(k+E_r+E_q-E_g) +\frac{\mathbf{r}\cdot\mathbf{q}}{E_r}(E_q-E_g) -\frac{\mathbf{k}\cdot\mathbf{l}}{k}(k+E_r+E_q-E_g) -\frac{(\mathbf{k}\cdot\mathbf{l})(\mathbf{r}\cdot\mathbf{q})}{kE_r}.
\label{eq:Ni_tilde}
\eea
Therefore, the three independent coefficient functions are
\bea
{\cal F}_1^{II} &=& F_1 \left[W(E_r+E_g)-W_q\right] \frac{2E_l{\cal N}_1} {\left[(E_r+E_g)^2-k^2\right]\left[(E_q-E_g)^2-E_l^2\right]}
+\frac{E_l}{E_q}\,\widetilde{\cal C}_1^{(k)}, \nonumber\\
{\cal F}_2^{II} &=& -F_2\left[W(E_g-E_r)+W_q\right] \frac{2E_l{\cal N}_2} {\left[(E_g-E_r)^2-k^2\right]\left[(E_g+E_q)^2-E_l^2\right]}
+\frac{E_l}{E_q}\,\widetilde{\cal C}_2^{(k)}, \nonumber\\
{\cal F}_3^{II} &=& -F_2\left[W(E_r-E_g)+W_q\right] \frac{2E_l{\cal N}_3} {\left[(E_r-E_g)^2-k^2\right]\left[(E_q-E_g)^2-E_l^2\right]} +\frac{E_l}{E_q}\,\widetilde{\cal C}_3^{(k)}.
\label{eq:topII_Fi_explicit}
\eea
Using the delta-function constraints, the products of thermal distribution functions appearing in Eq.~\eqref{eq:topII_Fi_explicit} can be simplified. For the unrelabelled parts, one obtains
\begin{subequations}
\label{eq:topII_balance_weights}
\bea
D_1:\qquad F_1\left[W(E_r+E_g)-W_q\right] &=& \left(e^{\beta p_0}-1\right)n_F(E_r)n_B(E_g)\left[1-n_F(E_q)\right], \\
D_2:\qquad F_2\left[W(E_g-E_r)+W_q\right] &=& \left(e^{\beta p_0}-1\right)\left[1-n_F(E_r)\right]n_B(E_g)n_F(E_q), \\
D_3:\qquad F_2\left[W(E_r-E_g)+W_q\right] &=& \left(e^{\beta p_0}-1\right)n_F(E_r)\left[1+n_B(E_g)\right]n_F(E_q).
\eea
\end{subequations}
The relabelled thermal factors obey the same detailed-balance relations on $D_1,D_2,D_3$, respectively, and therefore give the three weights in Eq.~\eqref{eq:topII_balance_weights}.
Here $E_g\equiv E_{k-r}$, and the three cut constraints are
\bea
D_1&=&\delta(E_r+E_g-E_q-p_0), \qquad  D_2=\delta(E_g+E_q-E_r-p_0), \qquad D_3=\delta(E_r+E_q-E_g-p_0).
\eea
Therefore, Eq.~\eqref{eq:topII_threecut} can be written in the detailed-balance form
\bea
\Im\Pi_{\mu}^{\mu\,II}(P) &=& -\frac{320\pi}{9}e^2g^2 \left(e^{\beta p_0}-1\right) \int\frac{d^3k}{(2\pi)^3} \int\frac{d^3r}{(2\pi)^3}
\frac{1}{8E_lE_g} \Big[ n_F(E_r)n_B(E_g)\left(1-n_F(E_q)\right) \nonumber\\
&\times& \delta(E_r+E_g-E_q-p_0)\,{\cal B}_1^{II} + \left(1-n_F(E_r)\right)n_B(E_g)n_F(E_q) \delta(E_g+E_q-E_r-p_0)\,{\cal B}_2^{II}
\nonumber\\
&+& n_F(E_r)\left(1+n_B(E_g)\right)n_F(E_q) \delta(E_r+E_q-E_g-p_0)\,{\cal B}_3^{II} \Big].
\label{eq:topII_thermal_simplified}
\eea
The coefficient functions ${\cal B}_1^{II}, {\cal B}_2^{II}$, and ${\cal B}_3^{II}$ are defined as
\bea
{\cal B}_1^{II} &=& {\cal K}_{1r}^{II} -\frac{E_l}{E_q}{\cal K}_{1k}^{II}, \qquad {\cal B}_2^{II} = -{\cal K}_{2r}^{II} +\frac{E_l}{E_q}{\cal K}_{2k}^{II}, \qquad {\cal B}_3^{II} = -{\cal K}_{3r}^{II} +\frac{E_l}{E_q}{\cal K}_{3k}^{II}.
\label{eq:topII_Bi_def}
\eea
Here
\bea
{\cal K}_{1r}^{II} &=& \frac{2E_l{\cal N}_1} {\left[(E_r+E_g)^2-k^2\right]\left[(E_q-E_g)^2-E_l^2\right]}, \qquad {\cal K}_{2r}^{II} = \frac{2E_l{\cal N}_2} {\left[(E_g-E_r)^2-k^2\right]\left[(E_g+E_q)^2-E_l^2\right]}, \nonumber\\
{\cal K}_{3r}^{II} &=& \frac{2E_l{\cal N}_3} {\left[(E_r-E_g)^2-k^2\right]\left[(E_q-E_g)^2-E_l^2\right]},
\label{eq:topII_Kir_def}
\eea
and
\bea
{\cal K}_{1k}^{II} &=& \frac{\widetilde{\cal N}_{1a}}{(E_l^2-(E_q-E_g)^2)(E_r+E_g-k)} +\frac{\widetilde{\cal N}_{1b}}{(E_l^2-(E_q-E_g)^2)(E_r+E_g+k)} +2\frac{E_q-E_g}{E_l^2-(E_q-E_g)^2}, \nonumber\\
{\cal K}_{2k}^{II} &=& \frac{\widetilde{\cal N}_{2a}}{(E_l^2-(E_q+E_g)^2)(E_r+k-E_g)} +\frac{\widetilde{\cal N}_{2b}}{(E_l^2-(E_q+E_g)^2)(E_r-k-E_g)} -2\frac{E_g+E_q}{E_l^2-(E_q+E_g)^2}, \nonumber\\
{\cal K}_{3k}^{II} &=& \frac{\widetilde{\cal N}_{3a}}{(E_l^2-(E_q-E_g)^2)(E_r-E_g-k)} +\frac{\widetilde{\cal N}_{3b}} {(E_l^2-(E_q-E_g)^2)(E_r+k-E_g)} -2\frac{E_q-E_g}{E_l^2-(E_q-E_g)^2},
\label{eq:topII_Kik_def}
\eea
The coefficient functions ${\cal B}_1^{II}, {\cal B}_2^{II}$, and ${\cal B}_3^{II}$ can be examined for the massless real photon. Using the cut constraints together with $2\,\bm k\cdot\bm l = E_r^2-p^2-E_g^2+E_q^2,$ the scalar prefactors appearing in the three unrelabelled numerator structures reduce to
$(E_r+E_{g}^{(1)})(E_q-E_{g}^{(1)})-\bm k\cdot\bm l = -\frac{P^2}{2}, \, (E_{g}^{(2)}-E_r)(E_{g}^{(2)}+E_q)+\bm k\cdot\bm l = \frac{P^2}{2}, \, (E_r-E_{g}^{(3)})(E_q-E_{g}^{(3)})+\bm k\cdot\bm l = \frac{P^2}{2}.$
Consequently, the numerator structures defined in Eq.~\eqref{eq:topII_N_def} become
\bea
{\cal{N}}_1 &=& -\frac{P^2}{2} \left( 1-\frac{\bm r\cdot\bm q}{E_rE_q} \right), \qquad {\cal{N}}_2 = \frac{P^2}{2} \left( 1-\frac{\bm r\cdot\bm q}{E_rE_q} \right), \qquad {\cal{N}}_3 = \frac{P^2}{2} \left( 1+\frac{\bm r\cdot\bm q}{E_rE_q} \right).
\label{eq:topII_N1_N2_N3_P2}
\eea
The relabelled numerator combinations entering in Eq.~\eqref{eq:topII_Kik_def} satisfy
\begin{subequations}
\bea 
\frac{\widetilde {\cal{N}}_{1a}}{E_r+E_g-k} + \frac{\widetilde {\cal{N}}_{1b}}{E_r+E_g+k} + 2(E_q-E_g) &=& \frac{2E_q {\cal{N}}_1}{(E_r+E_g)^2-k^2},
\label{eq:topII_tilde_identity_1}  \\
\frac{\widetilde {\cal{N}}_{2a}}{E_r+k-E_g} + \frac{\widetilde {\cal{N}}_{2b}}{E_r-k-E_g} - 2(E_g+E_q) &=& \frac{2E_q {\cal{N}}_2}{(E_g-E_r)^2-k^2},
\label{eq:topII_tilde_identity_2} \\
\frac{\widetilde {\cal{N}}_{3a}}{E_r-E_g-k} + \frac{\widetilde {\cal{N}}_{3b}}{E_r+k-E_g} - 2(E_q-E_g) &=& \frac{2E_q {\cal{N}}_3}{(E_r-E_g)^2-k^2}.
\label{eq:topII_tilde_identity_3}
\eea
\end{subequations}
For compactness, define $\Delta_1 \equiv (E_r+E_g)^2-k^2, \, \Delta_2 \equiv (E_g-E_r)^2-k^2, \, \Delta_3 \equiv (E_r-E_g)^2-k^2.$ Using Eqs.~\eqref{eq:topII_Bi_def}-\eqref{eq:topII_Kik_def} along with Eqs.~\eqref{eq:topII_tilde_identity_1}-\eqref{eq:topII_tilde_identity_3}, the three complete coefficient functions reduce to
\bea
{\cal B}^{II}_1 &=& -\frac{4E_l {\cal{N}}_1} {\Delta_1\left[E_l^2-(E_q-E_g)^2\right]},  \qquad  {\cal B}^{II}_2 = \frac{4E_l {\cal{N}}_2} {\Delta_2\left[E_l^2-(E_q+E_g)^2\right]},  \qquad {\cal B}^{II}_3 = \frac{4E_l{\cal{N}}_3} {\Delta_3\left[E_l^2-(E_q-E_g)^2\right]}.
\label{eq:topII_B1_B2_B3_N1_N2_N3}
\eea
Substituting the Eq.~\eqref{eq:topII_N1_N2_N3_P2} in Eq.~\eqref{eq:topII_B1_B2_B3_N1_N2_N3} gives the following structure of the coefficient functions ${\cal B}_1^{II}, {\cal B}_2^{II}$, and ${\cal B}_3^{II}$ as
\bea
{\cal B}^{II}_1 &=& \frac{2E_lP^2} {\Delta_1\left[E_l^2-(E_q-E_g)^2\right]} \left( 1-\frac{\bm r\cdot\bm q}{E_rE_q} \right), \nonumber \\
\quad {\cal B}^{II}_2 &=& \frac{2E_lP^2} {\Delta_2\left[E_l^2-(E_q+E_g)^2\right]} \left( 1-\frac{\bm r\cdot\bm q}{E_rE_q} \right),\nonumber\\ 
{\cal B}^{II}_3 &=& \frac{2E_lP^2} {\Delta_3\left[E_l^2-(E_q-E_g)^2\right]} \left( 1+\frac{\bm r\cdot\bm q}{E_rE_q} \right).
\label{eq:topII_B1_B2_B3__P2}
\eea
For an on-shell real photon, $ p_0 = p, \, P^2=0.$ Therefore, at fixed nonsingular kinematics,
\bea
{\cal B}^{II}_1 &=& {\cal B}^{II}_2 = {\cal B}^{II}_3 = 0.
\label{eq:topII_all_kernels_zero}
\eea
It follows that the retained massless three-propagator cuts corresponding to the physical $2\leftrightarrow2$ sector give
\bea
\left. \Im\Pi^{\mu II}_{\ \ \mu}(P) \right|_{2\leftrightarrow2,\;P^2=0} = 0. 
\label{eq:topII_real_photon_zero}
\eea
This result follows by taking the real-photon limit $P^2\to0$ at fixed nonsingular internal kinematics, with the denominators in Eq.~\eqref{eq:topII_B1_B2_B3__P2} kept nonzero. It does not apply to a simultaneous soft or collinear limit in which one of these denominators also vanishes. Thus, in topology $II$, after taking the Cutkosky thermal cuts, the Lorentz trace of the imaginary part of the photon self-energy vanishes in the real-photon limit $P^2\to 0$, when this limit is taken at fixed nonsingular kinematics; therefore, the production rate from it also vanishes, as can be seen following Eq.~\eqref{ph3}:
\bea
\left. E\frac{dR^{II}}{d^4x\,d^3{\mathbf p}} \right|_{2\leftrightarrow2,\;P^2=0} &=& 0.
\label{eq:topII_real_photon_rate_zero}
\eea
Accordingly, within this nonsingular real-photon limit and for the retained massless $2\leftrightarrow2$ sector, topology $II$ does not contribute to the LL or BLL real-photon production rate. This statement should not be interpreted as implying that topology $II$ is irrelevant in general: it is associated with a vertex correction and is required for the Ward identity vis-à-vis gauge invariance, as demonstrated in Appendix~\ref{ward_identity}.
\section{Photon production rate} \label{phot_rate}
In this section, we calculate the photon production rate from the imaginary part of the two-loop photon self-energy. We first evaluate the hard contribution in the leading-logarithmic and beyond-leading-logarithmic approximations. We then include the infrared-regulated soft contribution and obtain the total photon production rate.

\subsection{Leading logarithm contribution} \label{phot_rate_LL}
We first consider the leading-logarithmic contribution to the hard-photon production rate. The Compton and annihilation processes are obtained from the imaginary parts of the two-loop photon self-energy corresponding to the two topologies discussed in Sec.~\ref{qcd_2l}.
Combining Eq.~\eqref{ph3} and Eq.~\eqref{eq:topI_ImPi_Comp_LL}, the Compton leading-log contribution to the differential photon rate is obtained as
\bea
E\frac{dR_{\rm Comp}^{\rm LL}}{d^4x\,d^3{\mathbf p}} &=& \frac{80\alpha\alpha_s}{9\pi^4p} e^{-E/T} \left[ \frac{p\pi^2T^2}{64}L \right] =  \frac{5\alpha\alpha_s}{36\pi^2} T^2e^{-E/T} \ln\left(\frac{4ET}{k_c^2}\right), \label{rate_LL_Comp} 
\eea
where the standard relations $e^2=4\pi\alpha$ and $g^2=4\pi\alpha_s$ and  $p=E$ for real photon.

Similarly, the annihilation leading-log contribution to the differential photon rate is obtained by combining Eq.~\eqref{ph3} and Eq.~\eqref{eq:topI_ImPi_Ann_LL} as
\bea
E\frac{dR_{\rm Ann}^{\rm LL}}{d^4x\,d^3{\mathbf p}} &=& \frac{80\alpha\alpha_s}{9\pi^4p} e^{-E/T} \left[ \frac{p\pi^2T^2}{64}L \right] = \frac{5\alpha\alpha_s}{36\pi^2} T^2e^{-E/T} \ln\left(\frac{4ET}{k_c^2}\right). 
\eea
Adding both pieces, the total leading-logarithmic photon production rate is
\bea
E\frac{dR_{\rm hard}^{\rm LL}}{d^4x\,d^3{\mathbf p}} = \frac{5\alpha\alpha_s}{18\pi^2} T^2e^{-E/T} \ln\left(\frac{4ET}{k_c^2}\right). 
\label{eq_ll}
\eea
Eq.~\eqref{eq_ll} gives the leading-logarithmic contribution to the hard photon production rate. For comparison, the corresponding rates in the incoming-distribution MB approximation are
\begin{subequations}
\label{eq:MB_LL_rates}
\bea
E\frac{dR_{\rm Comp,MB}^{\rm LL}}{d^4x\,d^3\mathbf p} &=& \frac{80\alpha\alpha_s}{9\pi^4p} e^{-E/T} \left[ \frac{p\pi^2T^2}{96}L \right] =  \frac{5\alpha\alpha_s}{54\pi^2}T^2e^{-E/T}\ln\left(\frac{4ET}{k_c^2}\right), \\
E\frac{dR_{\rm Ann,MB}^{\rm LL}}{d^4x\,d^3\mathbf p} &=& \frac{80\alpha\alpha_s}{9\pi^4p} e^{-E/T} \left[ \frac{p\pi^2T^2}{48}L \right] = \frac{5\alpha\alpha_s}{27\pi^2}T^2e^{-E/T}\ln\left(\frac{4ET}{k_c^2}\right), \\
E\frac{dR_{\rm hard,MB}^{\rm LL}}{d^4x\,d^3\mathbf p} &=& \frac{5\alpha\alpha_s}{18\pi^2}T^2e^{-E/T}\ln\left(\frac{4ET}{k_c^2}\right).
\eea
\end{subequations}
Thus, exact quantum statistics gives equal Compton and annihilation LL contributions, whereas the MB approximation gives $R_{\rm Ann,MB}^{\rm LL}=2R_{\rm Comp,MB}^{\rm LL}$. Their total LL rates nevertheless agree. We now go beyond the leading-logarithmic approximation by retaining the finite terms accompanying the logarithmic contribution and evaluating the corresponding BLL corrections in the next subsection~\ref{phot_rate_BLL}.
\subsection{Beyond leading logarithm contribution}\label{phot_rate_BLL}
\begin{figure}[htb]
\centering
\includegraphics[width=0.78\linewidth]{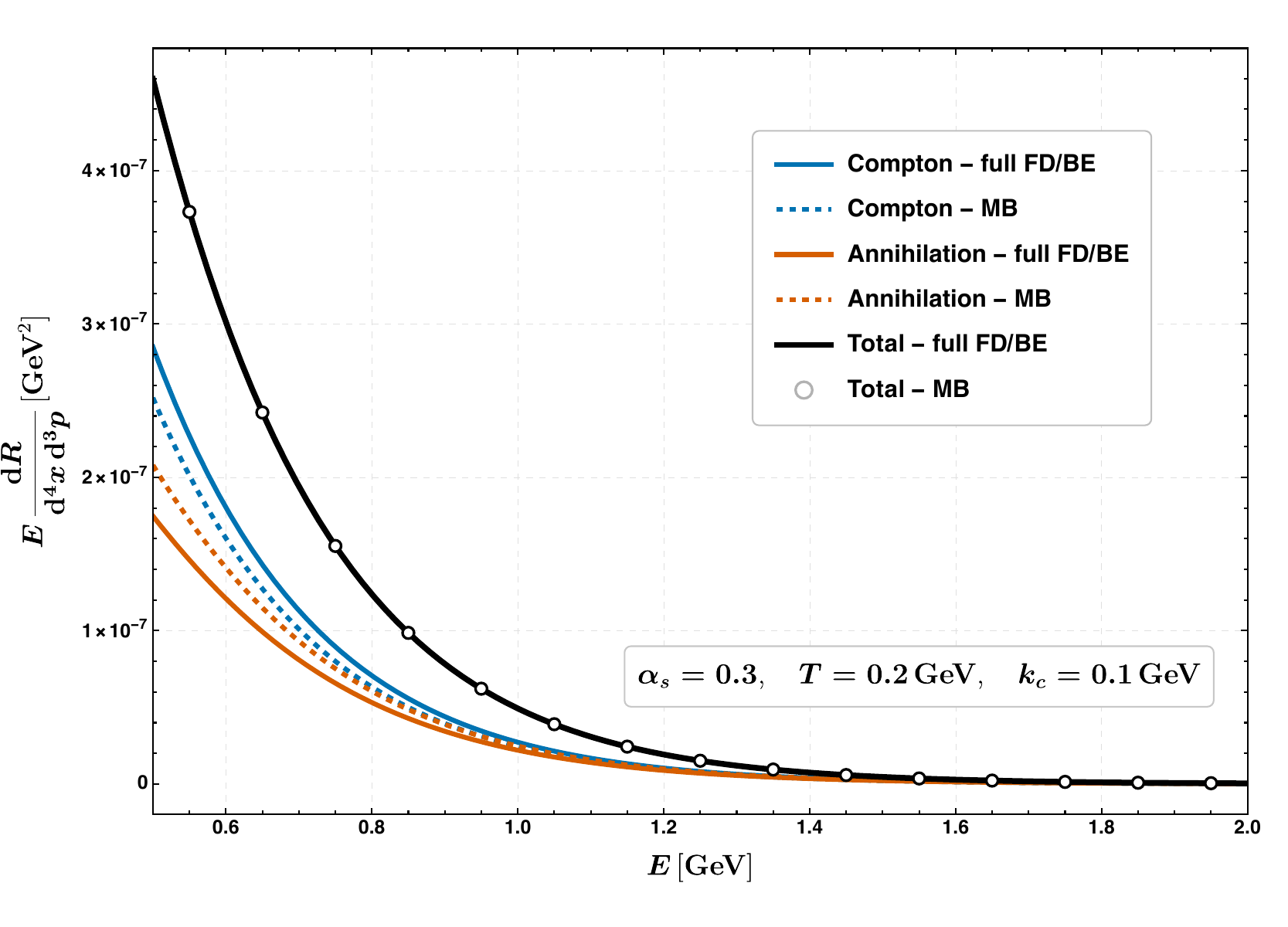}
\caption{Hard BLL photon production rate as a function of the photon energy $E$ for $T=0.2~\mathrm{GeV}$, $\alpha_s=0.3$, and $k_c=0.1~\mathrm{GeV}$. The solid blue and orange curves denote the Compton and annihilation contributions obtained with the full Fermi--Dirac and Bose--Einstein distributions. In contrast, the corresponding dotted curves represent results with MB distributions for incoming particles. The solid black curve is the total full-statistics rate, and the open circles denote the total MB rate. Their coincidence indicates that the summed hard rate remains unchanged under the MB approximation, up to BLL accuracy, even though the division between the two channels changes.}
\label{fig:BLL_rate_vs_E}
\end{figure}
\begin{figure}[htb]
\centering
\includegraphics[width=0.78\linewidth]{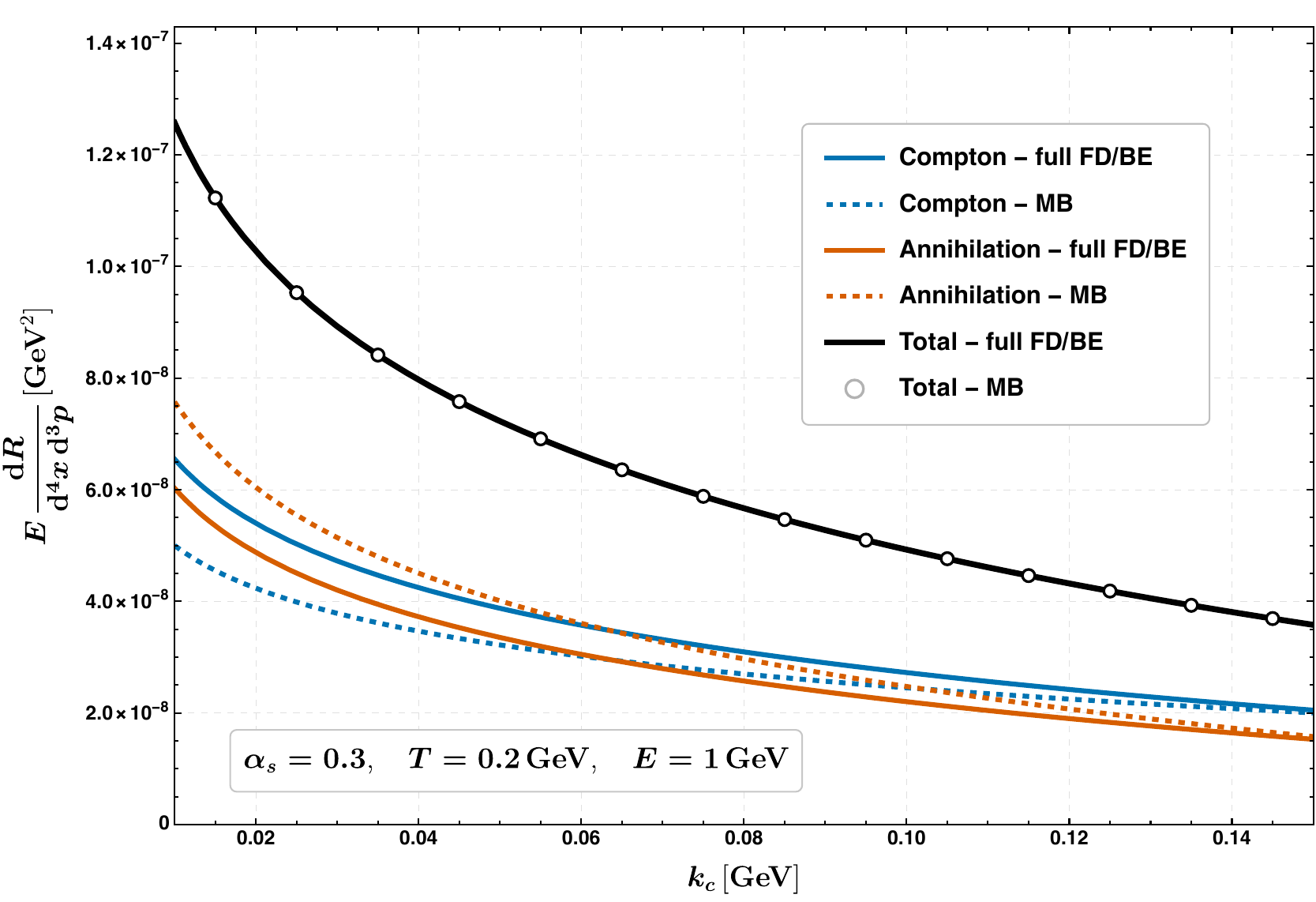}
\caption{Hard BLL photon production rate as a function of the hard--soft separation scale $k_c$ for $E=1~\mathrm{GeV}$, $T=0.2~\mathrm{GeV}$, and $\alpha_s=0.3$. The solid and dotted colored curves denote, respectively, the full quantum-statistical distributions for both incoming and outgoing particles and the MB distributions for incoming particles, resulting from the individual Compton and annihilation processes. The solid black curve gives the total full-statistics rate, while the open circles give the total MB rate. The coincidence of the total rates demonstrates their equality through BLL accuracy.}
\label{fig:BLL_rate_vs_cutoff}
\end{figure}
We next consider the BLL contributions to the hard-photon production rate. Combining Eq.~\eqref{ph3} and Eq.~\eqref{eq:topI_ImPi_Comp_BLL}, the Compton beyond-leading-log contribution to the differential photon rate is obtained as
\label{eq:full_BLL_individual_rates}
\bea
E\frac{dR_{\rm Comp}^{\rm BLL}}{d^4x\,d^3\mathbf p} &=& \frac{5\alpha\alpha_s}{36\pi^2}T^2e^{-E/T} \left[\ln\left(\frac{4ET}{k_c^2}\right)-\frac13\ln\left(\frac{E}{T}\right)+C_C \right],
\label{rate_BLL_Comp} 
\eea
Combining Eq.~\eqref{ph3} and Eq.~\eqref{eq:topI_ImPi_Ann_BLL}, the annihilation  beyond leading-log contribution to the differential photon rate is obtained as
\bea
E\frac{dR_{\rm Ann}^{\rm BLL}}{d^4x\,d^3\mathbf p} &=& \frac{5\alpha\alpha_s}{36\pi^2}T^2e^{-E/T} \left[ \ln\left(\frac{4ET}{k_c^2}\right) +\frac13\ln\left(\frac{E}{T}\right)+C_{\rm Ann} \right].
\label{rate_BLL_Ann}
\eea
Adding both pieces, the total beyond the leading-logarithmic photon production rate is
\bea
 E\frac{dR_{\rm hard}^{\rm BLL}}{d^4x\,d^3{\mathbf p}} = \frac{5\alpha\alpha_s}{18\pi^2} T^2e^{-E/T} \left(\ln\left(\frac{4ET}{k_c^2}\right)+\frac{1}{3}\ln 2 -\frac{1}{2} -\gamma_E + \frac{\zeta'(2)}{\zeta(2)} \right). 
\label{eq_bll}
\eea
For comparison, the final rates obtained in the incoming-distribution MB approximation are
\begin{subequations}
\label{eq:MB_BLL_rates}
\bea
E\frac{dR_{\rm Comp,MB}^{\rm BLL}}{d^4x\,d^3\mathbf p} &=& \frac{5\alpha\alpha_s}{54\pi^2}T^2e^{-E/T} \left[\ln\left(\frac{4ET}{k_c^2}\right) +\ln2+\frac12-\gamma_E+\frac{\zeta'(2)}{\zeta(2)} \right],
\label{rate_BLL_Comp_MB} \\
E\frac{dR_{\rm Ann,MB}^{\rm BLL}}{d^4x\,d^3\mathbf p} &=& \frac{5\alpha\alpha_s}{27\pi^2}T^2e^{-E/T} \left[\ln\left(\frac{4ET}{k_c^2}\right) -1-\gamma_E+\frac{\zeta'(2)}{\zeta(2)} \right],
\label{rate_BLL_Ann_MB} \\
E\frac{dR_{\rm hard,MB}^{\rm BLL}}{d^4x\,d^3\mathbf p} &=& \frac{5\alpha\alpha_s}{18\pi^2}T^2e^{-E/T} \left[\ln\left(\frac{4ET}{k_c^2}\right) +\frac13\ln2-\frac12-\gamma_E +\frac{\zeta'(2)}{\zeta(2)} \right].
\label{rate_BLL_total_MB}
\eea
\end{subequations}
The exact statistics and MB calculations, therefore, give the same total hard rate through BLL accuracy, although their individual Compton and annihilation contributions differ. Specifically, the channel-dependent $\mp\frac13\ln(E/T)$ and $\mp8{\cal S}_Q/\pi^2$ terms in the respective individual processes cancel between the exact-statistics channels. 

The numerical comparison of the full quantum-statistical distributions for both incoming and outgoing particles and the Maxwell--Boltzmann distributions for the incoming channel results is shown in Figs.~\ref{fig:BLL_rate_vs_E} and \ref{fig:BLL_rate_vs_cutoff}. We use $\alpha=1/137.036$ in both figures. \revision{In Fig.~\ref{fig:BLL_rate_vs_E}, the individual Compton and annihilation rates show modest differences at low photon energy. The dependence on the cut-off $k_c$ is more visible in Fig.~\ref{fig:BLL_rate_vs_cutoff}.} Nevertheless, the corresponding total rates agree through BLL accuracy as quantum corrections get canceled. Accordingly, the open circles representing the MB total rate lie on the solid black curve representing the full quantum-statistical total rate.

The \revision{resulting hard rates} are consistent with the photon production rates previously obtained using kinetic theory~\cite{Kapusta:1991qp,Baier:1991em} and closed-time-path formalisms~\cite{Baier:1997xc}. However, all of these rates exhibit a logarithmic infrared (IR) sensitivity, implying that they diverge logarithmically as the soft momentum cut-off $k_c \rightarrow 0$. A consistent treatment of these many-body effects is achieved through the Hard Thermal Loop (HTL) resummation framework developed by Braaten and Pisarski\footnote{Braaten and Pisarski demonstrated that propagators and vertices must be HTL-dressed whenever the momentum flowing through them is soft, i.e., much smaller than the temperature $T$. This is because the propagation of soft modes is associated with infrared divergences in loop diagrams.}~\cite{Braaten:1989mz}. This formalism was independently employed by Kapusta et al.~\cite{Kapusta:1991qp} and Baier et al.~\cite{Baier:1991em} to regulate the infrared divergence arising from the soft exchanged-quark region of hard-photon production from the quark--gluon plasma (QGP). In the following subsection~\ref{htl_ir_1loop}, we discuss the infrared-regulated soft contribution to the hard-photon production rate obtained from the one-loop HTL photon self-energy.
\subsection {Infrared-regulated soft contribution from \revision{the} one-loop HTL approximation}\label{htl_ir_1loop}
\begin{figure}[htb]
\begin{center}
\includegraphics [scale=0.28]{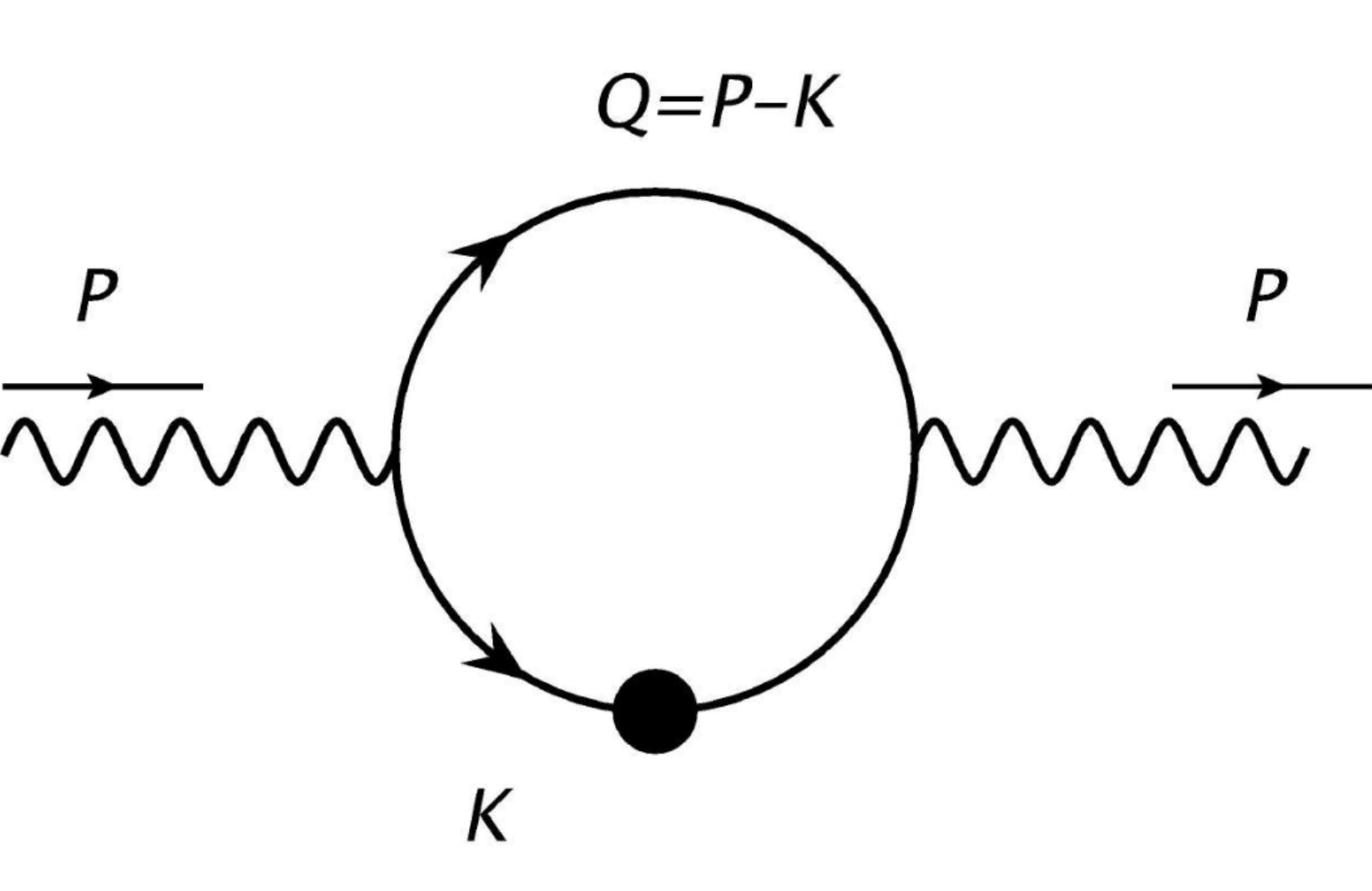}
\end{center}
\caption{One-loop photon self-energy in the HTL approximation. The internal quark line marked with a black blob represents the HTL-resummed soft quark propagator~\cite{Sumit:2022bor}, while the undressed line corresponds to a hard quark propagator. Cutting through the HTL-resummed (blobbed) propagator gives rise to the scattering processes illustrated in Fig.~\ref{fig:scatterings}. It is important to note that, at leading order in $\alpha_s$, infrared regulation of the soft contribution to the photon production rate requires HTL dressing of only one of the quark propagators~\cite{Kapusta:1991qp,Baier:1991em,Baier:1997xc}. Dressing both propagators or including HTL corrections to the quark--photon vertices is unnecessary, as such contributions yield only finite corrections that appear at higher orders in $\alpha_s$.}
\label{ir_soft_photon}
\end{figure}

In the previous subsection~\ref {topo1}, a cut-off $k^2_c$ on the momentum transfer was introduced to regulate infrared divergence, which forced the omission of a small region of phase space from the photon rate obtained in Eqs.~\eqref{eq_ll} and \eqref{eq_bll}. \revision{The omitted region corresponds to soft quark exchange and must be evaluated with an HTL-resummed quark propagator, as shown in Fig.~\ref{ir_soft_photon}. We work in the hierarchy $gT\ll k_c\ll T\ll E$ and impose $-k_c^2\le K^2=\omega^2-k^2\le0$. The corresponding soft photon rate is known~\cite{Kapusta:1991qp,Baier:1991em,Baier:1997xc} and is given by}
\bea
E\frac{dR_{\rm soft}}{d^4x d^3\mathbf{\mathbf p}} 
&=& \!\!\!\frac{5 \alpha\alpha_s}{18 \pi^ 2} \, \,T^2\,\, e^{-{E}/T} \, \, \ln\left (\frac{k_c^2}{M_\infty^2}\right ) \, ,\label{ph76}
\eea
where, $M_\infty =\sqrt{2({m^{q}_{\rm{th}}})^2} =gT/\sqrt{3}$ \revision{is the asymptotic quark mass, with $m_{\rm th}^q$ the thermal quark mass.}
Adding Eq.~\eqref{ph76} to the hard contribution obtained in Eqs.~\eqref{eq_ll} and \eqref{eq_bll} removes the dependence on the intermediate cut-off $k^2_c$, yielding the photon production rate restricted to only $2\leftrightarrow 2$ processes as done below in subsec.~\ref{phot_total}.
\subsection{Combination of soft and hard \revision{contributions}}
\label{phot_total}
\revision{Combining the hard contribution in Eq.~\eqref{eq_ll} with the soft contribution in Eq.~\eqref{ph76} gives the total LL photon rate}
\bea
 E\frac{dR_{\rm total}^{\rm LL}}{d^4x\,d^3{\mathbf p}} &=&  E\frac{dR_{\rm soft}}{d^4x\,d^3{\mathbf p}} +
 E\frac{dR_{\rm hard}^{\rm LL}}{d^4x\,d^3{\mathbf p}} = \frac{5\alpha\alpha_s}{18\pi^2} T^2e^{-E/T} \ln\left(\frac{4ET}{M_\infty^2}\right). 
\label{total_ll}
\eea

Similarly, adding the hard contribution from Eq.~\eqref{eq_bll} to the soft contribution from Eq.~\eqref{ph76}, the total BLL photon production rate becomes
\bea
 E\frac{dR_{\rm total}^{\rm BLL}}{d^4x\,d^3{\mathbf p}}&=&  E\frac{dR_{\rm soft}}{d^4x\,d^3{\mathbf p}} +
 E\frac{dR_{\rm hard}^{\rm BLL}}{d^4x\,d^3{\mathbf p}} = \frac{5\alpha\alpha_s}{18\pi^2} T^2e^{-E/T} \left(\ln\left(\frac{4ET}{M_\infty^2}\right)+\frac{1}{3}\ln 2 -\frac{1}{2} -\gamma_E + \frac{\zeta'(2)}{\zeta(2)} \right).
\label{total_bll}
\eea
\revision{The intermediate separation scale $k_c$ cancels between the hard and soft contributions, as expected~\cite{Braaten:1991dd}. The cancellation follows from the identical coefficient of the logarithm in the two regions and confirms the consistency of the HTL treatment of the soft contribution~\cite{Braaten:1989mz}.}

\section{Conclusions}
\label{conc}
The photon production rate can be calculated in the literature in two ways: one is from kinetic theory using the square of the scattering matrix element, and the other is from the imaginary part of the photon self-energy. They are related by the optical theorem in thermal field theory, as discussed. The first method has been rigorously used in the literature to calculate the photon rate. The second way has been used in the literature \revision{employing the} closed-time-path formalism~\cite{Baier:1997xc}. In this article, we used the second method, applying the imaginary-time formalism, to compute the photon rate. All these calculations are restricted to $2\leftrightarrow 2$ processes only. It is worth noting that the complete leading order rate is obtained by adding $2\leftrightarrow 3$ processes in Ref.~\cite{Arnold:2001ba,Arnold:2001ms} that include near-collinear bremsstrahlung and inelastic pair annihilation with Landau–Pomeranchuk–Migdal effect in HTL resummation. These contributions remain of the same order in the coupling for hard photons from $2\leftrightarrow 2$ processes and lie outside the scope of the present calculation.

At two-loop order, we consider two distinct topologies of the photon self-energy and perform the calculation in the Feynman gauge using imaginary time formalism. The topology $I$ contributes to the photon rate, whereas for the retained massless $2\leftrightarrow2$ cuts, the Lorentz trace of topology~$II$ vanishes in the real-photon limit at fixed nonsingular kinematics. However, we have explicitly shown that topology $II$ is essential to satisfy the Ward identity vis-à-vis gauge invariance, as it provides a vertex correction. We obtained the photon rate for topology $I$ from the imaginary part of the photon self-energy using the imaginary-time formalism, and it agrees with the results of kinetic theory and closed-time-path. To find the rate, we have considered two situations for $2\leftrightarrow 2$ processes: i) MB distributions for incoming particles while FD and BE distributions for outgoing particles, and ii) full quantum (FD and BE) statistics for both incoming and outgoing particles. The individual rate for $2\leftrightarrow 2$ processes acquires quantum corrections in the quantum-statistics case, whereas these corrections disappear within the MB approximation. When individual rates are added, the total rate is the same in both cases because the quantum corrections cancel out. In the interest of quantitative accuracy and reproducibility, we presented the detailed calculations underlying our analysis. \revision{Overall, our results demonstrate the consistency between the kinetic-theory and thermal-field-theory descriptions of hard-photon production and provide an explicit imaginary-time framework for evaluating the $2\leftrightarrow2$ contribution from two-loop photon self-energies in thermal QCD.}
Given the technical nature of the subject, a rigorous and transparent treatment was essential.

\begin{acknowledgments}
Sumit acknowledges partial support from the Beijing Natural Science Foundation under Grant No.~IS26016 and support from Beijing Institute of Technology through a postdoctoral fellowship. The work of R.G. was partly supported by Academia Sinica through Project No.~AS-CDA-114-M01 and additionally by an Academia Sinica postdoctoral fellowship. R. G. also acknowledges the Alexander von Humboldt Foundation for financial support through a Postdoctoral Research Fellowship. M.G.M. would like to acknowledge the financial support from the Department of Atomic Energy, Government of India, through the Raja Ramanna Chair scheme.
\end{acknowledgments}

\appendix
\section{Matsubara \revision{sums for Topology} $I$}
\label{matsubara_I}
\begin{figure}[h]
\begin{center}
\includegraphics[width=9cm,height=7cm]{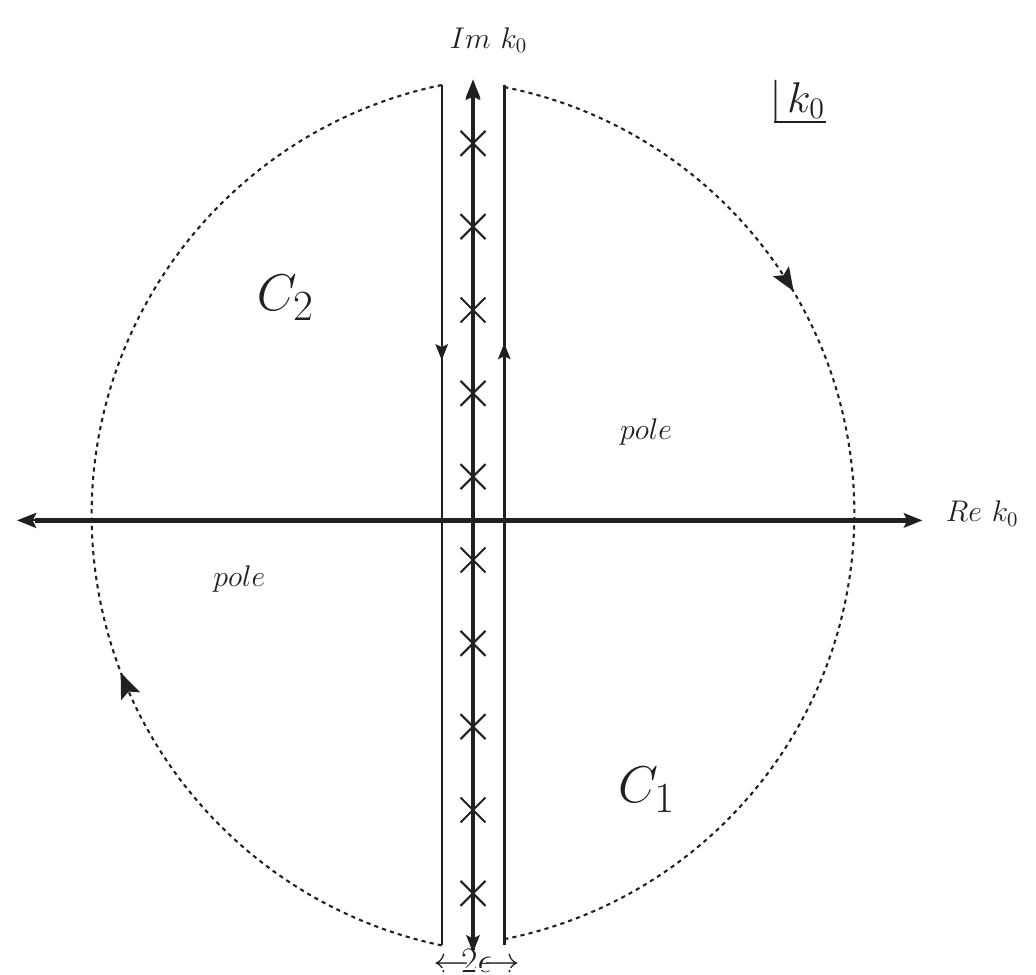}
\vspace*{.5in}
\caption{Contours  $C_1$ and $C_2$ that include the poles of the meromorphic function  $f(k_0)$ 
in complex $k_0$ plane for fermion with zero chemical potential ($\mu=0$). The contours are also shifted
by an amount  $\pm \epsilon$ from the  ${\rm{Im}} k_0$ line to exclude the poles of $\tanh(\beta k_0/2)$ at
$k_0=(2n+1)i\pi T; \, n\in\mathbb{Z}$. }
\label{tanh_poles}
\end{center}
\end{figure}

\revision{In this appendix, we evaluate the fermionic Matsubara sum over $k_0$ by contour integration~\cite{Kapusta:2006pm,Mustafa:2022got,Bellac:2011kqa}:}
\bea
\hspace{-.4cm} T\sum_{n=-\infty}^{\infty} f(k_0=i\omega_n=(2n+1)\pi i T) &=&\frac{1}{2\pi i} \oint_C  f(k_0) \frac {1}{2} \tanh\left (\frac{\beta k_0}{2} \right ) dk_0 = \frac{1}{2\pi i} \oint_{C_1\cup C_2}  f(k_0) \frac {1}{2} \tanh\left (\frac{\beta k_0}{2} \right ) dk_0  \nonumber\\
&=& \frac{1}{2\pi i}\times (-2\pi i) \sum {\rm{Residues}} , \label{eq7} 
\eea 
where $\beta=1/T$. \revision{We use the following properties of Eq.~\eqref{eq7}:}
\begin{enumerate}
\item[$\bullet$] The function $\tanh(\beta k_0/2)$  has simple poles at the fermionic Matsubara frequencies $k_0=i\omega_n=i(2n+1)\pi T,\, n\in\mathbb{Z}.$
\item[$\bullet$] We assume that $f(k_0)$ is a meromorphic function with isolated poles and no essential singularities or branch cuts in the region enclosed by the contours.
\item[$\bullet$] The poles of $f(k_0)$ should not coincide with the poles of $\tanh(\beta k_0/2)$.
Thus, the poles of $f(k_0)$ are separated from the fermionic Matsubara frequencies on the imaginary $k_0$ axis.
\item[$\bullet$] As illustrated in Fig.~\ref{tanh_poles}, the contour $C$ can be decomposed into two closed contours, $C_1$ and $C_2$~\cite{Kapusta:2006pm}. The contours are displaced from the imaginary axis by $\pm\epsilon$ and therefore avoid the poles of $\tanh(\beta k_0/2)$, while enclosing the poles of $f(k_0)$. The contour $C_1$ runs from $\epsilon-i\infty$ to $\epsilon+i\infty$, whereas $C_2$ runs from $-\epsilon+i\infty$ to $-\epsilon-i\infty$. Both contours are oriented clockwise. Assuming that the contribution from the large semicircles vanishes at $|k_0|\to\infty$, the residue theorem gives
\begin{equation}
\oint_{C_1\cup C_2}
f(k_0)\frac{1}{2}\tanh\left(\frac{\beta k_0}{2}\right)dk_0
=
-2\pi i\sum {\rm Residues}.
\end{equation}
\end{enumerate}

\revision{The following form of the $k_0$ sum is sufficient for the present calculation.} For example, 
\bea
T\sum_{k_0} \frac {k_0^nq_0^m}{K^{2l}Q^2 (k_0-A)} , \label{eq8}
\eea 
has poles $k_0=\pm k$ of order l, $k_0=p_0\pm E_q$ of first order and $k_0=A$ of first order. Not all poles would contribute. 
Only the relevant poles of the various terms in Eq.~\eqref{eq6} have been considered, and the corresponding results can be obtained using Eq.~\eqref{eq7}.
\revision{We list explicitly the residue terms that contribute:}
\footnotesize
\begin{subequations}
\begin{align}
\left.T\sum_{k_0} \frac {k_0^2q_0}{K^4Q^2 (k_0-E_r-E_{k-r})}\right |_{k_0=E_r+E_{k-r}} &=-\left. \rm{Res}\right |_{k_0=E_r+E_{k-r}} \nonumber \\
&=-\frac{(E_r+E_{k-r})^2(E_r+E_{k-r}-p_0) \big (\frac{1}{2}-n_F(E_r+E_{k-r})\big )}{\big( (E_r+E_{k-r})^2-E_k^2\big)^2(E_r+E_{k-r}+E_q-p_0)(E_r+E_{k-r}-E_q-p_0)} , \label{eq9a}\\
\left. T\sum_{k_0} \frac {k_0^2q_0}{K^4Q^2 (k_0+E_r-E_{k-r})}\right |_{k_0=E_{k-r}-E_r} 
&=-\frac{(E_{k-r}-E_r)^2(E_{k-r}-E_r-p_0) \big (\frac{1}{2}-n_F(E_{k-r}-E_r)\big )}{\big( (E_{k-r}-E_r)^2-E_k^2\big)^2(E_{k-r}-E_r+E_q-p_0)(E_{k-r}-E_r-E_q-p_0)} , \label{eq9b}\\
\left.T\sum_{k_0} \frac {k_0^2q_0}{K^4Q^2 (k_0-E_r+E_{k-r})}\right |_{k_0=E_r-E_{k-r}} 
&=-\frac{(E_r-E_{k-r})^2(E_r-E_{k-r}-p_0) \big (\frac{1}{2}-n_F(E_r-E_{k-r})\big )}{\big( (E_r-E_{k-r})^2-E_k^2\big)^2(E_r-E_{k-r}+E_q-p_0)(E_r-E_{k-r}-E_q-p_0)} , \label{eq9c}\\
\left.T\sum_{k_0} \frac {k_0q_0}{K^4Q^2 (k_0-E_r-E_{k-r})}\right |_{k_0=E_r+E_{k-r}} 
&=-\frac{(E_r+E_{k-r})(E_r+E_{k-r}-p_0) \big (\frac{1}{2}-n_F(E_r+E_{k-r})\big )}{\big( (E_r+E_{k-r})^2-E_k^2\big)^2(E_r+E_{k-r}+E_q-p_0)(E_r+E_{k-r}-E_q-p_0)} , \label{eq9d}\\
\left.T\sum_{k_0} \frac {k_0q_0}{K^4Q^2 (k_0-E_r+E_{k-r})}\right |_{k_0=E_r-E_{k-r}} 
&=-\frac{(E_r-E_{k-r})(E_r-E_{k-r}-p_0) \big (\frac{1}{2}-n_F(E_r-E_{k-r})\big )}{\big( (E_r-E_{k-r})^2-E_k^2\big)^2(E_r-E_{k-r}+E_q-p_0)(E_r-E_{k-r}-E_q-p_0)} , \label{eq9e}\\
\left. T\sum_{k_0} \frac {k_0q_0}{K^4Q^2 (k_0+E_r-E_{k-r})}\right |_{k_0=E_{k-r}-E_r} 
&=-\frac{(E_{k-r}-E_r)(E_{k-r}-E_r-p_0) \big (\frac{1}{2}-n_F(E_{k-r}-E_r)\big )}{\big( (E_{k-r}-E_r)^2-E_k^2\big)^2(E_{k-r}-E_r+E_q-p_0)(E_{k-r}-E_r-E_q-p_0)} , \label{eq9f}\\
\left.T\sum_{k_0} \frac {k_0}{K^4Q^2 (k_0-E_r-E_{k-r})}\right |_{k_0=E_r+E_{k-r}} 
&=-\frac{(E_r+E_{k-r}) \big (\frac{1}{2}-n_F(E_r+E_{k-r})\big )}{\big( (E_r+E_{k-r})^2-E_k^2\big)^2(E_r+E_{k-r}+E_q-p_0)(E_r+E_{k-r}-E_q-p_0)} , \label{eq9g}\\
\left. T\sum_{k_0} \frac {k_0}{K^4Q^2 (k_0+E_r-E_{k-r})}\right |_{k_0=E_{k-r}-E_r} 
&=-\frac{(E_{k-r}-E_r) \big (\frac{1}{2}-n_F(E_{k-r}-E_r)\big )}{\big( (E_{k-r}-E_r)^2-E_k^2\big)^2(E_{k-r}-E_r+E_q-p_0)(E_{k-r}-E_r-E_q-p_0)} , \label{eq9h}\\
\left.T\sum_{k_0} \frac {k_0}{K^4Q^2 (k_0-E_r+E_{k-r})}\right |_{k_0=E_r-E_{k-r}} 
&=-\frac{(E_r-E_{k-r}) \big (\frac{1}{2}-n_F(E_r-E_{k-r})\big )}{\big( (E_r-E_{k-r})^2-E_k^2\big)^2(E_r-E_{k-r}+E_q-p_0)(E_r-E_{k-r}-E_q-p_0)} , \label{eq9i}\\
\left.T\sum_{k_0} \frac {1}{K^4Q^2 (k_0-E_r-E_{k-r})}\right |_{k_0=E_r+E_{k-r}} 
&=-\frac{ \big (\frac{1}{2}-n_F(E_r+E_{k-r})\big )}{\big( (E_r+E_{k-r})^2-E_k^2\big)^2(E_r+E_{k-r}+E_q-p_0)(E_r+E_{k-r}-E_q-p_0)} , \label{eq9j}\\
\left.T\sum_{k_0} \frac {1}{K^4Q^2 (k_0-E_r+E_{k-r})}\right |_{k_0=E_r-E_{k-r}} 
&=-\frac{ \big (\frac{1}{2}-n_F(E_r-E_{k-r})\big )}{\big( (E_r-E_{k-r})^2-E_k^2\big)^2(E_r-E_{k-r}+E_q-p_0)(E_r-E_{k-r}-E_q-p_0)} ,\label{eq9k}\\
\left. T\sum_{k_0} \frac {1}{K^4Q^2 (k_0+E_r-E_{k-r})}\right |_{k_0=E_{k-r}-E_r} 
&=-\frac{ \big (\frac{1}{2}-n_F(E_{k-r}-E_r)\big )}{\big( (E_{k-r}-E_r)^2-E_k^2\big)^2(E_{k-r}-E_r+E_q-p_0)(E_{k-r}-E_r-E_q-p_0)} , \label{eq9l}\\
\left.T\sum_{k_0} \frac {q_0}{K^2Q^2 (k_0-E_r-E_{k-r})}\right |_{k_0=E_r+E_{k-r}} 
&=-\frac{ (E_r+E_{k-r}-p_0)\big (\frac{1}{2}-n_F(E_r+E_{k-r})\big )}{\big( (E_r+E_{k-r})^2-E_k^2\big)(E_r+E_{k-r}+E_q-p_0)(E_r+E_{k-r}-E_q-p_0)} , \label{eq9m}\\
\left. T\sum_{k_0} \frac {q_0}{K^2Q^2 (k_0+E_r-E_{k-r})}\right |_{k_0=E_{k-r}-E_r} 
&=-\frac{ (E_{k-r}-E_r-p_0)\big (\frac{1}{2}-n_F(E_{k-r}-E_r)\big )}{\big( (E_{k-r}-E_r)^2-E_k^2\big)(E_{k-r}-E_r+E_q-p_0)(E_{k-r}-E_r-E_q-p_0)} , \label{eq9n}\\
\left.T\sum_{k_0} \frac {q_0}{K^2Q^2 (k_0-E_r+E_{k-r})}\right |_{k_0=E_r-E_{k-r}} 
&=-\frac{(E_r-E_{k-r}-p_0) \big (\frac{1}{2}-n_F(E_r-E_{k-r})\big )}{\big( (E_r-E_{k-r})^2-E_k^2\big)(E_r-E_{k-r}+E_q-p_0)(E_r-E_{k-r}-E_q-p_0)} , \label{eq9o}\\
\left.T\sum_{k_0} \frac {1}{K^2Q^2 (k_0-E_r-E_{k-r})}\right |_{k_0=E_r+E_{k-r}} 
&=-\frac{ \big (\frac{1}{2}-n_F(E_r+E_{k-r})\big )}{\big( (E_r+E_{k-r})^2-E_k^2\big)(E_r+E_{k-r}+E_q-p_0)(E_r+E_{k-r}-E_q-p_0)} , \label{eq9p}\\
\left.T\sum_{k_0} \frac {1}{K^2Q^2 (k_0-E_r+E_{k-r})}\right |_{k_0=E_r-E_{k-r}} 
&=-\frac{ \big (\frac{1}{2}-n_F(E_r-E_{k-r})\big )}{\big( (E_r-E_{k-r})^2-E_k^2\big)(E_r-E_{k-r}+E_q-p_0)(E_r-E_{k-r}-E_q-p_0)} ,\label{eq9q}\\
\left. T\sum_{k_0} \frac {1}{K^2Q^2 (k_0+E_r-E_{k-r})}\right |_{k_0=E_{k-r}-E_r} 
&=-\frac{ \big (\frac{1}{2}-n_F(E_{k-r}-E_r)\big )}{\big( (E_{k-r}-E_r)^2-E_k^2\big)(E_{k-r}-E_r+E_q-p_0)(E_{k-r}-E_r-E_q-p_0)} ,\label{eq9r} \\
\left.T\sum_{k_0} \frac {k_0^2q_0}{K^4Q^2 (k_0-E_r-E_{k-r})}\right |_{k_0=p_0+E_q}
&=-\frac{1}{2}\frac{(p_0+E_q)^2 \big (\frac{1}{2}-n_F(E_q)\big )}{\big( (p_0+E_q)^2-E_k^2\big)^2(p_0+E_q-E_r-E_{k-r})} , \label{eq9s}\\
\left.T\sum_{k_0} \frac {k_0^2q_0}{K^4Q^2 (k_0+E_r-E_{k-r})}\right |_{k_0=p_0-E_q}
&=\frac{1}{2}\frac{(p_0-E_q)^2 \big (\frac{1}{2}-n_F(E_q)\big )}{\big( (p_0-E_q)^2-E_k^2\big)^2(p_0+E_r-E_q-E_{k-r})} , \label{eq9t}\\
\left.T\sum_{k_0} \frac {k_0^2q_0}{K^4Q^2 (k_0-E_r+E_{k-r})}\right |_{k_0=p_0-E_q}
&=\frac{1}{2}\frac{(p_0-E_q)^2 \big (\frac{1}{2}-n_F(E_q)\big )}{\big( (p_0-E_q)^2-E_k^2\big)^2(p_0+E_{k-r}-E_q-E_r)} , \label{eq9u}\\
\left.T\sum_{k_0} \frac {k_0q_0}{K^4Q^2 (k_0-E_r-E_{k-r})}\right |_{k_0=p_0+E_q}
&=-\frac{1}{2}\frac{(p_0+E_q) \big (\frac{1}{2}-n_F(E_q)\big )}{\big( (p_0+E_q)^2-E_k^2\big)^2(p_0+E_q-E_r-E_{k-r})} , \label{eq9v}\\
\left.T\sum_{k_0} \frac {k_0q_0}{K^4Q^2 (k_0-E_r+E_{k-r})}\right |_{k_0=p_0-E_q}
&=\frac{1}{2}\frac{(p_0-E_q) \big (\frac{1}{2}-n_F(E_q)\big )}{\big( (p_0-E_q)^2-E_k^2\big)^2(p_0+E_{k-r}-E_q-E_r)} , \label{eq9w}\\
\left.T\sum_{k_0} \frac {k_0q_0}{K^4Q^2 (k_0+E_r-E_{k-r})}\right |_{k_0=p_0-E_q}
&=\frac{1}{2}\frac{(p_0-E_q) \big (\frac{1}{2}-n_F(E_q)\big )}{\big( (p_0-E_q)^2-E_k^2\big)^2(p_0+E_r-E_q-E_{k-r})} , \label{eq9x}\\
\left.T\sum_{k_0} \frac {k_0}{K^4Q^2 (k_0-E_r-E_{k-r})}\right |_{k_0=p_0+E_q}
&=-\frac{1}{2E_q}\frac{(p_0+E_q) \big (\frac{1}{2}-n_F(E_q)\big )}{\big( (p_0+E_q)^2-E_k^2\big)^2(p_0+E_q-E_r-E_{k-r})} , \label{eq9y}\\
\left.T\sum_{k_0} \frac {k_0}{K^4Q^2 (k_0+E_r-E_{k-r})}\right |_{k_0=p_0-E_q}
&=-\frac{1}{2E_q}\frac{(p_0-E_q) \big (\frac{1}{2}-n_F(E_q)\big )}{\big( (p_0-E_q)^2-E_k^2\big)^2(p_0+E_r-E_q-E_{k-r})} , \label{eq9z}\\
\left.T\sum_{k_0} \frac {k_0}{K^4Q^2 (k_0-E_r+E_{k-r})}\right |_{k_0=p_0-E_q}
&=-\frac{1}{2E_q}\frac{(p_0-E_q) \big (\frac{1}{2}-n_F(E_q)\big )}{\big( (p_0-E_q)^2-E_k^2\big)^2(p_0+E_{k-r}-E_q-E_r)} , \label{eq9aa}\\
\left.T\sum_{k_0} \frac {1}{K^4Q^2 (k_0-E_r-E_{k-r})}\right |_{k_0=p_0+E_q}
&=-\frac{1}{2E_q}\frac{ \big (\frac{1}{2}-n_F(E_q)\big )}{\big( (p_0+E_q)^2-E_k^2\big)^2(p_0+E_q-E_r-E_{k-r})} , \label{eq9bb}\\
\left.T\sum_{k_0} \frac {1}{K^4Q^2 (k_0-E_r+E_{k-r})}\right |_{k_0=p_0-E_q}
&=-\frac{1}{2E_q}\frac{ \big (\frac{1}{2}-n_F(E_q)\big )}{\big( (p_0-E_q)^2-E_k^2\big)^2(p_0+E_{k-r}-E_q-E_r)} , \label{eq9cc}\\
\left.T\sum_{k_0} \frac {1}{K^4Q^2 (k_0+E_r-E_{k-r})}\right |_{k_0=p_0-E_q}
&=-\frac{1}{2E_q}\frac{ \big (\frac{1}{2}-n_F(E_q)\big )}{\big( (p_0-E_q)^2-E_k^2\big)^2(p_0+E_r-E_q-E_{k-r})} , \label{eq9dd}\\
\left.T\sum_{k_0} \frac {q_0}{K^2Q^2 (k_0-E_r-E_{k-r})}\right |_{k_0=p_0+E_q}
&=-\frac{1}{2}\frac{ \big (\frac{1}{2}-n_F(E_q)\big )}{\big( (p_0+E_q)^2-E_k^2\big)(p_0+E_q-E_r-E_{k-r})} , \label{eq9ee}\\
\left.T\sum_{k_0} \frac {q_0}{K^2Q^2 (k_0+E_r-E_{k-r})}\right |_{k_0=p_0-E_q}
&=\frac{1}{2}\frac{ \big (\frac{1}{2}-n_F(E_q)\big )}{\big( (p_0-E_q)^2-E_k^2\big)(p_0+E_r-E_q-E_{k-r})} , \label{eq9ff}\\
\left.T\sum_{k_0} \frac {q_0}{K^2Q^2 (k_0-E_r+E_{k-r})}\right |_{k_0=p_0-E_q}
&=\frac{1}{2}\frac{ \big (\frac{1}{2}-n_F(E_q)\big )}{\big( (p_0-E_q)^2-E_k^2\big)(p_0+E_{k-r}-E_q-E_r)} , \label{eq9gg}\\
\left.T\sum_{k_0} \frac {1}{K^2Q^2 (k_0-E_r-E_{k-r})}\right |_{k_0=p_0+E_q}
&=-\frac{1}{2E_q}\frac{ \big (\frac{1}{2}-n_F(E_q)\big )}{\big( (p_0+E_q)^2-E_k^2\big)(p_0+E_q-E_r-E_{k-r})} , \label{eq9hh}\\
\left.T\sum_{k_0} \frac {1}{K^2Q^2 (k_0-E_r+E_{k-r})}\right |_{k_0=p_0-E_q}
&=-\frac{1}{2E_q}\frac{ \big (\frac{1}{2}-n_F(E_q)\big )}{\big( (p_0-E_q)^2-E_k^2\big)(p_0+E_{k-r}-E_q-E_r)} , \label{eq9ii}\\
\left.T\sum_{k_0} \frac {1}{K^2Q^2 (k_0+E_r-E_{k-r})}\right |_{k_0=p_0-E_q}
&=-\frac{1}{2E_q}\frac{ \big (\frac{1}{2}-n_F(E_q)\big )}{\big( (p_0-E_q)^2-E_k^2\big)(p_0+E_r-E_q-E_{k-r})} , \label{eq9jj}
\end{align}
\end{subequations}
\section{Imaginary parts for Topology $I$}
\label{imaginary_I}
In this appendix~\ref{imaginary_I}, we \revision{extract} the imaginary part of the photon self-energy using the following identities
\begin{subequations}
\begin{align}
\lim_{\epsilon\to0^+}\Im\frac{1}{x\pm a\pm i\epsilon} &= \mp\pi\delta(x\pm a), \label{11a}\\
\lim_{\epsilon\to0^+}\Im\frac{x\pm a}{x\pm b+i\epsilon} &= \mp\pi(a-b)\delta(x\pm b),\label{11b}\\
\lim_{\epsilon\to0^+}\Im\bigg[ \frac{(x\pm a)^2} {\big((x\pm a+i\epsilon)^2-b^2\big)^2(x+c+i\epsilon)} \bigg]
&= -\frac{\pi(c\mp a)^2\delta(x+c)} {\left[(c\mp a)^2-b^2\right]^2} +\frac{\pi\delta'(x\pm a-b)} {4(b+c\mp a)}
+\frac{\pi\delta'(x\pm a+b)} {4(c-b\mp a)} \nonumber\\
& -\frac{\pi(c\mp a)\delta(x\pm a-b)} {4b(b+c\mp a)^2} +\frac{\pi(c\mp a)\delta(x\pm a+b)} {4b(c-b\mp a)^2},
\label{eq:11c}\\
\lim_{\epsilon\rightarrow 0^+} {{\Im}}\bigg[\frac{(x\pm a)}{\big((x\pm a+i\epsilon)^2-b^2\big)^2(x+c+i\epsilon)}\bigg]&= \frac{\pi(c\mp a)\delta(x+c)}{\big((c\mp a)^2-b^2\big)^2}
+\frac{\pi\delta^\prime(x\pm a-b)}{4b(b+c\mp a)}
-\frac{\pi\delta^\prime(x\pm a+b)}{4b(c-b\mp a)}\nonumber\\
&+\frac{\pi\delta(x\pm a-b)}{4b(b+c\mp a)^2}
-\frac{\pi\delta(x\pm a+b)}{4b(c-b\mp a)^2}, \label{11d}\\
\lim_{\epsilon\rightarrow 0^+} {{\Im}}\bigg[\frac{1}{\big((x\pm a+i\epsilon)^2-b^2\big)^2(x+c+i\epsilon)}\bigg]&= -\frac{\pi\delta(x+c)}{\big((c\mp a)^2-b^2\big)^2}
+\frac{\pi\delta^\prime(x\pm a-b)}{4b^2(b+c\mp a)}
+\frac{\pi\delta^\prime(x\pm a+b)}{4b^2(c-b\mp a)}\nonumber\\
&+\frac{\pi(2b+(c\mp a))\delta(x\pm a-b)}{4b^3(b+c\mp a)^2}
+\frac{\pi (2b-(c\mp a)) \delta(x\pm a+b)}{4b^3(c-b\mp a)^2}, \label{11e}\\
\lim_{\epsilon\rightarrow 0^+} {{\Im}}\bigg[\frac{1}{\big((x\pm a+i\epsilon)^2-b^2\big)(x+c+i\epsilon)}\bigg]&=
-\frac{\pi\delta(x+c)}{\big((c\mp a)^2-b^2\big)}
-\frac{\pi\delta(x\pm a-b)}{2b(b+c\mp a)}
+\frac{\pi\delta(x\pm a+b)}{2b(c-b\mp a)}. \label{11f}
\end{align}
\end{subequations}
\revision{For the retained physical cuts of topology I, we use the following elementary imaginary-part identities:}
\begin{subequations}
\begin{align}
{{\Im}}\frac{(E_r+E_{k-r}-p_0)}{(E_r+E_{k-r}+E_q-p_0)(E_r+E_{k-r}-E_q-p_0)}&=
{ {\Im}}\frac{(E_r+E_{k-r}-p_0)}{2E_q}\bigg[\frac{1}{E_r+E_{k-r}-E_q-p_0}-\frac{1}{E_r+E_{k-r}+E_q-p_0} \bigg] \nonumber \\
&=\frac{1}{2E_q}{ {\Im}}\frac{E_r+E_{k-r}-p_0}{E_r+E_{k-r}-E_q-p_0} = \frac{\pi}{2} \delta(E_r+E_{k-r}-E_q-p_0) \label{eq12a} \\
{{\Im}}\frac{(E_{k-r}-E_r-p_0)}{(E_{k-r}-E_r+E_q-p_0)(E_{k-r}-E_r-E_q-p_0)}&=\frac{\pi}{2} \delta(E_{k-r}+E_q-E_r-p_0) \label{eq12b} \\
{ {\Im}}\frac{(E_r-E_{k-r}-p_0)}{(E_r-E_{k-r}+E_q-p_0)(E_r-E_{k-r}-E_q-p_0)}&=\frac{\pi}{2} \delta(E_r+E_q-E_{k-r}-p_0) \label{eq12c}\\
{{\Im}}\frac{1}{(E_r+E_{k-r}+E_q-p_0)(E_r+E_{k-r}-E_q-p_0)}&=
{ {\Im}}\frac{1}{2E_q}\bigg[\frac{1}{E_r+E_{k-r}-E_q-p_0}-\frac{1}{E_r+E_{k-r}+E_q-p_0} \bigg] \nonumber \\
&=\frac{1}{2E_q}{ {\Im}}\frac{1}{E_r+E_{k-r}-E_q-p_0} =\frac{\pi}{2E_q}\delta (E_r+E_{k-r}-E_q-p_0) \label{eq12d} \\
{{\Im}}\frac{1}{(E_{k-r}-E_r+E_q-p_0)(E_{k-r}-E_r-E_q-p_0)}&=-\frac{\pi}{2E_q} \delta(E_{k-r}+E_q-E_r-p_0) \label{eq12e} \\
{{\Im}}\frac{1}{(E_r-E_{k-r}+E_q-p_0)(E_r-E_{k-r}-E_q-p_0)}&=-\frac{\pi}{2E_q} \delta(E_r+E_q-E_{k-r}-p_0) \label{eq12f}\\
{{\Im}}\frac{(p_0+E_q)^2}{\big((p_0+E_q)^2-E_k^2\big)^2
(p_0+E_q-E_r-E_{k-r})}&= - \pi \frac{(E_r+E_{k-r})^2}
{\big((E_r+E_{k-r})^2-E_k^2\big)^2}\delta(p_0+E_q-E_r-E_{k-r})
\label{eq12g}\\
{{\Im}}\frac{(p_0-E_q)^2}{\big((p_0-E_q)^2-E_k^2\big)^2
(p_0+E_r-E_q-E_{k-r})}&=-\pi \frac{(E_r-E_{k-r})^2}
{\big((E_r-E_{k-r})^2-E_k^2\big)^2}\delta(p_0+E_r-E_q-E_{k-r})
\label{eq12h}\\
{{\Im}}\frac{(p_0-E_q)^2}{\big((p_0-E_q)^2-E_k^2\big)^2
(p_0+E_{k-r}-E_q-E_r)}&=-\pi \frac{(E_{k-r}-E_r)^2}
{\big((E_{k-r}-E_r)^2-E_k^2\big)^2}\delta(p_0+E_{k-r}-E_q-E_r)
\label{eq12i}\\
{{\Im}}\frac{(p_0+E_q)}{\big((p_0+E_q)^2-E_k^2\big)^2
(p_0+E_q-E_r-E_{k-r})}&=-\pi \frac{(E_r+E_{k-r})}
{\big((E_r+E_{k-r})^2-E_k^2\big)^2}\delta(p_0+E_q-E_r-E_{k-r})
\label{eq12j}\\
{{\Im}}\frac{(p_0-E_q)}{\big((p_0-E_q)^2-E_k^2\big)^2
(p_0+E_{k-r}-E_q-E_r)}&=\pi \frac{(E_{k-r}-E_r)}
{\big((E_{k-r}-E_r)^2-E_k^2\big)^2}\delta(p_0+E_{k-r}-E_q-E_r)
\label{eq12k}\\
{{\Im}}\frac{(p_0-E_q)}{\big((p_0-E_q)^2-E_k^2\big)^2
(p_0+E_r-E_q-E_{k-r})}&=\pi \frac{(E_r-E_{k-r})}
{\big((E_r-E_{k-r})^2-E_k^2\big)^2}\delta(p_0+E_r-E_q-E_{k-r})
\label{eq12l}\\
{{\Im}}\frac{1}{\big((p_0+E_q)^2-E_k^2\big)^2
(p_0+E_q-E_r-E_{k-r})}&=- \frac{\pi}
{\big((E_r+E_{k-r})^2-E_k^2\big)^2}\delta(p_0+E_q-E_r-E_{k-r})
\label{eq12m}\\
{{\Im}}\frac{1}{\big((p_0-E_q)^2-E_k^2\big)^2
(p_0+E_{k-r}-E_q-E_r)}&=- \frac{\pi}
{\big((E_{k-r}-E_r)^2-E_k^2\big)^2}\delta(p_0+E_{k-r}-E_q-E_r)
\label{eq12n}\\
{{\Im}}\frac{1}{\big((p_0-E_q)^2-E_k^2\big)^2
(p_0+E_r-E_q-E_{k-r})}&=- \frac{\pi}
{\big((E_r-E_{k-r})^2-E_k^2\big)^2}\delta(p_0+E_r-E_q-E_{k-r})
\label{eq12o}\\
{{\Im}}\frac{1}{\big((p_0+E_q)^2-E_k^2\big)
(p_0+E_q-E_r-E_{k-r})}&=- \frac{\pi}
{\big((E_r+E_{k-r})^2-E_k^2\big)}\delta(p_0+E_q-E_r-E_{k-r})
\label{eq12p}\\
{{\Im}}\frac{1}{\big((p_0-E_q)^2-E_k^2\big)
(p_0+E_r-E_q-E_{k-r})}&=- \frac{\pi}
{\big((E_r-E_{k-r})^2-E_k^2\big)}\delta(p_0+E_r-E_q-E_{k-r})
\label{eq12q}\\
{{\Im}}\frac{1}{\big((p_0-E_q)^2-E_k^2\big)
(p_0+E_{k-r}-E_q-E_r)}&=- \frac{\pi}
{\big((E_{k-r}-E_r)^2-E_k^2\big)}\delta(p_0+E_{k-r}-E_q-E_r).
\label{eq12r}
\end{align}
\end{subequations}

\section{Matsubara \revision{sums for topology} $II$}
\label{matsubara_II}
\revision{In this appendix, we evaluate the $k_0$ sum for topology~$II$ using Eq.~\eqref{eq7}, following the procedure of Appendix~\ref{matsubara_I}. The internal quark carrying momentum $K$ is massless, so we use $E_k=|\mathbf k|=k$ and express the poles and denominators directly in terms of $k$. We list only the residue terms that generate physical $2\leftrightarrow2$ cuts; terms associated exclusively with unphysical cuts are omitted.} The relevant residues are quoted as follows
\footnotesize
\begin{subequations}
\begin{align}
\left. T\sum_{k_0} \frac{\mathcal M_+(k_0)} {K^2Q^2(k_0-p_0-E_g+E_l)(k_0-E_r-E_g)} \right|_{k_0=p_0+E_g-E_l} &= -\frac{ \mathcal M_+(p_0+E_g-E_l)W(E_g-E_l) } { \left[(p_0+E_g-E_l)^2-k^2\right] \left[(E_g-E_l)^2-E_q^2\right] (p_0-E_r-E_l) }, 
\label{eq87a}\\
\left. T\sum_{k_0} \frac{\mathcal M_-(k_0)} {K^2Q^2(k_0-p_0-E_g+E_l)(k_0+E_r-E_g)} \right|_{k_0=p_0+E_g-E_l} &= -\frac{ \mathcal M_-(p_0+E_g-E_l)W(E_g-E_l) } { \left[(p_0+E_g-E_l)^2-k^2\right] \left[(E_g-E_l)^2-E_q^2\right] (p_0+E_r-E_l) },
\label{eq87b}\\
\left. T\sum_{k_0} \frac{\mathcal M_+(k_0)} {K^2Q^2(k_0-p_0+E_g+E_l)(k_0-E_r+E_g)} \right|_{k_0=p_0-E_g-E_l} &=-\frac{ \mathcal M_+(p_0-E_g-E_l)W(-E_g-E_l) } { \left[(p_0-E_g-E_l)^2-k^2\right] \left[(-E_g-E_l)^2-E_q^2\right] (p_0-E_r-E_l) }, 
\label{eq87c}\\
\left. T\sum_{k_0} \frac{\mathcal M_-(k_0)} {K^2Q^2(k_0-p_0+E_g+E_l)(k_0+E_r+E_g)} \right|_{k_0=p_0-E_g-E_l}  &= -\frac{ \mathcal M_-(p_0-E_g-E_l)W(-E_g-E_l) } { \left[(p_0-E_g-E_l)^2-k^2\right] \left[(-E_g-E_l)^2-E_q^2\right] (p_0+E_r-E_l) }, 
\label{eq87d}\\
\left. T\sum_{k_0} \frac{\mathcal M_+(k_0)} {K^2Q^2(k_0-p_0+E_g-E_l)(k_0-E_r+E_g)} \right|_{k_0=p_0-E_g+E_l}  &= -\frac{ \mathcal M_+(p_0-E_g+E_l)W(-E_g+E_l) } { \left[(p_0-E_g+E_l)^2-k^2\right] \left[(-E_g+E_l)^2-E_q^2\right] (p_0+E_l-E_r) },
\label{eq87e}\\ 
\left. T\sum_{k_0} \frac{\mathcal M_-(k_0)} {K^2Q^2(k_0-p_0+E_g-E_l)(k_0+E_r+E_g)} \right|_{k_0=p_0-E_g+E_l}  &= -\frac{ \mathcal M_-(p_0-E_g+E_l)W(-E_g+E_l) } { \left[(p_0-E_g+E_l)^2-k^2\right] \left[(-E_g+E_l)^2-E_q^2\right] (p_0+E_r+E_l) }, 
\label{eq87f}\\
\left. T\sum_{k_0} \frac{\mathcal M_+(k_0)} {K^2Q^2(k_0-p_0-E_g+E_l)(k_0-E_r-E_g)} \right|_{k_0=E_r+E_g}  &= -\frac{ \mathcal M_+(E_r+E_g)W(E_r+E_g) } { \left[(E_r+E_g)^2-k^2\right] \left[(E_r+E_g-p_0)^2-E_q^2\right] (E_r+E_l-p_0) }, 
\label{eq87g}\\
\left. T\sum_{k_0} \frac{\mathcal M_-(k_0)} {K^2Q^2(k_0-p_0-E_g+E_l)(k_0+E_r-E_g)} \right|_{k_0=E_g-E_r} &= -\frac{ \mathcal M_-(E_g-E_r)W(E_g-E_r) } { \left[(E_g-E_r)^2-k^2\right] \left[(E_g-E_r-p_0)^2-E_q^2\right] (E_l-E_r-p_0) }, 
\label{eq87h}\\
\left. T\sum_{k_0} \frac{\mathcal M_+(k_0)} {K^2Q^2(k_0-p_0+E_g+E_l)(k_0-E_r+E_g)} \right|_{k_0=E_r-E_g} &= -\frac{ \mathcal M_+(E_r-E_g)W(E_r-E_g) } { \left[(E_r-E_g)^2-k^2\right] \left[(E_r-E_g-p_0)^2-E_q^2\right] (E_r+E_l-p_0) }, 
\label{eq87i}\\
\left. T\sum_{k_0} \frac{\mathcal M_+(k_0)} {K^2Q^2(k_0-p_0-E_g-E_l)(k_0-E_r-E_g)} \right|_{k_0=E_r+E_g} &= -\frac{ \mathcal M_+(E_r+E_g)W(E_r+E_g) } { \left[(E_r+E_g)^2-k^2\right] \left[(E_r+E_g-p_0)^2-E_q^2\right] (E_r-p_0-E_l) }, 
\label{eq87j}\\
\left. T\sum_{k_0} \frac{\mathcal M_-(k_0)} {K^2Q^2(k_0-p_0-E_g-E_l)(k_0+E_r-E_g)} \right|_{k_0=E_g-E_r}  &= -\frac{ \mathcal M_-(E_g-E_r)W(E_g-E_r) } { \left[(E_g-E_r)^2-k^2\right] \left[(E_g-E_r-p_0)^2-E_q^2\right] (-p_0-E_r-E_l) }, 
\label{eq87k}\\
\left. T\sum_{k_0} \frac{\mathcal M_+(k_0)} {K^2Q^2(k_0-p_0+E_g-E_l)(k_0-E_r+E_g)} \right|_{k_0=E_r-E_g} &= -\frac{
\mathcal M_+(E_r-E_g)W(E_r-E_g) } { \left[(E_r-E_g)^2-k^2\right] \left[(E_r-E_g-p_0)^2-E_q^2\right] (E_r-p_0-E_l) }, 
\label{eq87l}\\
\left. T\sum_{k_0} \frac{\mathcal M_+(k_0)} {K^2Q^2(k_0-p_0-E_g+E_l)(k_0-E_r-E_g)} \right|_{k_0=p_0+E_q} &= -\frac{ \mathcal M_+(p_0+E_q)W_q } { 2E_q[(p_0+E_q)^2-k^2] (E_q-E_g+E_l) (p_0+E_q-E_r-E_g) },  
\label{eq87m} \\ 
\left. T\sum_{k_0} \frac{\mathcal M_+(k_0)} {K^2Q^2(k_0-p_0-E_g-E_l)(k_0-E_r-E_g)} \right|_{k_0=p_0+E_q}  &= -\frac{ \mathcal M_+(p_0+E_q)W_q } { 2E_q[(p_0+E_q)^2-k^2] (E_q-E_g-E_l) (p_0+E_q-E_r-E_g) }, 
\label{eq87n} \\
\left. T\sum_{k_0} \frac{\mathcal M_-(k_0)} {K^2Q^2(k_0-p_0-E_g+E_l)(k_0+E_r-E_g)} \right|_{k_0=p_0-E_q} &= -\frac{ \mathcal M_-(p_0-E_q)W_q } { 2E_q[(p_0-E_q)^2-k^2] (-E_q-E_g+E_l) (p_0-E_q+E_r-E_g) }, 
\label{eq87o}\\
\left. T\sum_{k_0} \frac{\mathcal M_+(k_0)} {K^2Q^2(k_0-p_0+E_g+E_l)(k_0-E_r+E_g)} \right|_{k_0=p_0-E_q} &= -\frac{ \mathcal M_+(p_0-E_q)W_q } { 2E_q[(p_0-E_q)^2-k^2] (-E_q+E_g+E_l) (p_0-E_q-E_r+E_g) }, 
\label{eq87p}\\
\left. T\sum_{k_0} \frac{\mathcal M_-(k_0)} {K^2Q^2(k_0-p_0-E_g-E_l)(k_0+E_r-E_g)} \right|_{k_0=p_0-E_q} &= -\frac{ \mathcal M_-(p_0-E_q)W_q } { 2E_q[(p_0-E_q)^2-k^2] (-E_q-E_g-E_l) (p_0-E_q+E_r-E_g) }, 
\label{eq87q} \\
\left. T\sum_{k_0} \frac{\mathcal M_+(k_0)} {K^2Q^2(k_0-p_0+E_g-E_l)(k_0-E_r+E_g)} \right|_{k_0=p_0-E_q}  &= -\frac{ \mathcal M_+(p_0-E_q)W_q } { 2E_q[(p_0-E_q)^2-k^2] (-E_q+E_g-E_l) (p_0-E_q-E_r+E_g) }, 
\label{eq87r}\\
\left. T\sum_{k_0} \frac{\mathcal M_+(k_0)} {K^2Q^2(k_0-p_0-E_g+E_l)(k_0-E_r-E_g)} \right|_{k_0=k} &= -\frac{ \mathcal M_+(k)W(k) } { 2k \left[(k-p_0)^2-E_q^2\right] (k-p_0-E_g+E_l) (k-E_r-E_g) },
\label{eq87s}\\
\left. T\sum_{k_0} \frac{\mathcal M_-(k_0)} {K^2Q^2(k_0-p_0-E_g+E_l)(k_0+E_r-E_g)} \right|_{k_0=k} &= -\frac{ \mathcal M_-(k)W(k) } { 2k \left[(k-p_0)^2-E_q^2\right] (k-p_0-E_g+E_l) (k+E_r-E_g) },
\label{eq87t}\\
\left. T\sum_{k_0} \frac{\mathcal M_+(k_0)} {K^2Q^2(k_0-p_0+E_g+E_l)(k_0-E_r+E_g)} \right|_{k_0=-k} &= \frac{ \mathcal M_+(-k)W(-k) } { 2k \left[(-k-p_0)^2-E_q^2\right] (-k-p_0+E_g+E_l) (-k-E_r+E_g) }, 
\label{eq87u}\\
\left. T\sum_{k_0} \frac{\mathcal M_-(k_0)} {K^2Q^2(k_0-p_0+E_g+E_l)(k_0+E_r+E_g)} \right|_{k_0=-k} &= \frac{ \mathcal M_-(-k)W(-k) } { 2k \left[(-k-p_0)^2-E_q^2\right] (-k-p_0+E_g+E_l) (-k+E_r+E_g) }, 
\label{eq87v}\\
\left. T\sum_{k_0} \frac{\mathcal M_+(k_0)} {K^2Q^2(k_0-p_0+E_g-E_l)(k_0-E_r+E_g)} \right|_{k_0=k} &= -\frac{ \mathcal M_+(k)W(k) } { 2k \left[(k-p_0)^2-E_q^2\right] (k-p_0+E_g-E_l) (k-E_r+E_g) }, 
\label{eq87w}\\
\left. T\sum_{k_0} \frac{\mathcal M_-(k_0)} {K^2Q^2(k_0-p_0+E_g-E_l)(k_0+E_r+E_g)} \right|_{k_0=k} &= -\frac{ \mathcal M_-(k)W(k) } { 2k \left[(k-p_0)^2-E_q^2\right] (k-p_0+E_g-E_l) (k+E_r+E_g) }.
\label{eq87x}\\
\left. T\sum_{k_0} \frac{\mathcal M_+(k_0)} {K^2Q^2(k_0-p_0-E_g+E_l) (p_{0}-E_r-E_l)} \right|_{k_0=p_0+E_g-E_l} &= -\frac{ \mathcal M_+(p_0+E_g-E_l)W(E_g-E_l) } { [(p_0+E_g-E_l)^2-k^2][(E_g-E_l)^2-E_q^2] (p_{0}-E_r-E_l)}, 
\label{eq87y}\\
\left. T\sum_{k_0} \frac{\mathcal M_-(k_0)} {K^2Q^2(k_0-p_0-E_g+E_l)(p_{0}+E_r-E_l)} \right|_{k_0=p_0+E_g-E_l} &= -\frac{ \mathcal M_-(p_0+E_g-E_l)W(E_g-E_l) } { [(p_0+E_g-E_l)^2-k^2][(E_g-E_l)^2-E_q^2] (p_{0}+E_r-E_l)}, 
\label{eq87z}\\
\left. T\sum_{k_0} \frac{\mathcal M_+(k_0)} {K^2Q^2(k_0-p_0+E_g+E_l)(p_{0}-E_r-E_l)} \right|_{k_0=p_0-E_g-E_l}  &= -\frac{ \mathcal M_+(p_0-E_g-E_l)W(-E_g-E_l) } { [(p_0-E_g-E_l)^2-k^2][(-E_g-E_l)^2-E_q^2] (p_{0}-E_r-E_l)}, 
\label{eq87aa}\\
\left. T\sum_{k_0} \frac{\mathcal M_-(k_0)} {K^2Q^2(k_0-p_0+E_g+E_l)(p_{0}+E_r-E_l)} \right|_{k_0=p_0-E_g-E_l}  &= -\frac{ \mathcal M_-(p_0-E_g-E_l)W(-E_g-E_l) } { [(p_0-E_g-E_l)^2-k^2][(-E_g-E_l)^2-E_q^2] (p_{0}+E_r-E_l)}, 
\label{eq87bb}\\
 \left. T\sum_{k_0} \frac{\mathcal M_+(k_0)} {K^2Q^2(k_0-p_0+E_g-E_l)(p_{0}-E_r+E_l)} \right|_{k_0=p_0-E_g+E_l} &= -\frac{ \mathcal M_+(p_0-E_g+E_l)W(-E_g+E_l) } { [(p_0-E_g+E_l)^2-k^2][(-E_g+E_l)^2-E_q^2] (p_{0}-E_r+E_l)},
\label{eq87cc}\\
\left. T\sum_{k_0} \frac{\mathcal M_-(k_0)} {K^2Q^2(k_0-p_0+E_g-E_l)(p_{0}+E_r+E_l)} \right|_{k_0=p_0-E_g+E_l} &= -\frac{ \mathcal M_-(p_0-E_g+E_l)W(-E_g+E_l) } { [(p_0-E_g+E_l)^2-k^2][(-E_g+E_l)^2-E_q^2] (p_{0}+E_r+E_l)}
\label{eq87dd}\\
\left. T\sum_{k_0} \frac{\mathcal M_+(k_0)} {K^2Q^2(k_0-p_0-E_g+E_l)(p_{0}-E_r-E_l)} \right|_{k_0=k} &= -\frac{ \mathcal M_+(k)W(k) } { 2k \left[(k-p_0)^2-E_q^2\right] (k-p_0-E_g+E_l) (p_{0}-E_r-E_l) }, 
\label{eq87ee}\\
\left. T\sum_{k_0} \frac{\mathcal M_-(k_0)} {K^2Q^2(k_0-p_0-E_g+E_l)(p_{0}+E_r-E_l)} \right|_{k_0=k} &= -\frac{ \mathcal M_-(k)W(k) } { 2k \left[(k-p_0)^2-E_q^2\right] (k-p_0-E_g+E_l) (p_{0}+E_r-E_l)}, 
\label{eq87ff}\\
\left. T\sum_{k_0} \frac{\mathcal M_+(k_0)} {K^2Q^2(k_0-p_0+E_g+E_l)(p_{0}-E_r-E_l)} \right|_{k_0=-k} &= \frac{ \mathcal M_+(-k)W(-k) } { 2k \left[(-k-p_0)^2-E_q^2\right] (-k-p_0+E_g+E_l) (p_{0}-E_r-E_l)}, 
\label{eq87gg}\\
\left. T\sum_{k_0} \frac{\mathcal M_-(k_0)} {K^2Q^2(k_0-p_0+E_g+E_l)(p_{0}+E_r-E_l)} \right|_{k_0=-k} &= \frac{ \mathcal M_-(-k)W(-k) } { 2k \left[(-k-p_0)^2-E_q^2\right] (-k-p_0+E_g+E_l) (p_{0}+E_r-E_l)}, 
\label{eq87hh}\\
\left. T\sum_{k_0} \frac{\mathcal M_+(k_0)} {K^2Q^2(k_0-p_0+E_g-E_l)(p_{0}-E_r+E_l)} \right|_{k_0=k} &= -\frac{ \mathcal M_+(k)W(k) } { 2k \left[(k-p_0)^2-E_q^2\right] (k-p_0+E_g-E_l) (p_{0}-E_r+E_l)}, 
\label{eq87ii}\\
\left. T\sum_{k_0} \frac{\mathcal M_-(k_0)} {K^2Q^2(k_0-p_0+E_g-E_l)(p_{0}+E_r+E_l)} \right|_{k_0=k} &= -\frac{ \mathcal M_-(k)W(k) } { 2k \left[(k-p_0)^2-E_q^2\right] (k-p_0+E_g-E_l) (p_{0}+E_r+E_l)},
\label{eq87jj}\\
\left. T\sum_{k_0} \frac{A_2(k_0)} {K^2Q^2(k_0-p_0-E_g+E_l)} \right|_{k_0=p_0+E_g-E_l}  &= -\frac{ A_2(p_0+E_g-E_l)W(E_g-E_l) } { [(p_0+E_g-E_l)^2-k^2][(E_g-E_l)^2-E_q^2] }, 
\label{eq87kk}\\
\left. T\sum_{k_0} \frac{A_2(k_0)} {K^2Q^2(k_0-p_0+E_g-E_l)} \right|_{k_0=p_0-E_g+E_l} &= -\frac{ A_2(p_0-E_g+E_l)W(-E_g+E_l) } { [(p_0-E_g+E_l)^2-k^2][(-E_g+E_l)^2-E_q^2] }, 
\label{eq87ll}\\
\left. T\sum_{k_0} \frac{A_2(k_0)} {K^2Q^2(k_0-p_0+E_g+E_l)} \right|_{k_0=p_0-E_g-E_l}  &= -\frac{ A_2(p_0-E_g-E_l)W(-E_g-E_l) } { [(p_0-E_g-E_l)^2-k^2][(-E_g-E_l)^2-E_q^2] }, 
\label{eq87mm}\\
\left. T\sum_{k_0} \frac{A_2(k_0)} {K^2Q^2(k_0-p_0-E_g+E_l)} \right|_{k_0=k}  &= -\frac{ A_2(k)W(k) } { 2k \left[(k-p_0)^2-E_q^2\right] (k-p_0-E_g+E_l) }, 
\label{eq87nn}\\
\left. T\sum_{k_0} \frac{A_2(k_0)} {K^2Q^2(k_0-p_0+E_g+E_l)} \right|_{k_0=-k} &= +\frac{ A_2(-k)W(-k) } { 2k \left[(-k-p_0)^2-E_q^2\right] (-k-p_0+E_g+E_l) }, 
\label{eq87oo}\\
\left. T\sum_{k_0} \frac{A_2(k_0)} {K^2Q^2(k_0-p_0+E_g-E_l)} \right|_{k_0=k} &= -\frac{ A_2(k)W(k) } { 2k \left[(k-p_0)^2-E_q^2\right] (k-p_0+E_g-E_l) }.
\label{eq87pp}
\end{align}
\end{subequations}
\revision{Here we have used $W(p_0\pm x)=\pm W(x)$ for the discrete bosonic Matsubara frequency $p_0=2\pi i nT$, together with $W(-x)=-W(x)$.}

\section{Imaginary parts \revision{for} Topology $II$}
\label{imaginary_II}

 In this appendix~\ref{imaginary_II}, we obtain the imaginary parts of relevant terms of topology $II$ by following general identities: 
\bea
\lim_{\epsilon\to0^+} \Im\left[ \frac{N(x)} {\prod_{j=1}^{n}(a_j-x-i\epsilon)} \right] &=& \pi \sum_{j=1}^{n} \frac{N(a_j)\,\delta(a_j-x)} {\prod_{m\neq j}(a_m-a_j)}, \nonumber\\
\lim_{\epsilon\to0^+} \Im\left[ \frac{N(x)} {\prod_{j=1}^{n}(a_j+x+i\epsilon)} \right] &=& -\pi \sum_{j=1}^{n} \frac{N(-a_j)\,\delta(a_j+x)} {\prod_{m\neq j}(a_m-a_j)}.
\eea
 Using these identities, the imaginary parts of the residue terms appearing in 
 Eqs.~\eqref{eq87a}-\eqref{eq87pp} can be obtained compactly. 
 Now Eq.~\eqref{M_def} can be redefined as
\begin{subequations}
\begin{align}
\mathcal M_+(k_0) = (E_r-p_0)k_0q_0 -(\mathbf k\cdot\mathbf l)q_0 -\left(1-\frac{p_0}{E_r}\right)k_0(\mathbf r\cdot\mathbf q) +\frac{(\mathbf k\cdot\mathbf l)(\mathbf r\cdot\mathbf q)}{E_r}, \\
\mathcal M_-(k_0) = -(E_r+p_0)k_0q_0 -(\mathbf k\cdot\mathbf l)q_0 -\left(1+\frac{p_0}{E_r}\right)k_0(\mathbf r\cdot\mathbf q) -\frac{(\mathbf k\cdot\mathbf l)(\mathbf r\cdot\mathbf q)}{E_r}.
\end{align}
\end{subequations}
where $q_0=k_0-p_0$. The various imaginary parts are obtained as
\footnotesize
\begin{subequations}
\begin{align}
&\Im\left[ -\frac{ \left[\mathcal M_+(k_0)\right]_{k_0=p_0+E_g-E_l} \left(\frac12-n_F(E_g-E_l)\right) } { \left[(p_0+E_g-E_l)^2-k^2\right] \left[(E_g-E_l)^2-E_q^2\right] (p_0-E_r-E_l) } \right]= -\frac{\pi}{2} \bigg[ (k+E_l-E_g-E_r)(E_l-E_g) +\frac{\mathbf k\cdot\mathbf l}{k}(E_l-E_g) \nonumber\\
& +\frac{\mathbf r\cdot\mathbf q}{E_r}(k+E_l-E_g-E_r) +\frac{(\mathbf k\cdot\mathbf l)(\mathbf r\cdot\mathbf q)}{kE_r} \bigg] \frac{ \left(\frac12-n_F(E_g-E_l)\right) \delta(p_0-k-E_l+E_g) } { (k-E_r-E_g)\left[E_q^2-(E_l-E_g)^2\right] } ,
\label{91a}\\
&\Im\left[ -\frac{ \left[\mathcal M_-(k_0)\right]_{k_0=p_0+E_g-E_l} \left(\frac12-n_F(E_g-E_l)\right) } { \left[(p_0+E_g-E_l)^2-k^2\right] \left[(E_g-E_l)^2-E_q^2\right] (p_0+E_r-E_l) } \right] = -\frac{\pi}{2} \bigg[ (E_r+k+E_l-E_g)(E_l-E_g) +\frac{\mathbf k\cdot\mathbf l}{k}(E_l-E_g)\nonumber\\
& - \frac{\mathbf r\cdot\mathbf q}{E_r}(E_r+k+E_l-E_g) -\frac{(\mathbf k\cdot\mathbf l)(\mathbf r\cdot\mathbf q)}{kE_r} \bigg]  \frac{ \left(\frac12-n_F(E_g-E_l)\right) \delta(p_0-k-E_l+E_g) } { (k+E_r-E_g)\left[E_q^2-(E_l-E_g)^2\right] },
\label{91b}\\
&\Im\left[ -\frac{ \left[\mathcal M_+(k_0)\right]_{k_0=p_0-E_g-E_l} \left(\frac12-n_F(-E_g-E_l)\right) } { \left[(p_0-E_g-E_l)^2-k^2\right] \left[(E_g+E_l)^2-E_q^2\right] (p_0-E_r-E_l) } \right] = \frac{\pi}{2} \left[ (E_r-E_l-E_g+k)(E_l+E_g) +\frac{\mathbf k\cdot\mathbf l}{k}(E_l+E_g) \right. \nonumber\\
&+\left. \frac{\mathbf r\cdot\mathbf q}{E_r}(E_r-E_l-E_g+k) +\frac{(\mathbf k\cdot\mathbf l)(\mathbf r\cdot\mathbf q)}{kE_r} \right]
 \frac{ \left(\frac12-n_F(-E_g-E_l)\right) \delta(E_g+E_l-k-p_0) } { (E_g-E_r-k)\left[E_q^2-(E_l+E_g)^2\right] }, 
\label{91c}\\
&\Im \left[ -\frac{ \left[\mathcal M_-(k_0)\right]_{k_0=p_0-E_g-E_l} \left(\frac12-n_F(-E_g-E_l)\right) } { \left[(p_0-E_g-E_l)^2-k^2\right] \left[(E_g+E_l)^2-E_q^2\right] (p_0+E_r-E_l) } \right] = \frac{\pi}{2} \left[ -(E_r+E_l+E_g-k)(E_l+E_g) +\frac{\mathbf k\cdot\mathbf l}{k}(E_l+E_g) \right. \nonumber\\
&+\left. \frac{\mathbf r\cdot\mathbf q}{E_r}(E_r+E_l+E_g-k) -\frac{(\mathbf k\cdot\mathbf l)(\mathbf r\cdot\mathbf q)}{kE_r}
\right] \frac{ \left(\frac12-n_F(-E_g-E_l)\right) \delta(E_g+E_l-k-p_0) } { (E_g+E_r-k)\left[E_q^2-(E_l+E_g)^2\right] }, 
\label{91d}\\ 
&\Im \left[ -\frac{ \left[\mathcal M_+(k_0)\right]_{k_0=p_0-E_g+E_l} \left(\frac12-n_F(-E_g+E_l)\right) } { \left[(p_0-E_g+E_l)^2-k^2\right] \left[(E_l-E_g)^2-E_q^2\right] (p_0+E_l-E_r) } \right] = -\frac{\pi}{2} \left[ (E_r-k-E_g+E_l)(E_l-E_g) -\frac{\mathbf k\cdot\mathbf l}{k}(E_l-E_g) \right. \nonumber\\
&-\left. \frac{\mathbf r\cdot\mathbf q}{E_r}(E_r-k-E_g+E_l) +\frac{(\mathbf k\cdot\mathbf l)(\mathbf r\cdot\mathbf q)}{kE_r}
\right]  \frac{ \left(\frac12-n_F(-E_g+E_l)\right) \delta(p_0-k-E_g+E_l) } { (k+E_g-E_r)\left[E_q^2-(E_l-E_g)^2\right] }, 
\label{91e}\\
&\Im \left[ -\frac{ \left[\mathcal M_-(k_0)\right]_{k_0=p_0-E_g+E_l} \left(\frac12-n_F(-E_g+E_l)\right) } { \left[(p_0-E_g+E_l)^2- k^2\right] \left[(E_l-E_g)^2-E_q^2\right] (p_0+E_l+E_r) } \right] = -\frac{\pi}{2} \left[ -(E_r+k+E_g-E_l)(E_l-E_g) -\frac{\mathbf k\cdot\mathbf l}{k}(E_l-E_g) \right. \nonumber\\ 
&-\left. \frac{\mathbf r\cdot\mathbf q}{E_r}(E_r+k+E_g-E_l) -\frac{(\mathbf k\cdot\mathbf l)(\mathbf r\cdot\mathbf q)}{kE_r} \right]
 \frac{ \left(\frac12-n_F(-E_g+E_l)\right) \delta(p_0-k-E_g+E_l) } { (k+E_g+E_r)\left[E_q^2-(E_l-E_g)^2\right] }, 
\label{91f}\\
&\Im \left[ -\frac{ \left[\mathcal M_+(k_0)\right]_{k_0=E_r+E_g} \left(\frac12 -n_F(E_r+E_g)\right) } { \left[(E_r+E_g)^2-k^2\right] \left[(E_r+E_g-p_0)^2-E_q^2\right] (E_l+E_r-p_0) } \right] = -\frac{\pi}{2} \bigg[ (E_r+E_g)(E_q-E_g)-(\mathbf k\cdot\mathbf l)  \nonumber\\
&- \frac{(E_r+E_g)(\mathbf r\cdot\mathbf q)}{E_rE_q}(E_q-E_g)+\frac{(\mathbf k\cdot\mathbf l)(\mathbf r\cdot\mathbf q)}{E_rE_q} \bigg]\frac{ \left(\frac12-n_F(E_r+E_g)\right) \delta(E_r+E_g-E_q-p_0) } { \left[(E_r+E_g)^2-k^2\right](E_l+E_q-E_g) }, 
\label{91g}\\
&\Im \left[ -\frac{ \left[\mathcal M_+(k_0)\right]_{k_0=E_r+E_g} \left(\frac12 - n_F(E_r+E_g)\right) } { \left[(E_r+E_g)^2-k^2\right] \left[(E_r+E_g-p_0)^2-E_q^2\right] (E_r-p_0-E_l) } \right] = -\frac{\pi}{2} \bigg[ (E_r+E_g)(E_q-E_g) -(\mathbf k\cdot\mathbf l) \nonumber\\
&-\frac{(E_r+E_g)(\mathbf r\cdot\mathbf q)}{E_rE_q}(E_q-E_g) +\frac{(\mathbf k\cdot\mathbf l)(\mathbf r\cdot\mathbf q)}{E_rE_q} \bigg] \frac{ \left(\frac12-n_F(E_r+E_g)\right) \delta(E_r+E_g-E_q-p_0) } { \left[(E_r+E_g)^2-k^2\right](E_q-E_g-E_l) }, 
\label{91h}\\
&\Im \left[ -\frac{ \left[\mathcal M_-(k_0)\right]_{k_0=E_g-E_r} \left(\frac12-n_F(E_g-E_r)\right) } { \left[(E_g-E_r)^2-k^2\right] \left[(E_g-E_r-p_0)^2-E_q^2\right] (E_l-E_r-p_0) } \right] = \frac{\pi}{2} \bigg[ (E_g-E_r)(E_g+E_q) +(\mathbf k\cdot\mathbf l) \nonumber\\
&-\frac{(E_g-E_r)(\mathbf r\cdot\mathbf q)}{E_rE_q}(E_g+E_q) -\frac{(\mathbf k\cdot\mathbf l)(\mathbf r\cdot\mathbf q)}{E_rE_q} \bigg]  \frac{ \left(\frac12-n_F(E_g-E_r)\right) \delta(E_g+E_q-E_r-p_0) } { \left[(E_g-E_r)^2-k^2\right](E_l-E_g-E_q) }, 
\label{91i}\\
&\Im \left[ -\frac{ \left[\mathcal M_-(k_0)\right]_{k_0=E_g-E_r} \left(\frac12-n_F(E_g-E_r)\right) } { \left[(E_g-E_r)^2-k^2\right] \left[(E_g-E_r-p_0)^2-E_q^2\right] (-E_r-p_0-E_l) } \right] = -\frac{\pi}{2} \bigg[ (E_g-E_r)(E_g+E_q) +(\mathbf k\cdot\mathbf l)\nonumber\\
&-\frac{(E_g-E_r)(\mathbf r\cdot\mathbf q)}{E_rE_q}(E_g+E_q) -\frac{(\mathbf k\cdot\mathbf l)(\mathbf r\cdot\mathbf q)}{E_rE_q} \bigg]  \frac{ \left(\frac12-n_F(E_g-E_r)\right) \delta(E_g+E_q-E_r-p_0) } { \left[(E_g-E_r)^2-k^2\right](E_g+E_q+E_l) }, 
\label{91j}\\
&\Im \left[ -\frac{ \left[\mathcal M_+(k_0)\right]_{k_0=E_r-E_g} \left(\frac12-n_F(E_r-E_g)\right) } { \left[(E_r-E_g)^2-k^2\right] \left[(E_r-E_g-p_0)^2-E_q^2\right] (E_l+E_r-p_0) } \right] = \frac{\pi}{2} \bigg[ (E_r-E_g)(E_q-E_g) +(\mathbf k\cdot\mathbf l)\nonumber\\ 
&+\frac{(E_r-E_g)(\mathbf r\cdot\mathbf q)}{E_rE_q}(E_q-E_g) +\frac{(\mathbf k\cdot\mathbf l)(\mathbf r\cdot\mathbf q)}{E_rE_q} \bigg]  \frac{ \left(\frac12-n_F(E_r-E_g)\right) \delta(E_r+E_q-E_g-p_0) } { \left[(E_r-E_g)^2-k^2\right](E_l+E_g-E_q) }, 
\label{91k}\\
&\Im \left[ -\frac{ \left[\mathcal M_+(k_0)\right]_{k_0=E_r-E_g} \left(\frac12-n_F(E_r-E_g)\right) } { \left[(E_r-E_g)^2-k^2\right] \left[(E_r-E_g-p_0)^2-E_q^2\right] (E_r-p_0-E_l) } \right] = \frac{\pi}{2} \bigg[ (E_r-E_g)(E_q-E_g) +(\mathbf k\cdot\mathbf l)\nonumber\\ 
&+\frac{(E_r-E_g)(\mathbf r\cdot\mathbf q)}{E_rE_q}(E_q-E_g) +\frac{(\mathbf k\cdot\mathbf l)(\mathbf r\cdot\mathbf q)}{E_rE_q} \bigg]  \frac{ \left(\frac12-n_F(E_r-E_g)\right) \delta(E_r+E_q-E_g-p_0) } { \left[(E_r-E_g)^2-k^2\right](E_g-E_q-E_l) }
\label{91l}\\
&\Im\left[ -\frac{ \left[\mathcal M_+(k_0)\right]_{k_0=p_0+E_q} \left(\frac12-n_F(E_q)\right)} {2E_q\left[(p_0+E_q)^2-k^2\right](E_q-E_g+E_l)(p_0+E_q-E_r-E_g) } \right]  = \frac{\pi}{2} \bigg[ (E_r+E_g)(E_q-E_g)-(\mathbf k\cdot\mathbf l) \nonumber\\
&-\frac{(E_r+E_g)(\mathbf r\cdot\mathbf q)}{E_r\, E_q}(E_q-E_g) +\frac{(\mathbf k\cdot\mathbf l)(\mathbf r\cdot\mathbf q)}{E_r\, E_q} \bigg]  \frac{ \left(\frac12-n_F(E_q)\right)\delta(E_r+E_g-E_q-p_0) } { \left[(E_r+E_g)^2-k^2\right](E_q-E_g+E_l) }, 
\label{91m}\\
&\Im\left[ -\frac{ \left[\mathcal M_+(k_0)\right]_{k_0=p_0+E_q} \left(\frac12-n_F(E_q)\right) } { 2E_q\left[(p_0+E_q)^2-k^2\right](E_q-E_g-E_l)(p_0+E_q-E_r-E_g) } \right] = \frac{\pi}{2} \bigg[ (E_r+E_g)(E_q-E_g)-(\mathbf k\cdot\mathbf l) \nonumber\\
&- \frac{(E_r+E_g)(\mathbf r\cdot\mathbf q)}{E_r\, E_q}(E_q-E_g)+\frac{(\mathbf k\cdot\mathbf l)(\mathbf r\cdot\mathbf q)}{E_r\, E_q} \bigg]\frac{ \left(\frac12-n_F(E_q)\right)\delta(E_r+E_g-E_q-p_0) } { \left[(E_r+E_g)^2-k^2\right](E_q-E_g-E_l) }, 
\label{91n}\\
&\Im\left[ -\frac{ \left[\mathcal M_-(k_0)\right]_{k_0=p_0-E_q} \left(\frac12-n_F(E_q)\right) } { 2E_q\left[(p_0-E_q)^2-k^2\right](-E_q-E_g+E_l)(p_0+E_r-E_q-E_g) } \right] 
= \frac{\pi}{2} \bigg[ (E_g-E_r)(E_g+E_q)+(\mathbf k\cdot\mathbf l) \nonumber\\
&-\frac{(E_g-E_r)(\mathbf r\cdot\mathbf q)}{E_r\, E_q}(E_g+E_q)-\frac{(\mathbf k\cdot\mathbf l)(\mathbf r\cdot\mathbf q)}{E_r\, E_q} \bigg] \frac{ \left(\frac12-n_F(E_q)\right)\delta(E_g+E_q-E_r-p_0) } { \left[(E_g-E_r)^2-k^2\right](-E_q-E_g+E_l) }, 
\label{91o}\\
&\Im\left[ -\frac{ \left[\mathcal M_-(k_0)\right]_{k_0=p_0-E_q} \left(\frac12-n_F(E_q)\right) } { 2E_q\left[(p_0-E_q)^2-k^2\right](-E_q-E_g-E_l)(p_0+E_r-E_q-E_g) } \right]  = -\frac{\pi}{2} \bigg[ (E_g-E_r)(E_g+E_q)+(\mathbf k\cdot\mathbf l) \nonumber\\
&- \frac{(E_g-E_r)(\mathbf r\cdot\mathbf q)}{E_r\, E_q}(E_g+E_q)-\frac{(\mathbf k\cdot\mathbf l)(\mathbf r\cdot\mathbf q)}{E_r\, E_q} \bigg] \frac{ \left(\frac12-n_F(E_q)\right)\delta(E_g+E_q-E_r-p_0) } { \left[(E_g-E_r)^2-k^2\right](E_q+E_g+E_l) }, 
\label{91p}\\
&\Im\left[ -\frac{ \left[\mathcal M_+(k_0)\right]_{k_0=p_0-E_q} \left(\frac12-n_F(E_q)\right) } { 2E_q\left[(p_0-E_q)^2-k^2\right](-E_q+E_g+E_l)(p_0+E_g-E_q-E_r) } \right] 
= \frac{\pi}{2} \bigg[ (E_r-E_g)(E_q-E_g)+(\mathbf k\cdot\mathbf l)  \nonumber\\
&+\frac{(E_r-E_g)(\mathbf r\cdot\mathbf q)}{E_r\,E_q}(E_q-E_g)+\frac{(\mathbf k\cdot\mathbf l)(\mathbf r\cdot\mathbf q)}{E_r\,E_q} \bigg]\frac{ \left(\frac12-n_F(E_q)\right)\delta(E_r+E_q-E_g-p_0) } { \left[(E_r-E_g)^2-k^2\right](-E_q+E_g+E_l) }, 
\label{91q}\\
&\Im\left[ -\frac{ \left[\mathcal M_+(k_0)\right]_{k_0=p_0-E_q} \left(\frac12-n_F(E_q)\right) } { 2E_q\left[(p_0-E_q)^2-k^2\right](-E_q+E_g-E_l)(p_0+E_g-E_q-E_r) } \right]  = \frac{\pi}{2} \bigg[ (E_r-E_g)(E_q-E_g)+(\mathbf k\cdot\mathbf l) \nonumber\\
&+ \frac{(E_r-E_g)(\mathbf r\cdot\mathbf q)}{E_r \,E_q}(E_q-E_g)+\frac{(\mathbf k\cdot\mathbf l)(\mathbf r\cdot\mathbf q)}{E_r\, E_q} \bigg]\frac{ \left(\frac12-n_F(E_q)\right)\delta(E_r+E_q-E_g-p_0) } { \left[(E_r-E_g)^2-k^2\right](-E_q+E_g-E_l) },
\label{91r}\\
&\Im\left[ -\frac{ \left[\mathcal M_+(k_0)\right]_{k_0=k} \left(\frac12-n_F(k)\right) } { 2k\left[(k-p_0)^2-E_q^2\right](k-p_0-E_g+E_l)(k-E_r-E_g) } \right] = \frac{\pi}{2}\bigg[ (E_l-E_g)(k+E_l-E_g-E_r)+\frac{(\mathbf k\cdot\mathbf l)}{k}(E_l-E_g) \nonumber \\
&+\frac{(\mathbf r\cdot\mathbf q)}{E_r}(k+E_l-E_g-E_r)+\frac{(\mathbf k\cdot\mathbf l)(\mathbf r\cdot\mathbf q)}{k\,E_r} \bigg]\frac{\left(\frac12-n_F(k)\right)\delta(p_0-k-E_l+E_g)}{(k-E_r-E_g)\left[E_q^2-(E_l-E_g)^2\right]},
\label{91s}\\
&\Im\left[ -\frac{ \left[\mathcal M_-(k_0)\right]_{k_0=k} \left(\frac12-n_F(k)\right) } { 2k\left[(k-p_0)^2-E_q^2\right](k-p_0-E_g+E_l)(k+E_r-E_g) } \right] = \frac{\pi}{2}\bigg[ (E_l-E_g)(E_r+k+E_l-E_g)+\frac{(\mathbf k\cdot\mathbf l)}{k}(E_l-E_g)\nonumber\\
&-\frac{(\mathbf r\cdot\mathbf q)}{E_r}(E_r+k+E_l-E_g)-\frac{(\mathbf k\cdot\mathbf l)(\mathbf r\cdot\mathbf q)}{k\,E_r} \bigg]\frac{\left(\frac12-n_F(k)\right)\delta(p_0-k-E_l+E_g)}{(k+E_r-E_g)\left[E_q^2-(E_l-E_g)^2\right]},
\label{91t}\\
&\Im\left[ \frac{ \left[\mathcal M_+(k_0)\right]_{k_0=-k} \left(\frac12-n_F(-k)\right) } { 2k\left[(-k-p_0)^2-E_q^2\right](-k-p_0+E_g+E_l)(-k-E_r+E_g) } \right] = -\frac{\pi}{2}\bigg[ (E_l+E_g)(E_r-E_l-E_g+k)+\frac{(\mathbf k\cdot\mathbf l)}{k} (E_l+E_g) \nonumber\\
&+\frac{(\mathbf r\cdot\mathbf q)}{E_r}(E_r-E_l-E_g+k)+\frac{(\mathbf k\cdot\mathbf l)(\mathbf r\cdot\mathbf q)}{k\,E_r} \bigg]\frac{\left(\frac12-n_F(-k)\right)\delta(E_g+E_l-k-p_0)}{(E_g-E_r-k)\left[E_q^2-(E_l+E_g)^2\right]},
\label{91u}\\
&\Im\left[ \frac{ \left[\mathcal M_-(k_0)\right]_{k_0=-k} \left(\frac12-n_F(-k)\right) } { 2k\left[(-k-p_0)^2-E_q^2\right](-k-p_0+E_g+E_l)(-k+E_r+E_g) } \right] = -\frac{\pi}{2}\bigg[ -(E_l+E_g)(E_r+E_l+E_g-k)+\frac{(\mathbf k\cdot\mathbf l)}{k}(E_l+E_g) \nonumber\\
&+\frac{(\mathbf r\cdot\mathbf q)}{E_r}(E_r+E_l+E_g-k)-\frac{(\mathbf k\cdot\mathbf l)(\mathbf r\cdot\mathbf q)}{k\,E_r} \bigg]\frac{\left(\frac12-n_F(-k)\right)\delta(E_g+E_l-k-p_0)}{(E_g+E_r-k)\left[E_q^2-(E_l+E_g)^2\right]},
\label{91v}\\
&\Im\left[ -\frac{ \left[\mathcal M_+(k_0)\right]_{k_0=k} \left(\frac12-n_F(k)\right) } { 2k\left[(k-p_0)^2-E_q^2\right](k-p_0+E_g-E_l)(k-E_r+E_g) } \right] = \frac{\pi}{2}\bigg[ (E_l-E_g)(E_r-k-E_g+E_l)-\frac{(\mathbf k\cdot\mathbf l)}{k}(E_l-E_g) \nonumber\\
&-\frac{(\mathbf r\cdot\mathbf q)}{E_r}(E_r-k-E_g+E_l)+\frac{(\mathbf k\cdot\mathbf l)(\mathbf r\cdot\mathbf q)}{k\,E_r} \bigg]\frac{\left(\frac12-n_F(k)\right)\delta(p_0-k-E_g+E_l)}{(k-E_r+E_g)\left[E_q^2-(E_l-E_g)^2\right]},
\label{91w}\\
&\Im\left[ -\frac{ \left[\mathcal M_-(k_0)\right]_{k_0=k} \left(\frac12-n_F(k)\right) } { 2k\left[(k-p_0)^2-E_q^2\right](k-p_0+E_g-E_l)(k+E_r+E_g) } \right] = \frac{\pi}{2}\bigg[ -(E_l-E_g)(E_r+k+E_g-E_l)-\frac{(\mathbf k\cdot\mathbf l)}{k}(E_l-E_g)\nonumber\\
&-\frac{(\mathbf r\cdot\mathbf q)}{E_r}(E_r+k+E_g-E_l)-\frac{(\mathbf k\cdot\mathbf l)(\mathbf r\cdot\mathbf q)}{k\,E_r} \bigg]\frac{\left(\frac12-n_F(k)\right)\delta(p_0-k-E_g+E_l)}{(k+E_r+E_g)\left[E_q^2-(E_l-E_g)^2\right]}, 
\label{91x}\\
&\Im \left[ -\frac{ \left[\mathcal M_+(k_0)\right]_{k_0=p_0+E_g-E_l} \left(\frac12-n_F(E_g-E_l)\right) } { \left[(p_0+E_g-E_l)^2-k^2\right] \left[(E_g-E_l)^2-E_q^2\right] (p_0-E_r-E_l)} \right] = -\frac{\pi}{2} \left[ (E_l-E_g)(k+E_l-E_g-E_r) +\frac{(\mathbf k\cdot\mathbf l)}{k}(E_l-E_g) \right. \nonumber\\
&\left. +\frac{(\mathbf r\cdot\mathbf q)}{E_r}(k+E_l-E_g-E_r) +\frac{(\mathbf k\cdot\mathbf l)(\mathbf r\cdot\mathbf q)}{k\,E_r}
\right] \frac{ \left(\frac12-n_F(E_g-E_l)\right) \delta(p_0-k-E_l+E_g) } { \left[E_q^2-(E_l-E_g)^2\right] (k-E_g-E_r)}, 
\label{91y}\\
&\Im \left[ -\frac{ \left[\mathcal M_-(k_0)\right]_{k_0=p_0+E_g-E_l} \left(\frac12-n_F(E_g-E_l)\right) } { \left[(p_0+E_g-E_l)^2-k^2\right] \left[(E_g-E_l)^2-E_q^2\right] (p_0+E_r-E_l)} \right] = -\frac{\pi}{2} \left[ (E_l-E_g)(E_r+k+E_l-E_g) +\frac{(\mathbf k\cdot\mathbf l)}{k}(E_l-E_g) \right. \nonumber\\
&\left. -\frac{(\mathbf r\cdot\mathbf q)}{E_r}(E_r+k+E_l-E_g) -\frac{(\mathbf k\cdot\mathbf l)(\mathbf r\cdot\mathbf q)}{k\,E_r} \right]  \frac{ \left(\frac12-n_F(E_g-E_l)\right) \delta(p_0-k-E_l+E_g) } { \left[E_q^2-(E_l-E_g)^2\right] (k+E_r-E_g) }, 
\label{91z}\\
&\Im \left[ -\frac{ \left[\mathcal M_+(k_0)\right]_{k_0=p_0-E_g-E_l} \left(\frac12-n_F(-E_g-E_l)\right) } { \left[(p_0-E_g-E_l)^2-k^2\right] \left[(E_g+E_l)^2-E_q^2\right] (p_0-E_r-E_l)}  \right] = -\frac{\pi}{2} \left[ (E_l+E_g)(E_r-E_l-E_g+k) +\frac{(\mathbf k\cdot\mathbf l)}{k}(E_l+E_g) \right. \nonumber\\
&\left. +\frac{(\mathbf r\cdot\mathbf q)}{E_r}(E_r-E_l-E_g+k) +\frac{(\mathbf k\cdot\mathbf l)(\mathbf r\cdot\mathbf q)}{k\,E_r} \right]  \frac{ \left(\frac12-n_F(-E_g-E_l)\right) \delta(E_g+E_l-k-p_0) } { \left[E_q^2-(E_l+E_g)^2\right] (k+E_r-E_g)}, 
\label{91aa}\\
&\Im \left[ -\frac{ \left[\mathcal M_-(k_0)\right]_{k_0=p_0-E_g-E_l} \left(\frac12-n_F(-E_g-E_l)\right) } { \left[(p_0-E_g-E_l)^2-k^2\right] \left[(E_g+E_l)^2-E_q^2\right] (p_0+E_r-E_l)} \right] = -\frac{\pi}{2} \left[ -(E_l+E_g)(E_r+E_l+E_g-k) +\frac{(\mathbf k\cdot\mathbf l)}{k}(E_l+E_g) \right. \nonumber\\
&\left. +\frac{(\mathbf r\cdot\mathbf q)}{E_r}(E_r+E_l+E_g-k) -\frac{(\mathbf k\cdot\mathbf l)(\mathbf r\cdot\mathbf q)}{k\,E_r} \right]  \frac{ \left(\frac12-n_F(-E_g-E_l)\right) \delta(E_g+E_l-k-p_0) } { \left[E_q^2-(E_l+E_g)^2\right] (k-E_r-E_g)}, 
\label{91bb}\\
&\Im \left[ -\frac{ \left[\mathcal M_+(k_0)\right]_{k_0=p_0-E_g+E_l} \left(\frac12-n_F(-E_g+E_l)\right) } { \left[(p_0-E_g+E_l)^2-k^2\right] \left[(E_l-E_g)^2-E_q^2\right] (p_0-E_r+E_l)} \right] = -\frac{\pi}{2} \left[ (E_l-E_g)(E_r-k-E_g+E_l) -\frac{(\mathbf k\cdot\mathbf l)}{k}(E_l-E_g) \right. \nonumber\\
&\left. -\frac{(\mathbf r\cdot\mathbf q)}{E_r}(E_r-k-E_g+E_l) +\frac{(\mathbf k\cdot\mathbf l)(\mathbf r\cdot\mathbf q)}{k\,E_r} \right] \frac{ \left(\frac12-n_F(-E_g+E_l)\right) \delta(p_0-k-E_g+E_l) } { \left[E_q^2-(E_l-E_g)^2\right] (k+E_g-E_r)}, 
\label{91cc}\\
&\Im \left[ -\frac{ \left[\mathcal M_-(k_0)\right]_{k_0=p_0-E_g+E_l} \left(\frac12-n_F( -E_g+E_l)\right) } { \left[(p_0-E_g+E_l)^2-k^2\right] \left[(E_l-E_g)^2-E_q^2\right] (p_0+E_r+E_l)} \right] = -\frac{\pi}{2} \left[ -(E_l-E_g)(E_r+k+E_g-E_l) -\frac{(\mathbf k\cdot\mathbf l)}{k}(E_l-E_g) \right. \nonumber\\
&\left. -\frac{(\mathbf r\cdot\mathbf q)}{E_r}(E_r+k+E_g-E_l) -\frac{(\mathbf k\cdot\mathbf l)(\mathbf r\cdot\mathbf q)}{k\,E_r} \right] \frac{ \left(\frac12-n_F(-E_g+E_l)\right) \delta(p_0-k-E_g+E_l) } { \left[E_q^2-(E_l-E_g)^2\right] (k+E_r+E_g)}, 
\label{91dd}\\
&\Im \left[ -\frac{ \left[\mathcal M_+(k_0)\right]_{k_0=k} \left(\frac12-n_F(k)\right) } { 2k\left[(k-p_0)^2-E_q^2\right] (k-p_0-E_g+E_l) (p_0-E_r-E_l)} \right] = \frac{\pi}{2} \left[ (E_l-E_g)(k+E_l-E_g-E_r) +\frac{(\mathbf k\cdot\mathbf l)}{k}(E_l-E_g) \right. \nonumber\\
&\left. +\frac{(\mathbf r\cdot\mathbf q)}{E_r}(k+E_l-E_g-E_r) +\frac{(\mathbf k\cdot\mathbf l)(\mathbf r\cdot\mathbf q)}{ k \,E_r} \right] \frac{ \left(\frac12-n_F(k)\right) \delta(p_0-k-E_l+E_g) } { \left[E_q^2-(E_l-E_g)^2\right] (k-E_r-E_g)}, 
\label{91ee}\\
&\Im \left[ -\frac{ \left[\mathcal M_-(k_0)\right]_{k_0=k} \left(\frac12-n_F(k)\right) } { 2k\left[(k-p_0)^2-E_q^2\right] (k-p_0-E_g+E_l) (p_0+E_r-E_l)} \right] = \frac{\pi}{2} \left[ (E_l-E_g)(E_r+k+E_l-E_g) +\frac{(\mathbf k\cdot\mathbf l)}{k}(E_l-E_g) \right. \nonumber\\
&\left. -\frac{(\mathbf r\cdot\mathbf q)}{E_r}(E_r+k+E_l-E_g) -\frac{(\mathbf k\cdot\mathbf l)(\mathbf r\cdot\mathbf q)}{k \,E_r} \right] \frac{ \left(\frac12-n_F(k)\right) \delta(p_0-k-E_l+E_g) } { \left[E_q^2-(E_l-E_g)^2\right] (k-E_g+E_r)}, 
\label{91ff}\\
&\Im \left[ \frac{ \left[\mathcal M_+(k_0)\right]_{k_0=-k} \left(\frac12-n_F(-k)\right) } { 2k\left[(-k-p_0)^2-E_q^2\right] (-k-p_0+E_g+E_l) (p_0-E_r-E_l)} \right] = \frac{\pi}{2} \left[ (E_l+E_g)(E_r-E_l-E_g+k) +\frac{(\mathbf k\cdot\mathbf l)}{k}(E_l+E_g) \right. \nonumber\\
&\left. +\frac{(\mathbf r\cdot\mathbf q)}{E_r}(E_r-E_l-E_g+k) +\frac{(\mathbf k\cdot\mathbf l)(\mathbf r\cdot\mathbf q)}{k \,E_r} 
\right]  \frac{ \left(\frac12-n_F(-k)\right) \delta(E_g+E_l-k-p_0) } { \left[E_q^2-(E_l+E_g)^2\right] (k+E_r-E_g)}, 
\label{91gg}\\
&\Im \left[ \frac{ \left[\mathcal M_-(k_0)\right]_{k_0=-k} \left(\frac12-n_F(-k)\right) } { 2k\left[(-k-p_0)^2-E_q^2\right] (-k-p_0+E_g+E_l) (p_0+E_r-E_l)} \right] = \frac{\pi}{2} \left[ -(E_l+E_g)(E_r+E_l+E_g-k) +\frac{(\mathbf k\cdot\mathbf l)}{k}(E_l+E_g) \right. \nonumber\\
&\left. +\frac{(\mathbf r\cdot\mathbf q)}{E_r}(E_r+E_l+E_g-k) -\frac{(\mathbf k\cdot\mathbf l)(\mathbf r\cdot\mathbf q)}{k \,E_r} \right]\frac{ \left(\frac12-n_F(-k)\right) \delta(E_g+E_l-k-p_0) } { \left[E_q^2-(E_l+E_g)^2\right] (k-E_r-E_g)}, 
\label{91hh}\\
&\Im \left[ -\frac{ \left[\mathcal M_+(k_0)\right]_{k_0=k} \left(\frac12-n_F(k)\right) } { 2k\left[(k-p_0)^2-E_q^2\right] (k-p_0+E_g-E_l) (p_0-E_r+E_l)} \right] = \frac{\pi}{2} \left[ (E_l-E_g)(E_r-k-E_g+E_l) -\frac{(\mathbf k\cdot\mathbf l)}{k}(E_l-E_g) \right. \nonumber\\
&\left. -\frac{(\mathbf r\cdot\mathbf q)}{E_r}(E_r-k-E_g+E_l) +\frac{(\mathbf k\cdot\mathbf l)(\mathbf r\cdot\mathbf q)}{k \, E_r} \right] \frac{ \left(\frac12-n_F(k)\right) \delta(p_0-k-E_g+E_l) } { \left[E_q^2-(E_l-E_g)^2\right] (k+E_g-E_r)}, 
\label{91ii}\\
&\Im \left[ -\frac{ \left[\mathcal M_-(k_0)\right]_{k_0=k} \left(\frac12-n_F(k)\right) } { 2k\left[(k-p_0)^2-E_q^2\right] (k-p_0+E_g-E_l) (p_0+E_r+E_l)} \right] = \frac{\pi}{2} \left[ -(E_l-E_g)(E_r+k+E_g-E_l) -\frac{(\mathbf k\cdot\mathbf l)}{k}(E_l-E_g) \right. \nonumber\\
&\left. -\frac{(\mathbf r\cdot\mathbf q)}{E_r}(E_r+k+E_g-E_l) -\frac{(\mathbf k\cdot\mathbf l)(\mathbf r\cdot\mathbf q)}{k\, E_r} \right] \frac{ \left(\frac12-n_F(k)\right) \delta(p_0-k-E_g+E_l) } { \left[E_q^2-(E_l-E_g)^2\right] (k+E_r+E_g)}, 
\label{91jj}\\
&\Im \left[ -\frac{ \left[A_2(k_0)\right]_{k_0=p_0+E_g-E_l} \left(\frac12-n_F(E_g-E_l)\right) } { \left[(p_0+E_g-E_l)^2-k^2\right] \left[(E_g-E_l)^2-E_q^2\right] } \right] = \frac{\pi}{2}\, \frac{ (E_l-E_g) \left(\frac12-n_F(E_g-E_l)\right) \delta(p_0-k-E_l+E_g) } { E_q^2-(E_l-E_g)^2 }, 
\label{91kk}\\
&\Im \left[ -\frac{ \left[A_2(k_0)\right]_{k_0=p_0-E_g+E_l} \left(\frac12-n_F(-E_g+E_l)\right) } { \left[(p_0-E_g+E_l)^2-k^2\right] \left[(-E_g+E_l)^2-E_q^2\right] } \right] = -\frac{\pi}{2}\, \frac{ (E_l-E_g) \left(\frac12-n_F(-E_g+E_l)\right) \delta(p_0-k-E_g+E_l) } { E_q^2-(E_l-E_g)^2 }, 
\label{91ll}\\
&\Im \left[ -\frac{ \left[A_2(k_0)\right]_{k_0=p_0-E_g-E_l} \left(\frac12-n_F(-E_g-E_l)\right) } { \left[(p_0-E_g-E_l)^2-k^2\right] \left[(-E_g-E_l)^2-E_q^2\right] } \right] = \frac{\pi}{2}\, \frac{ (E_g+E_l) \left(\frac12-n_F(-E_g-E_l)\right) \delta(E_g+E_l-k-p_0) } { E_q^2-(E_g+E_l)^2 },
\label{91mm}\\
&\Im \left[ -\frac{ \left[A_2(k_0)\right]_{k_0=k} \left(\frac12-n_F(k)\right) } { 2k\left[(k-p_0)^2-E_q^2\right] (k-p_0-E_g+E_l) } \right] = -\frac{\pi}{2}\, \frac{ (E_l-E_g) \left(\frac12-n_F(k)\right) \delta(p_0-k-E_l+E_g) } { E_q^2-(E_l-E_g)^2 }, 
\label{91nn}\\
&\Im \left[ -\frac{ \left[A_2(k_0)\right]_{k_0=k} \left(\frac12-n_F(k)\right) } { 2k\left[(k-p_0)^2-E_q^2\right] (k-p_0+E_g-E_l) } \right] = \frac{\pi}{2}\, \frac{ (E_l-E_g) \left(\frac12-n_F(k)\right) \delta(p_0-k-E_g+E_l) } { E_q^2-(E_l-E_g)^2 }, 
\label{91oo}\\
&\Im \left[ \frac{ \left[A_2(k_0)\right]_{k_0=-k} \left(\frac12-n_F(-k)\right) } { 2k\left[(-k-p_0)^2-E_q^2\right] (-k-p_0+E_g+E_l) } \right]= -\frac{\pi}{2}\, \frac{ (E_g+E_l) \left(\frac12-n_F(-k)\right) \delta(E_g+E_l-k-p_0) } { E_q^2-(E_g+E_l)^2 }.
\label{91pp}
\end{align}
\end{subequations}

\section{Ward identity}
\label{ward_identity}
\revision{We finally verify the Ward identity for the complete two-loop photon self-energy. Gauge invariance requires $P_\mu\Pi^{\mu\nu}(P)=0$ after summing all diagrams at this order. Consider first the diagram shown in the left panel of Fig.~\ref{photon_self_1}:}
\begin{align}
iP_\mu\Pi_{Ia}^{\mu\nu}(P) =& -\int_K \mathrm{Tr}\Big[ \slashed{P}\, S(K)\Sigma(K)S(K) \gamma^\nu S(K-P) \Big](-ie)^2 \nonumber\\
=&- i\int_K \mathrm{Tr}\Big[ \big\{S^{-1}(K)-S^{-1}(K-P)\big\} S(K)\Sigma(K)S(K) \gamma^\nu S(K-P) \Big](-ie)^2 \nonumber\\
=& -i\int_K \mathrm{Tr}\Big[ \Sigma(K)S(K)\gamma^\nu \big\{S(K-P)-S(K)\big\} \Big](-ie)^2 \nonumber\\
=& -i\int_K \mathrm{Tr}\Big[ \Sigma(K)S(K)\gamma^\nu S(K-P) \Big](-ie)^2 + i\int_K \mathrm{Tr}\Big[ \Sigma(K)S(K)\gamma^\nu S(K) \Big](-ie)^2 ,
\end{align}
where the one-loop quark self-energy is given by 
\begin{eqnarray}
\Sigma(K)= \int_R\bigg\{(-ig \gm_\alpha t_a) S(R)(-i g \gm_\beta t_b)D^{\alpha\beta}(R-K)\bigg\}.
\end{eqnarray}
We have used the identity $\slashed{P}=i[S^{-1}(K)-S^{-1}(K-P)]$, which follows immediately from the inverse propagators $S^{-1}(K)=-i \,\slashed K$, together with cyclicity of the trace. The corresponding crossed two-loop diagram, right panel of Fig.~\ref{photon_self_1}, in which the self-energy insertion lies on the opposite side of the photon vertex, gives
\begin{align}
iP_\mu\Pi_{Ib}^{\mu\nu}(P) =&- i\int_K \mathrm{Tr}\Big[ \Sigma(K-P)S(K-P)\gamma^\nu S(K-P) \Big](-ie)^2 +i \int_K \mathrm{Tr}\Big[ \Sigma(K-P)S(K)\gamma^\nu S(K-P) \Big](-ie)^2 .
\end{align}
\revision{Adding the two contributions gives}
\begin{eqnarray}
iP_\mu[\Pi_{I}^{\mu\nu}]=   iP_\mu[\Pi_{Ia}^{\mu\nu}(P)+\Pi_{Ib}^{\mu\nu}(P)]=i\int_K \mathrm{Tr}\Big[ \Sigma(K-P)S(K)\gamma^\nu S(K-P) \Big](-ie)^2-i\int_K \mathrm{Tr}\Big[ \Sigma(K)S(K)\gamma^\nu S(K-P) \Big](-ie)^2.
\end{eqnarray}
The vertex-correction diagram Fig.~\ref{photon_self_2} produces:
\begin{align}
iP_\mu\Pi^{\mu\nu}_{II}(P) =& -i\int_{K,R} \mathrm{Tr}\Big[\Big\{ S^{-1}(K)-S^{-1}(K-P) \Big\}S(K) \gamma^\alpha S(R)\gamma^\nu
S(R-P)\gamma^\beta S(K-P) \Big] D_{\alpha\beta}(K-R)(-ie)^2(-ig)^2 \nonumber\\
=& -i\int_{K,R} \mathrm{Tr}\Big[ \gamma^\alpha S(R)\gamma^\nu S(R-P)\gamma^\beta S(K-P) \Big] D_{\alpha\beta}(K-R) (-ie)^2(-ig)^2 \nonumber\\
&+i \int_{K,R} \mathrm{Tr}\Big[ \gamma^\alpha S(R)\gamma^\nu S(R-P)\gamma^\beta S(K) \Big] D_{\alpha\beta}(K-R) (-ie)^2(-ig)^2 \nonumber\\
=& -i\int_{K,R} \mathrm{Tr}\Big[ S(R)\gamma^\nu S(R-P) \gamma^\beta S(K-P)\gamma^\alpha \Big] D_{\alpha\beta}(K-R) (-ie)^2(-ig)^2
\nonumber\\
&+ i\int_{K,R} \mathrm{Tr}\Big[ S(R)\gamma^\nu S(R-P) \gamma^\beta S(K)\gamma^\alpha \Big] D_{\alpha\beta}(K-R) (-ie)^2 (-ig)^2.
\end{align}
In the first term, changing the variable from $K\rightarrow K+P$, we get 
\begin{align}
iP_\mu\Pi^{\mu\nu}_{II}(P) =& -i\int_R \mathrm{Tr}\Big[ S(R)\gamma^\nu S(R-P)\Sigma(R-P) \Big](-ie)^2 +i \int_R \mathrm{Tr}\Big[ S(R)\gamma^\nu S(R-P)\Sigma(R) \Big](-ie)^2 .
\end{align}
Finally, by changing the variable  $R\rightarrow K$, and we get
\begin{eqnarray}
iP_\mu[\Pi_{I}^{\mu\nu}(P)+\Pi_{II}^{\mu\nu}(P)]=0.
\end{eqnarray}
\revision{The resulting cancellation confirms the Ward identity and provides a nontrivial consistency check of the two-loop calculation.
}

\end{document}